\documentstyle[equations,cltp]{article}
\def\hf{\hfill}
\def\ie{{\it i.e.}}

\begin{document}
\input psfig.sty
\pagestyle{empty}
\begin{titlepage}
\rightline{SCIPP 95/21}
\rightline{SLAC-PUB-95-6893   }
\rightline{BNL-61748       }
\rightline{LBL- 37186      }
\rightline{April 1995}
\vskip 1cm
\centerline{\Large\sc
Electroweak Symmetry Breaking and Beyond the Standard Model}
\vskip1pc
\centerline{\large\sc Timothy L. Barklow}
\centerline{\normalsize\it Stanford Linear Accelerator Center, Stanford, CA
94309}
\vspace{6pt}
\centerline{\large\sc Sally Dawson}
\centerline{\normalsize\it Brookhaven National Laboratory, Upton, NY 11973}
\vspace{6pt}
\centerline{\large\sc Howard E. Haber}
\centerline{\normalsize\it Santa Cruz Institute for Particle Physics,
University of California, Santa Cruz, CA 95064}
\vspace{6pt}
\centerline{\large\sc James L. Siegrist}
\centerline{\normalsize\it Lawrence Berkeley Laboratory, University of
California,
           Berkeley, CA 94720}

\vskip1pc
\begin{center}
{\large Working Group Members}\vspace*{1ex}
\begin{tabular}{llll}
H. Aihara, & {\it Lawrence Berkeley Laboratory}
& S.P. Martin, & {\it University of Michigan}  \\
H. Baer, & {\it Florida State University}
& H. Murayama, & {\it Lawrence Berkeley Laboratory}  \\
U. Baur, & {\it SUNY, Buffalo}
& J. Ng, &  {\it TRIUMF, Vancouver BC}        \\
R.S. Chivukula, & {\it Boston University}
& S.J. Parke, &  {\it Fermilab}                      \\
M. Cveti\v c, & {\it University of Pennsylvania}
& M.E. Peskin, &  {\it Stanford Linear Accelerator Center}  \\
A. Djouadi, & {\it University of Montreal}
& H. Pois, &  {\it University of California, Davis}
 \\
M. Drees, & {\it University of Wisconsin}
& T.G. Rizzo, &  {\it Stanford Linear Accelerator Center}   \\
S. Errede, & {\it University of Illinois}
& R. Rosenfeld, &  {\it Boston University}                \\
S. Godfrey, & {\it Carleton University}
& E.H. Simmons, &  {\it Boston University}                  \\
M. Golden, & {\it Harvard University}
& A. Stange, &  {\it Brookhaven National Laboratory}                      \\
J.F. Gunion, & {\it University of California, Davis}
& T. Takeuchi, &  {\it Fermilab}                          \\
T. Han, & {\it University of California, Davis}
& X. Tata, &  {\it University of Hawaii}                  \\
J.L.  Hewett, & {\it Stanford Linear Accelerator Center}
& J. Terning, &  {\it Boston University}                  \\
C.T. Hill, & {\it Fermilab}
& S. Thomas, &  {\it Stanford Linear Accelerator Center}  \\
D. London, & {\it University of Montreal}
& S. Willenbrock, &  {\it University of Illinois}         \\
T.W. Markiewicz, & {\it Stanford Linear Accelerator Center}
& D. Zeppenfeld, &  {\it University of Wisconsin}         \\
%
 \end{tabular}
\end{center}
\vskip1pc
\begin{table*}[hb] \large
\centerline{Abstract}
\vskip6pt
This is a report of the Electroweak Symmetry Breaking and Beyond the
Standard Model Working Group which was prepared for the Division of
Particles and Fields Committee for Long Term Planning.
We study the phenomenology of electroweak symmetry breaking
and attempt to quantify the ``physics reach" of present and
future colliders.  Our investigations encompass the
Standard Model (with one doublet of Higgs scalars) and
approaches to physics beyond the Standard Model.  These
include models of low energy
supersymmetry, technicolor and other approaches to
dynamical electroweak symmetry breaking, and
a variety of extensions of the Standard Model
with new particles and interactions
({\it e.g.}, non-minimal Higgs sectors, new
gauge bosons and/or exotic fermions, {\it etc.}).  Signals of
new physics in precision measurements
arising from virtual processes (which can
result, for example, in ``anomalous'' couplings of Standard Model
particles) are also considered.
Finally, we examine experimental issues associated with the study
of electroweak symmetry breaking and the search for new
physics at present and future hadron and $e^+e^-$ colliders.
\vskip-1pc
\end{table*}
 \end{titlepage}
\pagestyle{plain}
\title{Electroweak Symmetry Breaking and Beyond the Standard Model}

\author{Tim Barklow}
\address{Stanford Linear Accelerator Center, Stanford, CA 94309}

\author{Sally Dawson}
\address{Brookhaven National Laboratory, Upton, NY 11973}
\author{Howard E. Haber}
\address{Santa Cruz Institute for Particle Physics, Unversity of
         California, Santa Cruz, CA 95064}
\author{Jim Siegrist}
\address{Lawrence Berkeley Laboratory, University of California,
           Berkeley, CA 94720}

 \twocolumn[\maketitle\par
\begin{center}
{\large Working Group Members}\vspace*{1ex}
\begin{tabular}{@{}l@{}l@{}l@{}l@{}}
H. Aihara, & {\it Lawrence Berkeley Laboratory}
& S.P. Martin, & {\it University of Michigan}  \\
H. Baer, & {\it Florida State University}
& H. Murayama, & {\it Lawrence Berkeley Laboratory}  \\
U. Baur, & {\it SUNY, Buffalo}
& J. Ng, &  {\it TRIUMF, Vancouver BC}        \\
R.S. Chivukula, & {\it Boston University}
& S.J. Parke, &  {\it Fermilab}                      \\
M. Cveti\v c, & {\it University of Pennsylvania}
& M.E. Peskin, &  {\it Stanford Linear Accelerator Center}  \\
A. Djouadi, & {\it University of Montreal}
& H. Pois, &  {\it University of California, Davis}
 \\
M. Drees, & {\it University of Wisconsin}
& T.G. Rizzo, &  {\it Stanford Linear Accelerator Center}   \\
S. Errede, & {\it University of Illinois}
& R. Rosenfeld, &  {\it Boston University}                \\
S. Godfrey, & {\it Carleton University}
& E.H. Simmons, &  {\it Boston University}                  \\
M. Golden, & {\it Harvard University}
& A. Stange, &  {\it Brookhaven National Laboratory}                      \\
J.F. Gunion, & {\it University of California, Davis}
& T. Takeuchi, &  {\it Fermilab}                          \\
T. Han, & {\it University of California, Davis}
& X. Tata, &  {\it University of Hawaii}                  \\
J.L.  Hewett, & {\it Stanford Linear Accelerator Center}
& J. Terning, &  {\it Boston University}                  \\
C.T. Hill, & {\it Fermilab}
& S. Thomas, &  {\it Stanford Linear Accelerator Center}  \\
D. London, & {\it University of Montreal}
& S. Willenbrock, &  {\it University of Illinois}         \\
T.W. Markiewicz, & {\it Stanford Linear Accelerator Center}
& D. Zeppenfeld, &  {\it University of Wisconsin}         \\
\hspace*{1in}& \hspace*{2.8in}& \hspace*{1in}& \hspace*{2.4in}\\
 \end{tabular}
\end{center}
]

\def\mpl{M_{P}}
\def\eg{{\it e.g.}}
\def\etc{{\it etc.}}
\def\mw{m_W}
\def\mz{m_Z}

\section{Introduction}

The development of the Standard Model of particle physics is a
remarkable success story.  Its many facets have been tested at present day
accelerators; no significant
unambiguous deviations have yet been found.  In some
cases, the model has been verified at an accuracy of better than one
part in a thousand.  This state of affairs presents our field with a
challenge.  Where do we go from here?  What is our vision for
future developments in particle physics?  Are particle physicists'
recent successes a signal of the field's impending demise, or do real
long-term prospects exist for further progress?
We assert that the long-term health and intellectual vitality of
particle physics depends crucially on the development of a new
generation of particle colliders that push the energy frontier by an
order of magnitude beyond present capabilities.  In this report, we
address the scientific issues underlying this assertion.

Despite the success of the Standard Model as a description of the
properties of elementary particles, it is clear that the Standard
Model is not a fundamental theory.  Apart from any new physics
phenomena that might lie beyond the Standard Model, one knows that
the Standard Model neglects gravity.  A more complete theory of
elementary particle
interactions (including gravity) contains the Planck scale
($M_P\simeq 10^{19}$~GeV) as its basic energy scale.  The Standard
Model at best is a good low-energy approximation to this more fundamental
theory of particle interactions, valid in a limited domain of ``low-%
energies'' of order 100~GeV and below.  In order to extend the Standard
Model to higher energies, one must address three basic questions.

\begin{enumerate}
\item
{\sl What is the complete description of the effective theory of
fundamental particles at
and below the electroweak scale?}
\end{enumerate}

In the Standard Model with one physical Higgs scalar, the scale of
electroweak symmetry breaking is characterized by $v=246$~GeV,
the magnitude of the Higgs vacuum expectation value.  More
complicated models of particle physics can be easily constructed
which closely approximate the Standard Model at present day energies,
but have a much richer spectrum in the mass range above 100~GeV and
below a few TeV (henceforth referred to as the TeV scale).
If one wishes to extrapolate physics from the
electroweak scale and eventually deduce the nature of physics near the Planck
scale, one must know the complete structure of the particle theory at
the electroweak scale.  The Standard Model possesses three
generations of quarks and leptons (and no right-handed neutrinos) and
an SU(3)$\times$SU(2)$\times$U(1) gauge group.  Is that all?  Might there be
additional particles that populate the TeV scale?  Some possibilities
include: (i) an extended electroweak
gauge group (\eg, extra U(1) factors, left-right (LR) symmetry as in
SU(2)$_L\times$SU(2)$_R \times$U(1) models, \etc);
(ii) new fermions (\eg, a fourth
generation, mirror fermions,
vector-like fermions, massive neutrinos, \etc); (iii) new bosons
(CP-odd neutral and charged scalars in models with
extended Higgs sectors, Goldstone or pseudo-Goldstone bosons,
other exotic scalars, leptoquarks, vector resonances, \etc).
Clearly, in order to extrapolate our particle theories to high energy
with any confidence, one must know the complete TeV-scale spectrum.

\begin{description}{}{\itemsep 0pt \topsep 0pt}
\item{2.}
{\sl What is the mechanism for electroweak symmetry breaking?}
\end{description}

The Standard Model posits that electroweak symmetry breaking (ESB)
is the result of the dynamics of an elementary
complex doublet of scalar fields.  The
neutral component of the scalar doublet acquires a vacuum expectation
value and triggers ESB; the resulting particle spectrum contains three
massive gauge bosons and the photon, massive quarks and charged
leptons (and massless neutrinos) and a physical Higgs scalar.
However,
if one attempts to embed the Standard Model in a fundamental theory
with a high energy scale ($\mpl$), then the masses of elementary
scalars naturally assume values of order the high energy scale.
This is due to the quadratic sensitivity of squared scalar masses to the
high-energy scale of the underlying fundamental theory.  In
contrast, theoretical considerations ({\it e.g.} unitarity) require
the mass of the Standard Model Higgs boson to be of order the
electroweak scale.  To accomplish this, one has three choices:

\begin{description}{}{\itemsep 0pt \topsep 0pt}
\item{(i)}
``Unnaturally'' adjust the parameters of the high-energy theory
(a fine tuning of 34 orders of magnitude in the scalar squared mass
parameter is required) to produce the
required light elementary scalar field.

\item{(ii)} Invoke supersymmetry to cancel quadratic sensitivity of
the scalar squared  masses to $\mpl$.  Electroweak symmetry breaking
is triggered by the dynamics of a weakly-coupled Higgs sector.

\item{(iii)}
Elementary Higgs scalars do not exist.  A Higgs-like scalar
state (if it exists) would reveal its composite nature
at the TeV-scale, where new physics beyond the Standard
Model enters.  Electroweak symmetry breaking is triggered by
non-perturbative strong forces.
\end{description}

In order to implement (ii), one must first note that
supersymmetry is not an exact symmetry of
nature (otherwise, all known particles would have equal-mass supersymmetric
partners).  If supersymmetry is to explain why the
Higgs boson mass is of order the electroweak scale, then
supersymmetry-breaking effects which split the masses of particles
and their super-partners must be roughly of the same order
as the electroweak scale.
Thus, if supersymmetry is connected with the origin of
ESB, one expects to discover a new spectrum of
particles and interactions at the TeV-scale or below.
In addition, such models of ``low-energy" supersymmetry are
compatible with the existence of weakly-coupled elementary
scalars with masses of order the ESB-scale.
No direct experimental
evidence for the supersymmetric particle spectrum presently exists.
But there
is tantalizing indirect evidence.  Starting from the known values of
the SU(3), SU(2), and U(1) gauge couplings at $\mz$  and
extrapolating to high energies, one finds that the three gauge
couplings meet at a single point
if one includes the effects of
supersymmetric particles (with masses at or below 1~TeV)
in the running of the couplings.
Unification of couplings then takes place at around
$10^{16}$~GeV, only a few orders of magnitude below $\mpl$.
The three gauge couplings do not meet at a single
point if only Standard Model particles contribute to the running
of the couplings.
This could be a hint for low-energy supersymmetry, and
suggests that the theory
of fundamental particles remains weakly coupled and perturbative all
the way up to energies near $\mpl$.

New physics is also invoked to explain the origin of
electroweak symmetry breaking in choice (iii).
For example, in technicolor models (which make use of the
mechanism analogous to the one that is responsible for
chiral symmetry breaking in QCD),
electroweak symmetry breaking occurs when pairs of techni-fermions
condense in the vacuum.
One then identifies a new scale, $\Lambda_{ESB}\simeq 4\pi v\sim {\cal
O}(1 {\rm~TeV})$, where new physics beyond the Standard Model must
enter.
Other approaches, such as effective Lagrangian descriptions
of the strongly interacting Higgs sector,
preon models, top-mode condensate
models, composite Higgs models, \etc, also fall into this category.
Unfortunately, due to the presence of non-perturbative
strong forces, it is often difficult to make reliable
detailed computations in such models.
Moreover, phenomenological difficulties inherent in the simplest examples
often require additional structure
(\eg, an extended technicolor sector is
needed in technicolor models to generate fermion masses).
Unfortunately, a completely
phenomenologically viable fundamental model of strongly-coupled ESB
has not yet been constructed.

There is one important theoretical lesson that one
should draw from the above remarks.   Namely, {\it the nature of
ESB dynamics must be revealed at or below the 1~TeV energy scale.}
This is not a matter of conjecture.  The relevant scale is implicitly
built into the Standard Model through $v\equiv 2\mw/g$.  If one
wishes to take the most conservative approach,
one would say that the study of the TeV scale
must reveal the Higgs boson, but with no guarantees of anything
beyond.   However,
if one takes the naturalness argument seriously, it follows that
the study of the TeV scale must reveal
a new sector of physics beyond the Standard Model
responsible for ESB dynamics---either a
super-particle spectrum or a new strongly-interacting sector of
fermions, scalars, and/or vector resonances.

\begin{description}{}{\itemsep 0pt \topsep 0pt}
\item{3.}
{\sl What is the origin of the Standard Model parameters?}
\end{description}

I.I. Rabi asked the famous question concerning the
unexpected discovery of the muon: ``Who
ordered that?!''  In the context of the Standard Model, we ponder the
generalizations of Rabi's question: (i) Why are there three
generations? [or what is the origin of flavor?]; (ii) Where do the
quark and lepton masses come from? [or what is the origin of the
fermion mass matrices?]; (iii) What is the origin of CP-violation?
[or is the phase in the Cabibbo-Kobayashi-Maskawa (CKM) mass matrix the
only source of CP-violation? and why is $\theta_{QCD}$, which would
give rise to CP-violation from the strong interactions, so small?].
Note that although the minimal supersymmetric extension of the
Standard Model addresses the origin of electroweak symmetry breaking,
it provides little insight into any of the above questions.
In fact, one would simply add yet another question:
(iv) What is the origin of the (soft)-supersymmetry-breaking
parameters that characterize the model of low-energy supersymmetry?

In contrast to the problem of ESB discussed
above, the Standard Model does {\it not} provide any clues as to the
energy scale responsible for addressing the origin of the Standard
Model parameters (or a supersymmetric extension thereof).  The
relevant energy scale could lie anywhere from 1~TeV
to $\mpl$.  For example, in conventional theories of low-%
energy supersymmetry, the origin of the Standard Model and
supersymmetric parameters lies at or near the Planck scale.  In
contrast, theories of extended technicolor typically attempt to
solve the flavor problem at energy scales below 100~TeV.  The latter
approach is more ambitious; perhaps this is one reason why no
compelling model of this type has been constructed.
Nevertheless, any model of flavor and fermion masses requires a
definite framework for electroweak symmetry breaking.  Thus,
the elucidation of TeV-scale physics will have significant
ramifications for our understanding of physics at even higher energies.

Other future particle physics experiments may also address some of
these issues.    $B$-factories will
thoroughly test the CKM-parametrization of CP-violation.  Any
evidence for new sources of CP-violation would constitute physics
beyond the Standard Model.  The next generation of neutrino mixing
and mass experiments will try to verify recent experimental hints for
massive neutrinos.  If massive neutrinos are confirmed, then the
Standard Model will require an extension.  The simplest possibility is
to include right-handed neutrino states.
Majorana mass terms for the right-handed neutrinos are
SU(2)$\times$U(1) conserving and therefore constitute a completely
new energy scale ($M_R$)
characterizing the extended Standard Model.  At present, $M_R$ is
undetermined and can lie anywhere between the electroweak scale and
$\mpl$.  If $M_R\gg\mz$, then the effect of this new scale will
be simply to add the neutrino mass and mixing parameters to the list
of Standard Model parameters whose origin will be difficult to
discern directly from particle physics experiments in the foreseeable
future.

In this sense, the search for electroweak symmetry breaking presents
the unique guarantee---the dynamics that generates ESB must be
revealed by the next generation of colliders that can directly
explore the TeV-scale.  {\it Thus, it is crucial to explore
thoroughly the TeV energy scale in order to elucidate the
dynamics of electroweak symmetry breaking and
the associated fundamental particle spectrum.  This program must be
a primary focus of particle physics research in the long-term
future plans of the field.}

In this report, we explore the capabilities of the approved future
facilities and a variety of hypothetical future machines,
summarized in Table \ref{colliders}.  We have also begun to study
the physics opportunities at more speculative
future colliders, such as a 60 TeV hadron supercollider (which would
supersede the LHC) and a multi-TeV $\mu^+\mu^-$ collider,
although the overall capabilities of these machines have yet to be addressed
in a comprehensive fashion.

\begin{table}[htbp]
\begin{center}
\caption{}\label{colliders}
\setlength{\tabcolsep}{2.2pt}
\renewcommand\arraystretch{1.2}
\begin{tabular}{|lccc|}
\hline
\multicolumn{1}{|c}{Name}& Type& $\sqrt s$&  Yearly $\int{\cal L} dt$ \\
\hline
{\sl Approved Projects}&&& \\
LEP-II&    $e^+e^-$& $\sim$190 GeV& $\sim 100$ pb$^{-1}$    \\
Tevatron (Main &&&\\[-2pt]
\quad Injector)& $p\bar p$& 2 TeV& 1 fb$^{-1}$ \\
LHC&       $pp$&  14 TeV&  10--100 fb$^{-1}$ \\
\hline
\multicolumn{3}{|l}{{\sl Tevatron Upgrades (under consideration)}}& \\
TeV*& $p\bar p$& 2 TeV& 10 fb$^{-1}$ \\
Di-Tevatron& $p\bar p$& 4 TeV& 20 fb$^{-1}$ \\
\hline
{{\sl $e^+e^-$ Linear Collider}}& && \\
JLC, NLC, TESLA & $e^+e^-\ ^\dagger$& 0.5--1.5 TeV&  50--200 fb$^{-1}$ \\
&\multicolumn{3}{l|}{($^\dagger$ with $e\gamma,\gamma\gamma,e^-e^-$ options)}
\\
\hline
\end{tabular}
\end{center}
\end{table}

In this report, we do not directly address the origin of the Standard Model
parameters.  Nor do we address the possibility that new and
interesting physics beyond the Standard Model might populate the
energy interval {\it above} the TeV-scale.  We focus directly on
TeV-scale physics accessible to the next generation of colliders.
Namely, we examine the requirements for
completing the  TeV-scale particle spectrum, and
for unraveling the dynamics of electroweak symmetry breaking.
Particularly in the latter case, we expect that future colliders will
discover new physics beyond the Standard Model (BSM) associated with this
dynamics.  Ten ESB\&BSM subgroups were established to address these
issues.  Two additional subgroups were added to examine experimental
issues associated with the search for ESB\&BSM physics at future
hadron and $e^+e^-$ colliders, respectively.   Conveners of the
twelve subgroups prepared executive summaries which
form the basis of this document.  More detailed
expositions will be published separately in Ref.~\cite{wsbook},
which will also contain a more comprehensive list of references.

The subgroups and their conveners are listed below.

\noindent
Three subgroups addressed the case of weakly-coupled ESB
dynamics:

\begin{itemize}
\item
Weakly-Coupled Higgs Bosons (J.F. Gunion, A. Stange, and S. Willenbrock)

\item
Low-Energy Supersymmetry---Implications of Models (M. Drees, S.P.
Martin, and H. Pois)

\item
Low-Energy Supersymmetry Phenomenology (H. Baer, H. Murayama, and
X. Tata)
\end{itemize}

\noindent
Two subgroups addressed the case of strongly-coupled ESB dynamics:

\begin{itemize}
\setcounter{enumi}{3}
\item
Strongly-Coupled ESB Sector---Model Independent Study (M. Golden
and T. Han)

\item
Strongly-Coupled ESB Sector---Implications of Models (R.S. Chivukula,
R. Rosenfeld, E.H. Simmons, and J. Terning)
\end{itemize}

\noindent
Two subgroups addressed the question of identifying the
interactions and full spectrum of the TeV-scale particle theory:

\begin{itemize}
\setcounter{enumi}{5}
\item
New Gauge Bosons Beyond the $W^\pm$ and $Z$ (M. Cveti\v c and S.
Godfrey)

\item
New Particles and Interactions (A. Djouadi, J. Ng and T.G. Rizzo)
\end{itemize}

\noindent
Three subgroups addressed the search for ESB\&BSM physics via
precision measurements of Standard Model processes:

\begin{itemize}
\setcounter{enumi}{7}
\item
Anomalous Gauge Boson Couplings (H. Aihara, U. Baur, D. London and D.
Zeppenfeld)
\item
Top-Quarks as a Window to Electroweak Symmetry Breaking
(S.J. Parke, C.T. Hill and M.E. Peskin)

\item
Virtual Effects of New Physics (J.L. Hewett, T. Takeuchi, and S.
Thomas)
\end{itemize}

\noindent
Two subgroups focussed on the interplay of ESB\&BSM physics and
detector and machine issues for future colliders:
\begin{itemize}
\setcounter{enumi}{10}
\item
Experimental Issues in the Search for ESB\&BSM at Future Hadron Colliders
(S. Errede and J. Siegrist)

\item
Experimental Issues in the Search for ESB\&BSM at Future $e^+e^-$ Colliders
(T. Barklow and T.W. Markiewicz)
\end{itemize}



\def\epem{e^+e^-}
\def\etal{{\it et al.}}
\def\epem{e^+e^-}
\def\tanb{\tan\beta}
\def\gam{\gamma}
\def\lsim{\mathrel{\raise.3ex\hbox{$<$\kern-.75em\lower1ex\hbox{$\sim$}}}}
\def\gsim{\mathrel{\raise.3ex\hbox{$>$\kern-.75em\lower1ex\hbox{$\sim$}}}}
\def\nsd{N_{SD}}
\def\anti{\overline}
\def\tanb{\tan\beta}
\def\hsm{H_{SM}}
\def\mhsm{m_{\hsm}}
\def\hl{h^0}
\def\ha{A^0}
\def\hh{H^0}
\def\hpm{H^\pm}
\def\mhl{m_{\hl}}
\def\mhh{m_{\hh}}
\def\mha{m_{\ha}}
\def\anti{\overline}
\def\gam{\gamma}
\def\rta{\rightarrow}
\def\hp{H^+}
\def\hm{H^-}
\def\fbi{~{ fb}^{-1}}
\def\fb{~{ fb}}
\def\pb{~{ pb}}
\def\gev{\,{\rm GeV}}
\def\tev{\,{ \rm TeV}}
\def\wh{\widehat}
\def\wt{\widetilde}
\def\lra{\leftrightarrow}
\def\rta{\rightarrow}
\def\mstop{m_{\widetilde t}}
\def\ie{{\it i.e.}}
\def\hf{}
\section{Weakly-Coupled Higgs Bosons}


\subsection{The Standard Model and MSSM Higgs Sectors \protect\cite{hhg}}

In the Standard Model (SM), the electroweak
symmetry is broken by an elementary weak doublet of scalar fields,
whose neutral component
acquires a non-zero vacuum expectation value.
This model predicts the existence of a single CP-even neutral
scalar called the Higgs
boson, $\hsm$, of unknown mass but with fixed couplings to other particles
proportional to their masses.  This is the minimal model of electroweak
symmetry breaking.  Detailed studies of the phenomenological profile of
$\hsm$ are important, since they provide significant benchmarks for
any experimental search for electroweak symmetry breaking.

The underlying physics responsible for electroweak symmetry breaking
is only weakly constrained by present-day experimental data.  Thus,
in the search for electroweak symmetry breaking, one must be
open to more complicated possibilities.
For example, it is easy to generalize the minimal
Higgs model by simply adding additional scalar doublets.   Models of this
type possess additional CP-even neutral scalars, new CP-odd
neutral scalars, and charged scalars.  One of the simplest extensions of
the SM Higgs sector is the two-Higgs-doublet model (with a CP-invariant
scalar potential).  The scalar spectrum of this model contains two
CP-even neutral scalars, $\hl$ and $\hh$ (with $m_{\hl}\leq m_{\hh}$), one
CP-odd neutral scalar, $\ha$, and a charged Higgs pair, $H^\pm$.
The importance of this model will be clarified below.

In Section 1, a theoretical overview was presented of
the possible mechanisms that could be responsible for electroweak symmetry
breaking.
The Standard Model with its minimal Higgs sector or with
an extended Higgs sector (as described above) is theoretically
unsatisfactory, since there is no ``natural'' way of understanding
the origin of the electroweak symmetry breaking scale.
Two generic
solutions to this theoretical problem have been put forward.  In the first
solution, the Standard Model is extended to include supersymmetric partners.
In the second solution, new strongly-interacting forces
are invoked, whose dynamics are responsible for triggering electroweak symmetry
breaking.  In the former case, the Higgs bosons are elementary
weakly-coupled scalars, while in the latter case, any physical scalar
state is probably strongly-coupled and composite. The supersymmetric
approach is treated in Sections 3 and 4, while the strongly-coupled
scalar sector is considered in Sections 5 and 6.

The focus of this section is the search for
 weakly-coupled elementary Higgs bosons.
In the Standard Model, the Higgs self-coupling is proportional to the
square of the Higgs mass.  For Higgs masses below (roughly)
700 GeV, $\hsm$ is a weakly-coupled scalar.
In supersymmetric models, the Higgs bosons are also elementary and
weakly-coupled.  Moreover, as reviewed in Section 3, the Higgs
sector of the minimal supersymmetric extension of the Standard
Model (MSSM) consists of two scalar doublets
(interacting via a CP-conserving potential).  Thus, the
MSSM Higgs spectrum consists of $\hl$, $\hh$, $\ha$, and $H^\pm$,
introduced above.  As a result, the phenomenology of this Higgs
sector is particularly important, since it introduces
new kinds of Higgs phenomena that can be observed in
future experiments, while being a theoretically
well-motivated model for the scalar sector associated with electroweak
symmetry breaking.

This section considers the prospects for the discovery of
weakly-coupled Higgs bosons at present and future colliders.
Both the SM Higgs boson, $\hsm$,  and the scalars of the MSSM Higgs sector
are considered.  Some aspects of the phenomenology of
$\hl$, $\hh$, $\ha$, and $H^\pm$ do not strongly depend on
the fact that they arise in a supersymmetric model.  However, the
general parameter space of an arbitrary two-Higgs doublet model
is quite large (with four independent mass parameters and at least two other
additional dimensionless parameters).
In contrast, the two-doublet Higgs sector of the MSSM
possesses numerous relations among Higgs masses and couplings due to the
underlying supersymmetry.  In particular,
the MSSM tree-level Higgs masses and couplings are determined
in terms of two free parameters:  $\mha$ and $\tan\beta\equiv v_2/v_1$
[where $v_2$ $(v_1)$ is the vacuum expectation value
of the Higgs field that couples
to up-type (down-type) fermions].  One remarkable consequence of supersymmetry
is that the mass of the lightest CP-even neutral scalar, $\hl$,
is bounded from above by a mass value of order $m_Z$.
However, in obtaining the precise upper bound, there are important
radiative corrections that cannot be neglected.
Through the radiative corrections, the Higgs masses also acquire dependence
on the details of the MSSM spectrum.  The most important piece of
these corrections depends primarily
on the masses of the top-quark (taken here to be $m_t=175$~GeV)
and its supersymmetric partners \cite{radcorrect}.
To be conservative, in this section the
effects of the radiative corrections have been maximized
by assuming that all supersymmetric particle masses (and mass mixing
parameters) are of order 1~TeV.
In this case, the light CP-even
Higgs boson mass is bounded by $\mhl \leq \mhl^{\rm max} \simeq 120$ GeV.
The following properties of the MSSM Higgs sector are noteworthy.
If $\mha \gg m_Z$, then $\mha \simeq \mhh \simeq m_{H^\pm}$
(in addition, if $\tan\beta \gg 1$ then $\mhl \simeq \mhl^{\rm max}$).
Moreover, the couplings of $\hl$ to fermions and gauge bosons
are indistinguishable from those of the Standard Model Higgs boson,
while the couplings $\hh W^+W^-$, $\hh
ZZ$ and $Z\ha\hl$ are severely suppressed.  Finally,
assuming $\tan\beta > 1$, the couplings of $\ha$ and $\hh$
to $b\bar b$ and $\tau^+\tau^-$ ($t\bar t$) are
enhanced (suppressed).  In contrast,
for the case $\mha \lsim {\cal O}(m_Z)$ and
$\tan\beta > 1$, $\ha$ and $\hl$ exhibit the latter coupling
pattern to fermions.  The couplings $\hl W^+W^-$, $\hl
ZZ$ and $Z\ha\hh$ are severely suppressed, while the $\hh$ couplings
to fermions and gauge bosons are SM-like.

\begin{table*}[hp]
\begin{center}
\vskip3pc
\caption{%
Standard Model Higgs boson discovery modes.
All masses are specified in GeV.}
\label{discovery2}
\vskip1pc
\renewcommand{\arraystretch}{1.3}
\begin{tabular}{|l|c|c|}
\hline
{\bf Machine ($\sqrt{s},\,\int {\cal L}\,dt$)} & {\bf Mode} &
{\bf Discovery Region}
\\ \hline
LEP-I  & $\epem\rta Z^* \hsm$ & $\mhsm\lsim 65$
\\ \hline
LEP-II, ($176$ GeV, $500$ pb$^{-1}$)  & $\epem\rta Z\hsm$ &
$\mhsm\lsim 80$
\\ \hline
LEP-II, ($190$ GeV, $500$ pb$^{-1}$)  & $\epem\rta Z\hsm$ &
$\mhsm\lsim 95 $
\\ \hline
Tevatron, ($2$ TeV, $5$ fb$^{-1}$)  & $p\bar p\rta W\hsm$; $\hsm\rta b\anti
b$ & $\mhsm\lsim 60-80$
\\ \hline
TeV$^*$, ($2$ TeV, $30$ fb$^{-1}$)  & $p\bar p\rta W\hsm$; $\hsm\rta b\anti
b$ & $\mhsm\lsim 95$ \\
  & $p\bar p\rta W\hsm$; $\hsm\rta \tau^+\tau^-$ & $110\lsim\mhsm\lsim 120$
\\ \hline
DiTeV, ($4$ TeV, $30$ fb$^{-1}$)  & $p\bar p\rta W\hsm$; $\hsm\rta b\anti
b$ & $\mhsm\lsim 95$ \\
  & $p\bar p\rta ZZ\rta 4\ell$ & $\mhsm\sim 200$
\\ \hline
LHC, ($14$ TeV, $100$ fb$^{-1}$)  & $pp\rta \hsm \rta ZZ^{(*)}\rta 4\ell$ &
$120\lsim\mhsm\lsim 700$ \\
  & $pp\rta \hsm \rta \gam\gam$ & $80\lsim\mhsm\lsim 150$ \\
  & $pp\rta t\anti t \hsm,W\hsm$; $\hsm\rta\gam\gam$ & $80\lsim\mhsm\lsim
120-130$ \\
  & $pp\rta t\anti t \hsm$; $\hsm\rta b\anti b$ & $\mhsm\lsim
100-110$ \\ \hline
NLC, ($ 500$ GeV, $50$ fb$^{-1}$)  & $\epem\rta Z\hsm$ & $\mhsm\lsim
350$ \\
  & $WW\rta \hsm$ & $\mhsm\lsim 300$ \\
  & $\epem\rta t\anti t \hsm$ & $\mhsm\lsim 120$ \\ \hline
NLC, ($ 1$ TeV, $200$ fb$^{-1}$)  & $\epem\rta Z\hsm$ & $\mhsm\lsim
800$ \\
 &  $WW\rta\hsm$ & $\mhsm\lsim 700$ \\ \hline
\end{tabular}
\end{center}
\vskip3pc
\end{table*}

\begin{table*}[hp]
\begin{center}
\begin{minipage}{16.5cm}
\caption{%
MSSM Higgs boson discovery modes.  All masses are specified in GeV.
The ordered pair ($\mha$, $\tan\beta$) fixes the MSSM Higgs sector
masses and couplings.  The parameter regime $\mha\leq 1000$~GeV and
$1\leq\tan\beta\leq 60$ is surveyed.
If a range of $\tan\beta$ values is specified below,
then the first (second)
number in the range corresponds to the appropriate minimal (maximal) value of
$\mha$.  See text for further clarifications.}
\label{discovery3}
\vskip.5pc
\renewcommand{\arraystretch}{1.3}
\begin{tabular}{|l|c|c|}
\hline
{\bf Machine($\sqrt{s},\,\int {\cal L}\,dt$)}& {\bf Mode} &
         \protect\boldmath$(\mha,\tanb)$ {\bf Discovery Region}
\\ \hline
LEP-I
 & $\epem\rta Z^* \hl,\hl\ha$ & $(\lsim 25,\gsim 1)~{\rm
or}~(\lsim 45,\gsim 2)$
\\ \hline
LEP-II ($176$ GeV, $500$ pb$^{-1}$)
 & $\epem\rta Z\hl,\hl\ha$ & $(\lsim 80,\gsim 1)$~or~$(\lsim 150, \lsim 4$--1)
\\ \hline
LEP-II ($190$ GeV, $500$ pb$^{-1}$)
 & $\epem\rta Z\hl,\hl\ha$ & $(\lsim 85,\gsim 1)~{\rm or}~(\gsim 85,
\lsim 5$--1.5)
\\ \hline
TeV$^*$ ($2$ TeV, $30$ fb$^{-1}$)
 & $p\bar p\rta W\hl$; $\hl\rta b\anti b$; $2b$-tag & $(\gsim 130$--$150,\geq
1)$ \\ \hline
DiTeV ($4$ TeV, $30$ fb$^{-1}$)
 & $p\bar p\rta W\hl$; $\hl\rta b\anti b$; $2b$-tag & $(\gsim 130$--$150,\geq
1)$ \\ \hline
LHC ($14$ TeV, $100$ fb$^{-1}$)
   & $pp\rta \!W\hl\!,t\anti t \hl$; $\hl\!\rta b\anti b$; $2,3b$-tag &
  $(\gsim 130$--$150,\geq 1)$ \\
 & $pp\rta t\anti t$; $t\rta \hp b$; $1b$-tag & $(\lsim 130,\geq 1)$ \\
 & $pp\rta \hh$; $\hh\rta ZZ^{(*)}\rta 4\ell$ & $(\lsim 350,\lsim 10$--$2)$ \\
 & $pp\rta \hl,W\hl,t\anti t \hl$; $\hl\rta \gam\gam$ &
    $(\gsim180,\geq 1)$\\
 & $pp\rta b\anti b \ha,\hh$; $\ha,\hh\rta \tau^+\tau^-$ &
   $(\gsim 100,\gsim 5$--50) \\
 & $pp\rta b\anti b \ha,\hh$; $\ha,\hh\rta \mu^+\mu^-$ &
    $(\gsim 100,\gsim 5$--50) \\
 & $pp\rta b\anti b \hl$; $\hl\rta b\anti b$; 3$b$-tag
    & $(\lsim 125,\gsim 5$--10) \\
 & $pp\rta b\anti b \hh$; $\hh\rta b\anti b$; 3$b$-tag
 & $(\gsim 125,\gsim 10$--60) \\
 & $pp\rta b\anti b \ha$; $\ha\rta b\anti b$; 3$b$-tag
    & $(\lsim 125,\gsim 5$--10$)~{\rm or}~(\gsim 125,\gsim 10$--60)
 \\ \hline
NLC ($500$~GeV, $50$ fb$^{-1}$)
            & $\epem\rta Z\hl$ & visible unless $(\lsim 90,\gsim 7)$ \\
 & $WW\rta \hl$ & visible unless $(\lsim 80,\gsim 12)$ \\
 & $\epem\rta \hl\ha$ & $(\lsim 120, \geq 1)$ \\
 & $\epem\rta Z\hh,WW\rta \hh$ & $(\lsim 140,\geq 1)$ \\
 & $\epem\rta \hh\ha$ & $(\lsim 220,\geq 1)$, unless $(\lsim 90,\gsim 7)$ \\
& $\epem\rta \hp\hm$ & $(\lsim 230,\geq 1)$ \\ \hline
NLC ($ 1$ TeV, $200$ fb$^{-1}$)
 & $\epem\rta \hh\ha$ & $(\lsim 450,\geq 1)$, unless $(\lsim 90,\gsim 7)$ \\
 & $\epem\rta \hp\hm$ & $(\lsim 450,\geq 1)$ \\ \hline
\end{tabular}
\end{minipage}
\end{center}
\end{table*}

We now outline the capabilities of present and proposed future colliders to
search for the SM Higgs boson and the MSSM Higgs bosons \cite{hhg}.
The SM Higgs discovery limits for the most useful discovery modes
are summarized in
Table \ref{discovery2}.  The corresponding MSSM Higgs discovery limits
are summarized in Table \ref{discovery3}.
For $b$-tagged modes, a tagging efficiency of 30\% and purity of 1\% are
assumed.  Since the MSSM Higgs discovery limits are rather complex,
additional information is provided here to assist the reader in
interpreting the results presented in Table \ref{discovery3}.
The MSSM Higgs sector is most
conveniently parametrized in terms of $\mha$ and $\tan\beta$.
All values of $\mha\leq 1000$~GeV and
$1\leq\tan\beta\leq 60$ have been surveyed (as preferred in the MSSM).
There is also some dependence of the discovery limits on the
radiative corrections, the main effect of which is to increase
the mass of $\hl$ above its tree-level value.  The discovery
regions of Table \ref{discovery3} have been obtained assuming that the
effects of the radiative corrections are maximal, as noted above.

A few examples are now given to clarify the meaning of the discovery
regions presented in Table \ref{discovery3}.
At LEP-II at $\sqrt{s}=176$~GeV
and $500~{\rm pb}^{-1}$ integrated luminosity, either $\hl$ or
$\ha$ (or both) can be discovered via $\epem\rta Z\hl,\hl\ha$ if
$\mha\lsim 80$~GeV and $\tan\beta\gsim 1$.  A second discovery
region also exists, beginning at $\tan\beta\lsim 4$ (at $\mha=80$~GeV)
and ending at $\mha\lsim 150$~GeV (at $\tan\beta=1$).
If $\sqrt{s}$ is increased
to 190~GeV, both discovery regions become larger; in particular, the
latter region now begins at
$\tan\beta\lsim 5$ (at $\mha=85$~GeV) and ends at $\tan\beta\lsim 1.5$
for the maximal value of $\mha$ considered.
At the NLC-500, $\hl$
can be detected via $\epem\rta Z\hl$ unless $\tan\beta\gsim 7$ and
$\mha\lsim 90$~GeV.  A similar discovery region exception appears for
$WW\rightarrow \hl$.
The reason such restrictions arise is that in the indicated
region of parameter space, $\hh$ is SM-like
in its couplings, while the $\hl ZZ$ and $\hl W^+W^-$ couplings
are very suppressed.  The $Z\hh\ha$ coupling is likewise
suppressed in the same parameter regime, which explains the other
two discovery region exceptions listed in Table \ref{discovery3}.

\subsection{A Tour of Higgs Search Techniques at Future \hfill\break
Colliders}

The goal of the Higgs search at
present and future colliders is to examine the full mass range of
the SM Higgs boson, and the full parameter
space of the MSSM Higgs sector.  The LHC can
cover the entire range of SM Higgs boson masses from the upper limit of
LEP-II ($\mhsm = 80$--$95$~GeV, depending on machine energy)
to a Higgs mass of 700 GeV (and beyond).  The most difficult region
for high luminosity hadron colliders is the intermediate-mass range
$\mhsm = 80$--$130$~GeV.  The LHC detectors are being designed with the
capability of fully covering the intermediate Higgs mass region.
The TeV$^*$ and DiTevatron  may also have some discovery
potential in this mass region.   In contrast, the intermediate mass
Higgs search at the NLC (which makes
use of the same search techniques employed at LEP-II) is straightforward.
At design luminosity,
the NLC discovery reach is limited only by the center-of-mass energy of the
machine.

In the search for the Higgs bosons of the MSSM, two objectives are:
(i) the discovery of $\hl$ and (ii)
the discovery of the non-minimal Higgs states
($\hh$, $\ha$, and $H^\pm$).  Two theoretical results play a key role
in the MSSM Higgs search.  First, the mass of
$\hl$ is bounded by $\mhl^{\rm max}\simeq 120$~GeV, as noted above.  Second,
if the properties of $\hl$ deviate significantly from the SM Higgs boson,
then $\mha\lsim{\cal O}(m_Z)$ and the $\hh$ and $H^\pm$ masses must
lie in the intermediate Higgs mass region.
As a result, experiments that are sensitive to the intermediate Higgs
mass region have the potential for detecting
at least one of the MSSM Higgs bosons
over the entire MSSM Higgs parameter space
(parametrized by $\tan\beta$ and $\mha$, as noted above).
LEP-II does not have sufficient energy to cover the entire MSSM Higgs
parameter space, since $\mhl^{\rm max}$ lies above the LEP-II Higgs
mass reach.  The TeV$^*$, DiTevatron and LHC all possess the
capability of detecting Higgs bosons in the intermediate mass range,
and consequently can probe regions of the MSSM Higgs parameter space
not accessible to LEP-II.
As of this writing, no set of experiments at any future hadron
collider considered in this report can guarantee the discovery (or exclusion)
of at least one MSSM Higgs boson for {\it all} values of $\mha$ and $\tan\beta$
not excluded by the LEP-II search.  However, it may be possible to
close this gap in the MSSM parameter space if sufficiently reliable
$b$ quark identification in hadron collider events is possible.
On the other
hand, because of the relative simplicity of the NLC
Higgs search in the intermediate mass region, the
NLC is certain to discover at least one MSSM Higgs boson
(either $\hl$ or $\hh$) if the supersymmetric approach is correct.
If $\hl$ is discovered and proves to be SM-like in its properties,
then one must be in the region of MSSM Higgs parameter space where
the non-minimal Higgs states are rather heavy and approximately
degenerate in mass.  In this case, the LHC may not be
capable of discovering any Higgs bosons beyond
$\hl$, while $\hh$, $\ha$ and
$\hpm$ can be detected at the NLC only if its center-of-mass energy
$\sqrt{s}\gsim 2\mha$.

Below is a detailed description of the Higgs potential of present and
future colliders. A Higgs boson is deemed observable if a $5\sigma$-excess
of events can be detected in a given search channel.

\noindent
$\bullet$ LEP-I --- The current lower bound on the SM Higgs mass is
64.5 GeV, and will increase by at most a few GeV.
The lower bounds on the MSSM Higgs masses are
$\mhl > 44.5$ GeV, $\mha > 24.3$ GeV (for $\tan\beta > 1$),
and $m_{H^\pm} > 45$ GeV.

\noindent
$\bullet$ LEP-II --- For $\sqrt s=176$ GeV,
the SM Higgs is accessible via $Z\hsm$ production up
to $80$ GeV for an integrated luminosity of
$500$ pb$^{-1}$ \cite{janotlep}. For $\sqrt s=190$~GeV,
$\mhsm$ values as high as $m_Z$ can be probed with $200$ pb$^{-1}$
of data, using
$b$-tagging in the region $\mhsm\sim m_Z$. The reach for $\sqrt s=205$~GeV is
about 105 GeV.  In the MSSM, the
same mass reaches apply to $\hl$ if $\mha\gsim m_Z$,
since in this case $Z\hl$ is produced at about the same rate as $Z\hsm$.
If $\mha\lsim m_Z$, then the $\hl ZZ$ coupling (which controls the $Z\hl$
cross section) becomes suppressed while the $Z\ha\hl$ coupling becomes
maximal.  In the latter
parameter region, $e^+e^-\rightarrow\ha\hl$ can be detected
for all values of $\mha\lsim\sqrt{s}/2-10$~GeV, assuming that $\tan\beta>1$.
Since $\mhl\lsim 120$ GeV, increasing the LEP-II energy to roughly
$\sqrt s = 220$ GeV while maintaining the luminosity would be
be sufficient to guarantee the detection of at least one Higgs boson
(via $Z\hl$ and/or $\ha\hl$ final states) over the entire MSSM
Higgs parameter space \cite{janotlep,gunerice}.

\noindent
$\bullet$ Tevatron ($\sqrt s = 2$ TeV, ${\cal L} =
10^{32}~{\rm cm}^{-2}~{\rm s}^{-1}$  with the
Main Injector) --- The most promising mode for the SM Higgs
is $W\hsm$ production, followed by $\hsm \to b\anti b$ \cite{zeuthen}.
With 5~fb$^{-1}$ of integrated luminosity, this mode could potentially
explore a Higgs mass region of 60--80 GeV, a region which will already
have been covered by LEP II via the $Z\hsm$ process.

\noindent
$\bullet$ TeV$^*$ ($\sqrt s = 2$ TeV, ${\cal L}
\ge 10^{33}~{\rm cm}^{-2}~{\rm s}^{-1}$) ---
The Higgs mass reach in the $W\hsm$ mode, with $\hsm \to b\anti b$, is extended
over that of the Tevatron \cite{zeuthen,MK}.
With 30~fb$^{-1}$ of integrated luminosity, a Higgs of mass 95 GeV
is potentially accessible, but the peak will
not be separable from the $WZ$, $Z
\to b\bar b$ background.
The mode $(W,Z)\hsm$, followed by $\hsm \to \tau^+ \tau^-$ and $W,Z \to jj$
(where $j$ stands for a hadronic jet), is
potentially viable for Higgs masses sufficiently far above the $Z$ mass
\cite{MK}.
With 30~fb$^{-1}$ of integrated luminosity, a Higgs in the mass range $110-120$
GeV may be detectable. However, the Higgs peak will be difficult to
recognize on the slope of the much larger $Zjj$, $Z \to \tau^+\tau^-$
background.
Both of these modes are of particular significance for $\hl$.
For $\tanb>1$ and $\mha\lsim 1.5 m_Z$,
the enhanced coupling of the $\hl$ to $b$'s and $\tau$'s makes it
unobservable at the LHC
via $\hl \to \gamma \gamma$ or $\hl \to ZZ^* \to 4\ell$.

\noindent
$\bullet$ DiTevatron ($\sqrt s = 4$ TeV, ${\cal L}
\ge 10^{33}~{\rm cm}^{-2}~{\rm s}^{-1}$) ---
For the SM Higgs, the ``gold-plated'' mode, $\hsm \to ZZ \to 4 \ell$,
requires about 30~fb$^{-1}$
for the optimal Higgs mass in this mode, $\mhsm = 200$ GeV \cite{zeuthen}.
The $4\ell$ mode is not useful for any of the MSSM Higgs bosons.
The Higgs mass reach in the $W\hsm$, $\hsm
\to b\anti b$ mode is only marginally
better than at the TeV$^*$, due to the increase in the top-quark backgrounds
relative to the signal \cite{zeuthen,MK}.
The mode $(W,Z)\hsm$, with $\hsm \to \tau^+ \tau^-$ and
$W,Z \to jj$, has less promise than at the TeV$^*$ due to the relative increase
in the background \cite{MK}.

\noindent
$\bullet$ LHC ($\sqrt s = 14$ TeV, ${\cal L} =  10^{33}$%
--$10^{34}~{\rm cm}^{-2}~{\rm s}^{-1}$) ---
%
For the SM Higgs boson,
the ``gold-plated'' mode, $\hsm \to ZZ^{(*)} \to 4 \ell$,
including the case where
one $Z$ boson is virtual, covers the range of Higgs masses 130--700 GeV and
beyond with 100~fb$^{-1}$ \cite{ATLAS,CMS}.
For $\mhsm > 700$ GeV, the Higgs is no longer ``weakly coupled'', and
search strategies become more complex --- see Section 5.

The CMS detector is planning an exceptional
electromagnetic calorimeter (PbWO$_4$ crystal) which will enable
the decay mode $\hsm \to \gamma\gamma$ to cover the Higgs mass range 85--150
GeV
with 100 fb$^{-1}$ of integrated luminosity \cite{CMS}. This range
overlaps the reach of LEP II (with $\sqrt s = 190$ GeV)
and the lower end of the range covered by the gold-plated mode. The ATLAS
detector covers the range 110-140 GeV with this mode \cite{ATLAS}.
Both CMS and ATLAS find
that the modes $t\anti t \hsm$ and $W\hsm$, with $\hsm \to \gamma\gamma$, are
 viable
in the mass range 80--120 GeV with 100~fb$^{-1}$ of integrated
luminosity \cite{ATLAS,CMS}.
Since backgrounds are smaller, these modes do not require
such excellent photon resolutions and jet-photon discrimination
as does the inclusive $\gam\gam$ mode. CMS has also studied the production of
the Higgs in association with two jets, followed by $\hsm\to \gamma\gamma$,
and concludes that this mode covers the Higgs mass range 70--130 GeV
\cite{CMS}.

The modes $t\anti t \hsm$ and $WH$, with $\hsm \to b\anti b$, are
potentially useful
in the intermediate-mass region \cite{zeuthen,MK}.
The reach with 100~fb$^{-1}$ of integrated
luminosity is 100 GeV, reduced to 80 GeV with 30~fb$^{-1}$ \cite{ATLAS}.
Overall, ATLAS will cover the Higgs mass region 80--140 GeV with 100
fb$^{-1}$ using a combination of $\hsm \to \gamma\gamma$;
$t\bar t \hsm$, $W\hsm$ with $\hsm\to \gamma\gamma$; and
$t\bar t \hsm$ with $\hsm \to b\bar b$.

The main search mode for the lightest MSSM Higgs boson is $\hl\to\gamma\gamma$,
which is viable when $\mhl$ is near $\mhl^{\rm max}$
\cite{ATLAS,CMS,gunerice,baererice,zerice}.
The processes $t\anti t \hl$ and $W\hl$,
with $\hl \to b\anti b$, can potentially
extend $\hl$ detection to the somewhat lower $\mhl$ values (corresponding to
somewhat lower $\mha$) that would assure that at least one
MSSM Higgs boson can be detected over all of the MSSM parameter
space \cite{zeuthen}.
Adapting the ATLAS study for the SM Higgs to $\hl$ suggests
that the required sensitivity could be achieved in the $t\anti t \hl$ mode.
Due to the large top-quark background, the $W\hl$ with $\hl\to b\bar b$
mode at the LHC
has little advantage over this mode at the TeV$^*$ or the DiTevatron for the
same integrated luminosity.

The other MSSM Higgs bosons are generally more elusive. There are small
regions of parameter space in which $\hh$ can be observed decaying to
$\gamma\gamma$ or $ZZ^{(*)}$ \cite{ATLAS,CMS}.
The possibility exists that the charged Higgs can be discovered
in top decays \cite{ATLAS,CMS}.
For $\mha>2 m_Z$ and large $\tan\beta$, $\hh$ and $\ha$ can have sufficiently
enhanced $b$ quark couplings that
they would be observed when produced in association with
$b\anti b$ and decaying via $\hh, \ha \to \tau^+\tau^-$
\cite{ATLAS,CMS,zerice,zeuthen}.
For very enhanced couplings of the Higgs bosons to $b$ quarks
(which occurs for very large $\tanb$), the modes
$b\anti b (\hl, \hh, \ha)$, with $\hl, \hh, \ha \to b\anti b$,
and $\anti t b H^+$, with $H^+
\to t \anti b$, are potentially viable \cite{zeuthen}.

Let us suppose that
the modes involving $b$-tagging (especially $t\anti t \hl$ production
with $\hl\rta b\anti b$) eventually prove to be viable.
Then, the LHC Higgs search using the $b$-tagging modes
in combination with the $4\ell$ and $\gam\gam$ final state modes
should be capable of discovering at least one of the MSSM Higgs
bosons (or in absence of a discovery rule out the MSSM) \cite{zeuthen}.
For much of parameter space only the $\hl$
will be detectable.  Detection of all the MSSM Higgs bosons at the
LHC is only possible for very small regions of parameter space at
moderate values of $\mha$.

\noindent
$\bullet$ $e^+e^-$ linear collider ($\sqrt s = 500$ GeV to 1 TeV,
${\cal L} \ge 5 \times 10^{33}~{\rm cm}^{-2}~{\rm s}^{-1}$) --- At $\sqrt s
 =500$ GeV
with an integrated luminosity of 50~fb$^{-1}$, the SM Higgs boson is observable
via the $Z\hsm$ process up to $\mhsm=350$ GeV \cite{desyworkshop}.
Employing the same process, a Higgs boson with $\mhsm=130$~GeV
(200~GeV) would be discovered with 1~fb$^{-1}$ (10~fb$^{-1}$) of
data.  Other Higgs production mechanisms are also useful.
With 50
fb$^{-1}$ of data, the $WW$-fusion process is observable up to
$\mhsm = 300$ GeV \cite{desyworkshop,hawaii,Janot}.
The $t\anti t \hsm$ process is accessible for $\mhsm \lsim 120$ GeV,
thereby allowing a direct determination of the $t\anti t \hsm$ Yukawa
coupling \cite{hawaii}.
The $\gamma\gamma$ collider mode of operation does not extend the reach of
the machine for the SM Higgs, but does allow a measurement of the
$\hsm \to \gamma\gamma$ partial width up to
$\mhsm = 300$ GeV~\cite{hawaii}.

In the MSSM,
for $\mha\lsim m_Z$, the $\hl$ and $\ha$ will be detected in the $\hl\ha$ mode,
the $\hh$ will be found via $Z\hh$ and $WW$-fusion production, and $H^+H^-$
pair production will be kinematically allowed and easily
observable \cite{desyworkshopsusy,hawaii,Janot}.
At higher values of the parameter $\mha$,
the lightest Higgs boson is accessible in both the $Z\hl$ and
$WW$-fusion modes.  However, the search for the heavier
Higgs boson states is limited by machine energy.
The mode $\hh\ha$ is observable up to
$\mhh \sim \mha \sim 220$ GeV, and $H^+H^-$ can be detected up to
$m_{H^{\pm}} =230$ GeV \cite{desyworkshopsusy,hawaii,Janot}.
The $\gamma\gamma$ collider mode could potentially
extend the reach for $\hh$ and $\ha$ up to 400 GeV if $\tanb$
is not large \cite{gunerice}.

At $\sqrt s = 1$ TeV with an integrated luminosity of 200~fb$^{-1}$,
the SM Higgs boson can be detected via the $WW$-fusion process up to
$\mhsm=700$
GeV \cite{hawaii,Janot}.  In the MSSM,
$\hh\ha$ and $H^+H^-$ detection would be extended
to $\mhh\sim \mha\sim m_{H^{\pm}}\sim 450$ GeV
\cite{gunerice,desyworkshopsusy,hawaii,Janot}.
Measurements of cross sections, branching ratios, and angular distributions
of the Higgs events will determine model parameters and test the
underlying theory of the scalar sector \cite{hawaii}.

\noindent
$\bullet$ Influence of MSSM Higgs decays into supersymmetric particle
final states --- For some MSSM parameter choices,
the $\hl$ can decay primarily to invisible modes, including
a pair of the lightest supersymmetric neutralinos
or a pair of invisibly decaying sneutrinos \cite{zeuthen}.
The $\hl$ would still be easily discovered
at $e^+e^-$ colliders in the $Z\hl$ mode using missing-mass techniques
\cite{hawaii}.
At hadron colliders, the $W\hl$ and $t\anti t \hl$ modes may
be detectable via large missing energy and lepton plus invisible energy
signals, but determination of $\mhl$ would be difficult \cite{zeuthen}.
The $\hh$, $\ha$, and $H^\pm$ decays can be dominated by
chargino and neutralino pair final states and/or slepton pair final states
\cite{hhg,gunerice,baererice,zeuthen}. Such modes can either
decrease or increase the chances of detecting these heavier MSSM
Higgs bosons
at a hadron collider, depending upon detailed MSSM parameter choices
\cite{gunerice,baererice,zeuthen}.
At $e^+e^-$ colliders $\hh$, $\ha$ and $H^\pm$ detection up to the earlier
quoted (largely kinematical) limits should in general remain possible.

\noindent
$\bullet$ Hadron collider beyond the LHC ---
It could provide increased event rates and more overlap of discovery
modes for the SM Higgs boson.
However, its primary impact would be to extend sensitivity
to high mass signals associated with
strongly-coupled electroweak symmetry breaking scenarios (see Section 5).
In the case of the MSSM Higgs bosons, the increased
event rates would significantly extend the overlap of the various
discovery modes, expanding the parameter regions where several and perhaps
all of the MSSM Higgs bosons could be observed.

\noindent
$\bullet$ $e^+e^-$ collider beyond $\sqrt s =1$ TeV --- Such an extension in
energy is not required for detecting and studying a
weakly-coupled SM Higgs bosons,
but could be very important for a strongly-coupled
electroweak-symmetry-breaking scenario. Detection of $\hh\ha$ and $H^+H^-$
production in the MSSM model would be extended to higher masses.

\def\qq{$q\overline{q}\;$}
\def\bb{$b\overline{b}\;$}
\def\cc{$c\overline{c}\;$}
\def\uds{$u\overline{u},\;d\overline{d},\;s\overline{s}\;$}
\def\qqg{q$\overline{\rm q}$g$\;$}
\def\z0{$Z^0$}
\def\h0{$h^0$}
\def\H0{$H^0$}
\def\A0{$A^0$}
\def\hbbar{h^0 \rightarrow b \overline{b}}
\def\hccbar{h^0 \rightarrow c \bar{c}}
\def\huds{h^0 \rightarrow u\bar{u},\ d\bar{d},\ s\bar{s}}
\def\zudsc{Z^0 \rightarrow udsc}
\def\zhad{Z^0 \rightarrow hadrons}

\subsection{Precision Measurements of Higgs Properties}

Detailed studies of the
properties of Higgs bosons
can be carried out at $e^+e^-$ colliders.
The measurements of cross sections, branching ratios, and angular
distributions of Higgs events will determine model parameters and
test the underlying theory of the scalar sector.
In this discussion, $h$ denotes the lightest CP-even
neutral Higgs.  It may be $\hsm$ or $\hl$ (of the MSSM),
or perhaps a scalar arising from a completely different model.
Once discovered, determining the identity of $h$
is precisely what one hopes to be able to address
based on the considerations presented here.
Below
we list some examples of the measurements that can be made
at $\sqrt{s}=500$~GeV with an integrated luminosity of
50~fb$^{-1}$, assuming that the Higgs mass is less than
350~GeV and that the process
$e^+e^-\rightarrow Zh$
occurs at approximately the
Standard Model rate.

The single event Higgs
mass resolution using
$e^+e^-\rightarrow Zh$
is approximately 4 GeV \cite{Janot}.  The number of Higgs
events provided by an integrated luminosity of 50 fb$^{-1}$ will then
allow the determination of the central value
of the Higgs mass with a statistical accuracy of $\pm$ 180 MeV.
Since most Standard Model parameters only have logarithmic
dependence on the Higgs mass, this accuracy is
far greater than what is needed to check the
self-consistency of the Standard Model.
One could use this mass determination to make
accurate predictions for the Higgs
production cross sections and branching fractions.

For a SM Higgs boson,
signals can be selected with small backgrounds and one can
easily measure the angular
distributions of the Higgs production direction in the process
$e^+e^-\rightarrow Zh$
and the directions of the outgoing
fermions in the $Z^0$ rest frame \cite{Janot}, and verify the expectations
for a CP-even Higgs boson \cite{barger94}.
For a more general Higgs eigenstate,
since only the CP-even part couples
at tree-level to $ZZ$,  these measurements would still agree with the CP-even
predictions unless the Higgs is primarily
(or purely) CP-odd, in which case the cross section will be much smaller
and the event rate inadequate to easily verify the mixed-CP or CP-odd
distribution forms. Analysis of photon
polarization asymmetries in $\gamma\gamma\rta h$ production rates,
and of angular correlations among secondary
decay products arising from primary
$h\rta\tau\tau$ or $h\rta t\overline t$ final state decays in
$Zh$ production, can both provide much better sensitivity to a CP-odd
Higgs component in many cases \cite{bohguncpasym}.

By choosing a decay channel whose reconstruction is independent
of the Higgs mass and
decay mode, such
as $e^+e^-\rightarrow Zh$, where $Z \rightarrow \ell^+ \ell^-$,
one can measure the
total cross section
for Higgs production. For an integrated luminosity
of 50 fb$^{-1}$, this can be determined with a statistical accuracy of
better than 10\% \cite{Janot}.
This is one possible technique for measuring the $hZZ$ coupling.

A specific study \cite{PRDHiggs} has been conducted
to consider the experimental problems of
determining the production cross section $\sigma$
times the branching fractions of a discovered
Higgs boson into all of its possible
(Standard Model) final states.  The resulting
expected statistical precisions
are strong functions of the mass of the
Higgs, as the branching fractions change dramatically
in the mass range 100 GeV$ < m_h < $ 180 GeV.
For a Standard Model Higgs of mass 120 GeV, it is found that
statistical precisions of 7\%, 14\%, 39\%, and 48\% can
be achieved for
$\sigma \times BR(h \rightarrow b \bar b)$,
$\sigma \times BR(h \rightarrow \tau^+ \tau^-)$,
$\sigma \times BR(h \rightarrow c \bar c+\ gg)$, and
$\sigma \times BR(h \rightarrow W W^*)$ respectively.

Reasonably precise measurements of certain $\sigma \times BR$
combinations will also be possible at hadron colliders.  For example,
at the LHC a good determination of $\sigma(gg\rightarrow
h)\times BR(h\rightarrow \gamma\gamma)$ should be possible for a SM-like
Higgs in the 80 to 150 GeV mass range, and
of $\sigma (gg\rightarrow h)\times BR(h\rightarrow ZZ)$ for $m_h > 130$ GeV.



\def\dt{m_Z^2 \cos \! 2 \beta}
\newcommand{\mgl}{\mbox{$M_{\tilde{g}}$}}
\newcommand{\mhalf}{\mbox{$m_{1/2}$}}
\newcommand{\mx}{\mbox{$M_X$}}
\newcommand{\eplem}{\mbox{$e^+e^-$}}
\newcommand{\een}{\end{subequations}}
\newcommand{\ben}{\begin{subequations}}

\section{Low Energy Supersymmetry:  Implications of Models}

\subsection{Elements of Minimal Low-Energy Supersymmetry}

A truly fundamental theory of particles and their
interactions must account for the large hierarchy of energy
scales inherent in the theory---from
the scale of electroweak symmetry breaking [characterized
by $(\sqrt{2}G_F)^{-1/2}=246$~GeV] up to energies as large as the
Planck scale (of order $10^{19}$~GeV).  Supersymmetry (SUSY)
provides a mechanism for stabilizing the large hierarchy
of scales, and thus is a natural framework for addressing such
problems.  The simplest example of such an approach is the minimal
supersymmetric extension of the Standard Model (MSSM).   The MSSM
possesses a new class of particles---supersymmetric partners of the
quarks, leptons, gauge and Higgs bosons of the Standard Model.  A
particle and its supersymmetric partner would be degenerate in mass
if supersymmetry were an exact symmetry of nature.   Since this
degeneracy is not realized in nature, supersymmetry must be broken.
But, if the breaking is ``soft'' (with supersymmetry-breaking masses of order
1~TeV or less), then the stability of the hierarchy can be successfully
maintained.

Unification of gauge couplings provides an intriguing hint for an
underlying supersymmetric theory of particle interactions.
If one extrapolates the SU(3)$\times$SU(2)$\times$U(1)
gauge couplings from their low-energy values measured at LEP to very
large energy scales, one finds that the three couplings meet at
approximately the same energy scale if the renormalization group
equations (RGE's) of the MSSM are used \cite{dmpunification}.
In contrast, if Standard Model RGE's are
used, unification of gauge couplings fails by more then seven standard
deviations!  As a result, it is possible to construct a model of
particle interactions such that: (i) the strong and electroweak
forces are unified at very large energies (at a scale $M_X\sim 2\times
10^{16}$~GeV, which is not too far from the Planck
scale, $M_P=10^{19}$~GeV);
(ii) the couplings of all particles with non-trivial
SU(3)$\times$SU(2)$\times$U(1) quantum numbers are perturbative in
strength at all energies below the Planck scale;
and (iii) the stability of the large hierarchy between the
electroweak and grand unified energy scales is guaranteed by the
underlying supersymmetry.
In this approach, the origin of the electroweak scale is tied to the
scale of supersymmetry breaking.  In particular, in the models
considered later in this section, electroweak symmetry breaking is
generated through radiative corrections.  Briefly stated, the
squared Higgs mass parameter is positive at the
high energy scale, $M_X$.  However, due to the one-loop radiative
effects generated by the large top quark Higgs Yukawa coupling ({\it i.e.},
as a consequence of the large top quark mass),  the squared Higgs
mass parameter is driven negative in its evolution from $M_X$ down
to the low energy electroweak scale \cite{dmprewsb}.
These effects trigger electroweak
symmetry breaking at the proper mass scale (of order $v=246$~GeV) and
establish the large hierarchy of scales between $M_X$ and $m_Z$.

Although the electroweak symmetry breaking scale has now been related
to the supersymmetry breaking scale, one has not yet explained
where the supersymmetry-breaking masses come from.   As noted above,
these masses must be of order 1~TeV or less in order to generate the
electroweak symmetry breaking scale without an excessive
(``unnatural'') fine-tuning of model parameters.   The fundamental origin of
supersymmetry breaking is one of the most important unsolved problems
of supersymmetry theory.  Nevertheless, one can take a purely
phenomenological approach and parameterize the MSSM
in terms of a general set of a priori unknown
supersymmetry breaking parameters.
Ultimately, these parameters would be measured in
experiment once supersymmetric particles are discovered.
A detailed knowledge of these parameters could provide important
clues as to their origins.

A brief synopsis of the parameters of the MSSM is now presented.
The SUSY-conserving part of the MSSM is
fixed by the three gauge coupling constants of the Standard Model,
the Higgs-fermion Yukawa couplings (denoted by $h_U$, $h_D$ and
$h_E$ below), and one new supersymmetric Higgs mass parameter $\mu$.
The particle content of the MSSM consists of three families of
quarks and leptons and their scalar partners;
the vector gauge bosons and their fermionic
gaugino partners corresponding to the gauge group
SU(3)$\times$SU(2)$\times$U(1); and {\it two}
SU(2)-doublet Higgs scalars and their fermionic higgsino partners.
The introduction
of a second Higgs doublet is necessary in order for the gauge anomalies
associated with the higgsinos to cancel,
and to give masses to both ``up-type'' quarks
and ``down-type'' quarks and charged leptons
(the latter cannot be done with one Higgs scalar without violating SUSY).
As a result, the Higgs sector of the MSSM is a two-Higgs doublet model,
whose phenomenology was discussed in great detail in Section 2.
The colorless gauginos and higgsinos with the same electric charge can mix.
The resulting mass eigenstates are called charginos and neutralinos,
with corresponding electric charges $\pm 1$  and $0$, respectively.
With the particle content and the Higgs-fermion Yukawa couplings
fixed as described above,  all SUSY-preserving interactions
(which correspond to all the dimensionless couplings of the model)
are determined by the SU(3)$\times$SU(2)$\times$U(1)
gauge invariance and/or by the supersymmetry.

However, in contrast to the Standard Model, the Higgs-fermion Yukawa
couplings do not exhaust all possible scalar--fermion interactions.
The MSSM possesses new scalars---squarks and sleptons---which carry
baryon number ($B$) and lepton number ($L$), respectively.
Consequently, it is possible to write down additional scalar--fermion
interaction terms which violate either $B$ or $L$ (but preserve
supersymmetry).
In the MSSM, one simply assumes that these terms are
not present.  This can be implemented with a discrete symmetry,
called $R$-parity which assigns $(-1)^{2S+3(B-L)}$
to particles of spin $S$ [{\it i.e.}, $+1$
to all Standard Model particles and $-1$ to their supersymmetric
partners], and is multiplicatively conserved by all
interactions of the model.   $R$-parity conservation guarantees the
existence of a stable lightest supersymmetric particle (LSP), which is an
excellent candidate for dark matter.   Although there are some
strong phenomenological constraints on $R$-parity violating interactions, it
is possible to construct $R$-violating models which are not
inconsistent with present data.  Such models lie outside the scope of
the MSSM, and will be briefly mentioned at the end of this section.

The SUSY-breaking part of the MSSM is where most of the new
parameters of the MSSM lie.  In addition to breaking the mass
degeneracy between particles and their supersymmetric partners,
the SUSY-breaking terms are required in order to break the
SU(2)$\times$U(1) electroweak gauge symmetry.  The Higgs potential,
with soft-SUSY breaking mass terms included, can have a non-trivial
minimum.  Both neutral Higgs fields acquire vacuum expectation
values; this introduces an important new parameter, $\tan\beta\equiv
v_2/v_1$, where $v_i\equiv\langle H_i^0 \rangle$.  Note that
after electroweak symmetry breaking,
the Yukawa coupling matrices are proportional to the corresponding quark and
lepton mass matrices.  (In the limit where only third generation
Yukawa couplings are considered,  $h_\tau=\sqrt 2m_\tau/(v\cos\beta)$,
$h_b=\sqrt 2m_b/(v\cos\beta)$, and $h_t=\sqrt 2m_t/(v\sin\beta)$,
where $v=(v_1^2+v_2^2)^{1/2}=246$~GeV.)

Supersymmetric breaking parameters are
not completely arbitrary, since they must preserve the stability of
the hierarchy discussed above.  Such terms, called ``soft'',
were classified in ref.~\cite{dmpsoft}, and consist of
masses for scalars (but not their fermionic partners), masses for
the gauginos, and trilinear scalar couplings.
If there were no reason for the scalar masses and trilinear
couplings to be diagonal in the same basis as the
Yukawa couplings, then there would be a huge number ($>100$) of possible
new SUSY-breaking parameters.
In models based on supergravity \cite{dmpsupergravity},
a dramatic simplification
occurs. SUSY is presumed to be broken
in a ``hidden sector" of particles which have
only gravitational interactions with the ``visible sector" of quarks,
leptons, gauge and Higgs bosons, and their
supersymmetric partners. The SUSY breaking is then transmitted
to the visible sector by gravitational interactions, which results in
the soft-SUSY breaking terms of the MSSM, characterized by a mass scale
$M_{\rm SUSY}\lsim{\cal O}(1~{\rm TeV})$.
Moreover, in the minimal approach,
the soft-SUSY breaking terms are taken to be flavor independent and
can be summarized in terms of a few universal parameters,
regardless of the unknown physics in the hidden sector.
The SUSY-breaking scalar interactions which arise take the form
\begin{eqnarray}
{\cal L}_{\rm soft}& = & m_0^2 \Bigl [ |H_1|^2 + |H_2|^2 \nonumber \\
& & +\! \sum_{\rm families}\! (|Q_L|^2+ |U_R|^2+ |D_R|^2+ |L_L|^2+ |E_R|^2)
\Bigr ]\nonumber  \\
& & + A_0 (
h_U Q_L H_2 U_R + h_D Q_L H_1 D_R + h_E L_L H_1 E_R ) \nonumber \\[4pt]
& & + B \mu H_1 H_2 +\ {\rm h.c.} \label{edmp3}
\end{eqnarray}
This simple form of the SUSY-breaking Lagrangian, which depends on three
new SUSY-breaking parameters: $m_0^2$, $A$, and $B$, holds at a
mass scale close to the Planck mass.  In addition, three soft-SUSY-breaking
Majorana mass terms can be introduced for the fermionic gaugino
partners of the SU(3)$\times$SU(2)$\times$U(1) gauge bosons, respectively.
If one adds the additional assumption of minimal grand
unification at a high energy scale $M_X$, then one finds
a further reduction of parameters.
Both the gauge couplings and the gaugino masses unify at $M_X$.
There may be additional relations among the Yukawa couplings,
although these depend on the details of the superheavy particle
spectrum and their interactions associated with the grand unified
gauge group.
The supersymmetry model parameters can then be taken to be:
(i) the common mass parameter $m_{1/2}$ for the
gauginos,
(ii) a common
scalar mass parameter $m_0$,
(iii) a common trilinear scalar
coupling parameter $A_0$,
(iv) the ratio of the Higgs vacuum expectation values,
$\tan\beta$, and
(v) the sign of $\mu$. (The parameters $|\mu|$ and $B$ can be
eliminated in favor of
$\tan \beta$ and $m_Z$ by minimizing the scalar potential.)
Here $m_{1/2}$, $m_0$, and $A_0$ are all typically of order $M_{\rm SUSY}$.

To find the Lagrangian relevant for physics at
the electroweak scale, one must evolve
the parameters of the model according to their renormalization
group equations from $M_X$ down to $m_Z$.
This splits the gaugino masses and the
various scalar masses at low energies.
The most striking result of this running is that radiative corrections due to
the large top Yukawa coupling push
the squared mass of $H_2$ negative at scales far below $M_X$,
resulting in electroweak symmetry
breaking \cite{dmprewsb}.   This is a remarkable effect, connected
with the large top quark mass, and exhibits the intimate relation
between low-energy supersymmetry and electroweak symmetry breaking.
The large top quark mass also has an impact on a number of other MSSM
parameters via the effects of the large top-quark Yukawa coupling in
the RGE's.  For example, one finds that
the squared masses of the top squarks and left-handed bottom squark
are reduced relative to other squark masses, while the mass of the
lightest CP-even neutral Higgs boson may be substantially increased
above $m_Z$.

\subsection{Features of the Superpartner Spectrum}

With the masses of all the superpartners determined only by the parameters
($m_{1/2}$, $m_0$, $A_0$, $\tan\beta$, sign $\mu$), the spectrum is
highly constrained. With a universal $m_0$, squarks
(or sleptons) with the same gauge quantum numbers will
remain degenerate even at low energies, unless they have large Yukawa
couplings.  This degeneracy is a success of the model,
since it automatically maintains the suppression of FCNCs in low-energy
physics.
(This remains true even if one relaxes the usual minimal assumptions
by introducing a larger gauge group
at some intermediate scale or allowing the gaugino mass parameters to be
independent.)
So the superpartner spectrum will contain seven distinct
groups of
degenerate scalar states
$(\tilde u_L, \tilde c_L)$;
$(\tilde d_L, \tilde s_L)$;
$(\tilde u_R, \tilde c_R )$;
$(\tilde d_R, \tilde s_R, \tilde b_R)$;
$(\tilde e_L, \tilde \mu_L, \tilde \tau_L)$;
$(\tilde \nu_e, \tilde \nu_\mu, \tilde \nu_\tau)$;
$(\tilde e_R, \tilde \mu_R, \tilde \tau_R)$.  If $\tan \beta
\gg 1$, then the bottom and tau Yukawa couplings are non-negligible,
so that $\tilde b_R$, $\tilde \tau_L$, $\tilde \tau_R$, and
$\tilde\nu_\tau$ need
not be degenerate with their colleagues.
With the minimal assumptions, one
finds for the physical squark and slepton masses of the first two families
approximately
\ben \label{en1} \begin{eqnarray}
m^2_{\tilde{e}_R} &=&m_0^2 + 0.15 m^2_{1/2} -0.23 \dt; \label{en1a}\\[2pt]
m^2_{\tilde{e}_L} &=&m_0^2 + 0.52 m^2_{1/2} - 0.27 \dt; \label{en1b} \\[2pt]
m^2_{\tilde{\nu}} &=&m_0^2 + 0.52 m^2_{1/2} +0.50 \dt; \label{en1c} \\ [2pt]
m^2_{\tilde q} &=&m_0^2 + (4.5 \ {\rm to}\ 6.5) m^2_{1/2}. \label{en1d}
\end{eqnarray} \een
Eq.~(\ref{en1d}) holds for first and second generation squark
mass. The range in the coefficient of $m^2_{1/2}$ is partly due to
the uncertainty in $\alpha_3$; moreover, this coefficient is
smaller for larger squark masses.
Eqs.~(\ref{en1}) imply $m_{\tilde q}>
m_{\tilde{e}_L} > m_{\tilde{e}_R}$. At least one of the top squark
mass eigenstates and usually a bottom squark are below the main
squark band implied by (\ref{en1d}),
because of radiative effects from the top Yukawa coupling.
This ordering of squark masses should lead to enhanced production of third
generation squarks at hadron colliders, and to enhanced branching ratios of
gluinos, charginos and neutralinos into  third generation quarks.

In the minimal approach, the physical
gluino mass is related to \mhalf\ by
\begin{equation}
\label{en2} \mgl = (2.2\ {\rm to} \ 3.5) \mhalf
\end{equation}
Therefore, the squarks whose corresponding Higgs--quark Yukawa couplings
are small cannot be much lighter than the gluino.  When correlations among
parameters are taken into account, one finds $m_{\tilde q} > 0.8 \mgl$.
With $m_t\simeq 175$~GeV,
radiative electroweak breaking requires that
$|\mu|$ be large compared to the electroweak
gaugino masses. The lighter two neutralinos and
the lighter chargino will then be gaugino--like, and the heavier neutralinos
and chargino will only rarely be produced in the decays of squarks and
gluinos. This also implies that the heavier
Higgs mass eigenstates $A,\ H^0$ and
$H^{\pm}$ will be quite heavy and almost degenerate; they
will usually also be able to decay into
charginos and neutralinos.
The lightest Higgs scalar, $h^0$, will have
couplings close to those of the Standard Model Higgs boson. There is an
upper bound on the $h^0$ mass of about 120~GeV in the MSSM.
A similar $h^0$ mass bound of about 150~GeV is found in
a completely general class of low-energy supersymmetric
models \cite{dmphiggsbound}
(assuming all model couplings are perturbative up to $M_X$).


\subsection{Beyond the Minimal Planck Scale Assumptions}

Many model builders have begun to
explore the consequences of relaxing the minimal assumptions outlined
above.  For example,
the boundary condition of universal scalar masses at $M_X$ can be
modified by non-trivial string effects, threshold effects
associated with the grand unified model, evolution effects between $M_X$ and
$M_P$, or additional contributions arising from the effects of
a larger gauge symmetry group which is broken at a scale above $M_X$
\cite{dmpdterms}.
Generally, small or moderate changes of the boundary conditions at
$M_X$ will only be important near
the border of the allowed region of parameter space. For example, if the mass
squared of $H_1$ is larger than that of
$H_2$ at the scale \mx, $\tan\beta$ of order  $m_t/m_b$ or larger
becomes possible
without strong fine-tuning of soft breaking parameters.
Non-universal scalar masses may also reduce $|\mu|$,
possibly allowing for light higgsinos.

\par\eject
One can even entertain the possibility of large non-universalities
among sfermions with the same electroweak quantum numbers.
One often finds the statement in the literature that
near--degeneracy of sfermion masses at the scale $\mx$
is required by bounds on
FCNC. This statement is not entirely correct;
the potentially most dangerous gluino loop contributions will
always vanish if quark and squark mass matrices are
aligned \cite{dmpalignment},
i.e.~can be diagonalized simultaneously. This is true {\em independent} of the
(differences between) squark masses. The same is also true for
neutralino loops. In such models only chargino loops contribute to FCNC.
Notice that in general these do {\em not} constrain masses
of $SU(2)$ singlet squarks.
The most sensitive bound
comes from $K^0$--$\overline{K^0}$ mixing.
Demanding that the chargino loop contribution to $\Delta
m_K$ be smaller than the experimental value of $3.5\times 10^{-15}$ GeV gives
\begin{equation}\label{edmp9}
\left|\delta_{\tilde{u} \tilde{c}} \right| \leq 0.8
{\rm TeV}^{-1}  ( 1 + 6 m^2_{1/2} / { m_0^2 } )
\sqrt{ 4.3 m_{1/2}^2 + 0.4 {m_0^2}},
\end{equation}
\vskip1pc\noindent
where $m_0^2$ is the average and $\delta_{\tilde{u} \tilde{c}} \cdot m_0^2$
the difference between the squared $\tilde{u}_L$ and $\tilde{c}_L$ masses at
scale \mx. We see that $\delta_{\tilde{u}\tilde{c}} \simeq 1$ is allowed
even for $m_0 = \mhalf = 100$ GeV.
Highly non--degenerate squarks are therefore allowed {\em if}
quark and squark mass matrices are (approximately) aligned.
This might make squark hunting at hadron
colliders difficult, since the signal from any one squark
state might be too weak to be visible above backgrounds.

One should also consider extensions of the MSSM that go
beyond simple modifications of the MSSM
boundary conditions at the scale \mx. The ``Next--to--Minimal" Supersymmetric
extension of the
Standard Model introduces a gauge--singlet Higgs chiral supermultiplet in
addition to the usual MSSM particles. This variation does not seem to have a
major effect on the spectrum of the MSSM particles \cite{dmpnmssm}, but it does
impact on discovery channels for the Higgs and neutralinos. In
particular, one may have a light Higgs scalar with reduced
couplings to the $Z$ boson.   One can also contemplate low-energy
supersymmetric models in which $R$-parity conservation is absent.
For example, $R$-parity could be broken spontaneously by
a sneutrino vacuum expectation value $\langle \tilde \nu \rangle \not=0$.
One can also replace $R$-parity by an alternative discrete
symmetry \cite{dmpir}
which allows either baryon number or lepton number violation (but not
both!) among the renormalizable interactions.
The phenomenology of $R$-parity violating models is quite different from that
of the MSSM.  This is mainly due to the fact that lightest supersymmetric
particle is no longer stable, but can decay via an $R$-parity violating
interaction.   No systematic study of $R$-parity violating models in the
context of supergravity and super-unified theories has yet been undertaken.

\par\eject



\def\gsim{\lower3pt\hbox{$\buildrel>\over\sim$}}
\def\eslt{E\llap/_T}
\def\etmiss{E\llap/_T}
\def\tg{\tilde g}
\def\tq{\tilde q}
\def\tl{\tilde \ell}
\def\tz{\widetilde\chi^0}
\def\tw{\widetilde\chi^\pm}

\section{Low Energy Supersymmetry:  Phenomenology}

If supersymmetry is associated with the origin of the electroweak
symmetry breaking scale, then low-energy supersymmetry must exist
at a scale of order 1 TeV or below.  That is, supersymmetry must
be discovered at either upgrades of existing colliders or at the
LHC.  Due to its mass reach,
the LHC is the definitive machine for discovering or
excluding low-energy supersymmetry.   Section 4.1 summarizes
the supersymmetry mass discovery potential for hadron colliders.
A variety of signatures are explored.  The conventional models
of low-energy supersymmetry (which are $R$-parity conserving)
possess a stable lightest
supersymmetric particle (LSP) which is electrically (and color)
neutral and therefore interacts very weakly in matter.  It follows that
the LSP behaves like a heavy neutrino which does not interact
in collider detectors.  Its presence would be signaled by excess
missing transverse energy (which could not be attributed to
neutrinos).  Many of the supersymmetry searches rely on the missing
energy signature as an indication of new physics beyond the
Standard Model.  Multi-leptonic signatures also play an
important role in supersymmetry searches at hadron colliders.
(Such signals can also be exploited in the search for
$R$-parity-violating low-energy supersymmetry.)

Suppose that a signal is observed in one of the expected channels.
This would not be
a confirmation of low-energy supersymmetry, unless
there is confirming evidence from other expected signatures.
This presents a formidable challenge to experimenters at the LHC.
Can they prove that a set of signatures of new physics is low-energy
supersymmetry?  Can they extract parameters of the supersymmetric
models with any precision and test the details of the theory?
These are questions that have only recently attracted seriously
study.  It is in this context that a future $e^+e^-$ collider
can be invaluable. As shown in Section 4.2.
if the lightest supersymmetric particles were
produced at LEP-II or the NLC, precision measurement could begin
to map out in detail the parameter space of the supersymmetric
model.  In particular, beam polarization at the NLC provides an
invaluable tool for studying the relation between chirality and
the properties of supersymmetric particles.  For example, one
can begin to prove that there is a correlation between the
left and right-handed electrons and their slepton partners as
expected in supersymmetry.  The elucidation of the details of
the light chargino and neutralino spectrum at the NLC could
also play a pivotal role in untangling complex decay chains of
heavier supersymmetric particles produced at the LHC.

\subsection{Supersymmetry at Hadron Colliders}

\begin{table*}[ht]
\begin{center}
\caption{Discovery reach of various options of future hadron colliders.
The numbers are subject to $\pm$15\% ambiguity. Also, the clean
trilepton signals are sensitive to other model parameters; we show
representative ranges from ref.~\protect\cite{bcpt}
where $|\mu |$ is typically much larger
than the soft-breaking electroweak gaugino masses.
For $\mu>0$, the leptonic decay of $\tz_2$ may be strongly suppressed
so that $3\ell$ signals may not be observable even if charginos
are just above the LEP bound ($m_{\tg}\sim 150$~GeV).}
\protect\label{discovery}
\vskip 0.5\baselineskip
\renewcommand\tabcolsep{10.5pt}
\renewcommand\arraystretch{1.2}
\begin{tabular}{|c|cccccc|}
\hline
&Tevatron I&Tevatron II&Main Injector&Tevatron$^*$&DiTevatron&LHC\\[-2pt]
Signal&0.01~fb$^{-1}$&0.1~fb$^{-1}$&1~fb$^{-1}$&10~fb$^{-1}$&
1~fb$^{-1}$&10~fb$^{-1}$\\[-2pt]
&1.8~TeV&1.8~TeV&2~TeV&2~TeV&4~TeV&14~TeV\\
\hline
$\etmiss (\tq \gg \tg)$ & $\tg(150)$ & $\tg(210)$ & $\tg(270)$ &
$\tg(340)$ & $\tg(450)$ & $\tg(1300)$ \\
$\ell^\pm \ell^\pm (\tq \gg \tg)$ & --- & $\tg(160)$ & $\tg(210)$ &
$\tg(270)$ & $\tg(320)$ & $\tg(1000)$ \\
$all\to 3\ell$ $(\tq \gg \tg)$ & --- & $\tg(150$--$180)$ &
$\tg(150$--$260)$ & $\tg(150$--$430)$ & $\tg(150$--$320)$ & \\
$\etmiss (\tq \sim \tg)$ & $\tg(220)$ & $\tg(300)$ & $\tg(350)$ &
$\tg(400)$ & $\tg(580)$ & $\tg(2000)$ \\
$\ell^\pm \ell^\pm (\tq \sim \tg)$ & --- & $\tg(180$--$230)$ & $\tg(325)$ &
$\tg(385$--$405)$ & $\tg(460)$ & $\tg(1000)$ \\
$all\to 3\ell$ $(\tq \sim \tg)$ & --- & $\tg(240$--$290)$ &
$\tg(425$--$440)$ & $\tg(550)$ & $\tg(550)$ & $\gsim\tg(1000)$ \\
$\tilde{t}_1 \rightarrow c \tz_1$ & --- & $\tilde{t}_1(80$--$100)$ &
$\tilde{t}_1 (120)$ & & & \\
$\tilde{t}_1 \rightarrow b \tw_1$ & --- & $\tilde{t}_1(80$--$100)$ &
$\tilde{t}_1 (120)$ & & & \\
$\Theta(\tilde{t}_1 \tilde{t}_1^*)\rightarrow \gamma\gamma$ &
--- & --- & --- & --- & --- & $\tilde{t}_1 (250)$\\
$\tl \tl^*$ & --- & $\tl(50)$ & $\tl(50)$ & $\tl(100)$ & &
$\tl(250$--$300)$ \\
\hline
\end{tabular}
\end{center}
\end{table*}

The detection of supersymmetry at a hadron collider is a complex
task.  It requires the ability to extract signals of new physics
out of data samples with potentially large Standard Model
backgrounds.  Realistic simulations are required that can accurately
model the physics of QCD jets and the theoretical predictions of
low-energy supersymmetry.  The physics of low-energy supersymmetry
can be treated either in the context of specific theoretical
models such as those discussed in Section 3, or in a more
phenomenological approach which makes as few theoretical assumptions
as is practical.  Both approaches have been followed in this section.
The minimal supersymmetric model (MSSM) \cite{mssm}
has recently been embedded in the
event generator program ISAJET \cite{isajet}, so that all lowest order
sparticle production processes and decay chains can be adequately simulated.
In addition, ISAJET now allows the input of a mixed GUT and weak scale
parameter set
($m_0,m_{1/2},A_0,\tan\beta ,sgn(\mu ),m_t$), for which a complete
superparticle spectrum is generated using renormalization group equations
with radiative electroweak symmetry breaking, under the assumption of
gauge coupling unification at $M_{GUT}\sim 2\times 10^{16}$ GeV,
and universal GUT scale soft supersymmetry breaking terms. In addition,
versions of ISAJET past 7.11 include all lowest order
$e^+e^-\rightarrow$ SUSY reactions, so the effect of cascade decays can be
explored at future $e^+e^-$ colliders~\cite{bbkmt}.

Much effort has gone into examining the superparticle mass reach and
discovery potential of various proposed future colliders, to aid in decision
making and prioritization for future collider facilities.
A summary of these results can be found in Table \ref{discovery}.
First, we consider the case of gluino and squark production,
followed by their cascade decays
via three-body $\tw$ and $\tz$ decay modes.
Kamon {\it et al.}~\cite{kamon} have examined the mass
reach in $m_{\tilde g}$ at the Tevatron and DiTevatron colliders
using the missing transverse energy ($\etmiss$)
signature, and found gluino mass reaches of $m_{\tg}\sim 270(450)$
GeV for the Tevatron Main injector (DiTevatron),
assuming that $m_{\tq}\gg m_{\tg}$. In addition,
ref.~\cite{bkt} has examined
the $\tilde g$ mass reach via {\it multi-lepton} topologies.
The LHC reach for gluinos and squarks has recently been examined
in considerable detail by ATLAS studies \cite{ATLAS}.
It is generally expected that the lighter of the two top squarks
($\tilde t_1$) is
significantly lighter than the other squarks, due to the large
top quark Yukawa coupling which drives diagonal stop masses to small
values and enhances the off-diagonal mixing
of $\tilde t_L$ and $\tilde t_R$.   In this case, the
present squark mass collider limits do not apply to $\tilde t_1$.
In ref.~\cite{bst},
it is shown that Tevatron collider experiments with 100 pb$^{-1}$ of data
can explore $m_{\tilde t_1}\sim 80$--100 GeV, considerably
further than the best current bounds of $\sim 45$ GeV from LEP.

Next, we examine neutralino and chargino production.  A significant
portion of the  parameter space of minimal supergravity
models can be explored at the Tevatron and DiTevatron colliders
by searching for the clean trilepton plus $\etmiss$ signal from
$\tw_1\tz_2\rightarrow 3\ell$
production \cite{bcpt}.
In Table \ref{discovery}, we present the mass reach of the neutralino
and chargino searches in terms of an equivalent gluino mass.
The latter is obtained under the assumption of unified soft
gaugino masses at the GUT scale (although there is some
additional dependence on other supersymmetric model parameters).
Note that
the
DiTevatron does not significantly increase the mass reach in
the $3\ell$ channel beyond
that of the Tevatron Main Injector, although
further optimization of cuts may be possible.
This is in contrast to the mass reach for gluinos and squarks via
$\etmiss$ searches, which is significantly higher for the DiTevatron.
The clean trilepton signal has also been
examined in ref.~\cite{bcpt} for the LHC.
Note that it is not possible to probe $m_{\tilde g} >\sim$600--700 GeV
via $3\ell+\etmiss$ mode at the LHC due to the turn on
of the $\tz_2\rightarrow\tz_1 h$
or $\tz_2\rightarrow \tz_1 Z$ decay modes which
spoil the signal. However,
the dilepton spectrum from $\tz_2\to\tz_1\ell\bar \ell$ decay can be
used at the LHC
to measure $m_{\tz_2}-m_{\tz_1}$ with significant precision.
Finally, if
$R$-violating $\tz_1$ decays are present (via baryon or lepton number
violating interactions), then sparticle mass reaches can change
significantly to either higher or lower values \cite{barger}.

The search for sleptons at hadron colliders was addressed
in ref.~\cite{slepton}.
It was shown that it would be difficult
to search at the Tevatron for $m_{\tilde \ell}$
much beyond 50 GeV (and beyond 100~GeV at TeV$^*$), due to low signal rates
and irreducible backgrounds. In contrast, at
the LHC stiffer cuts may be made; in this case slepton masses
up to $250$--$300$ GeV can be detected.

Tevatron signals within the constrained supergravity framework have been
discussed in ref.~\cite{bcmpt}. Because all signals are determined by just
four parameters, the various SUSY processes are strongly correlated,
and may provide an experimental
probe of GUT scale assumptions. For example, ref.~\cite{bgkp} argues
that in the parameter regime where sleptons are much lighter
than squarks, the Main Injector discovery limit
may  reach  $m_{\tg}\sim 500$ GeV via the
enhanced multi-lepton signals from supersymmetric particle
production and their subsequent decay chains.


\subsection{Supersymmetry at e$^+$ e$^-$ Colliders}

At an $e^+ e^-$ collider, discovery of superparticles is relatively easy
up to the kinematic limit. There is a vast literature concerning the
discovery reach at LEP-II \cite{LEP-II}. Recent works discuss
measurement of SUSY parameters rather than discoveries alone.
For instance with a chargino pair production alone, one can put stringent
constraints on the parameter space $(M_1, M_2, \mu, \tan\beta,
m_{\tilde{l}})$ using the cross section, forward-backward asymmetry
and branching fractions, where the sensitivity on particular parameters
depends on the actual parameter values \cite{Jonathan}.

Experimentation at linear $e^+ e^-$ colliders would be richer due to the
higher energies and beam polarization with the following characteristics:
(i) small beam energy spread (better than 1~\%) even
after including initial state radiation and beamstrahlung effects,
(ii) clean environment basically without underlying events even
with photon induced hadronic processes,
(iii) high beam polarization of 90\% and
beyond. The virtues of polarization are two-fold. First, use of
polarization can suppress the background substantially by as much as two
orders of magnitudes, resulting in a very pure signal sample
appropriate for precision studies.
Second, one basically doubles the
number of experimental observables by using both polarization states,
which enables the efficient measurements of various parameters.

A detector simulation of $e^+e^- \to \tilde\mu^+\tilde\mu^-$
based on an ALEPH-type detector at $\sqrt{s} = 500$~GeV and
$\int {\cal L} dt = 20~\mbox{fb}^{-1}$ gives a 5~$\sigma$ signal for
$\tilde{\mu}$ for $m_{\tilde\mu}$
below 230~GeV \cite{Becker}. A mass difference of
25~GeV between slepton and the LSP is enough for detection even without
employing beam polarization. The case for the chargino search
is similar. It is
noteworthy that both purely hadronic and mixed hadronic-leptonic modes
can be used for discovery and study of charginos \cite{Grivaz}.

\begin{figure}[hp]
\centering{%
\centerline{\psfig{file=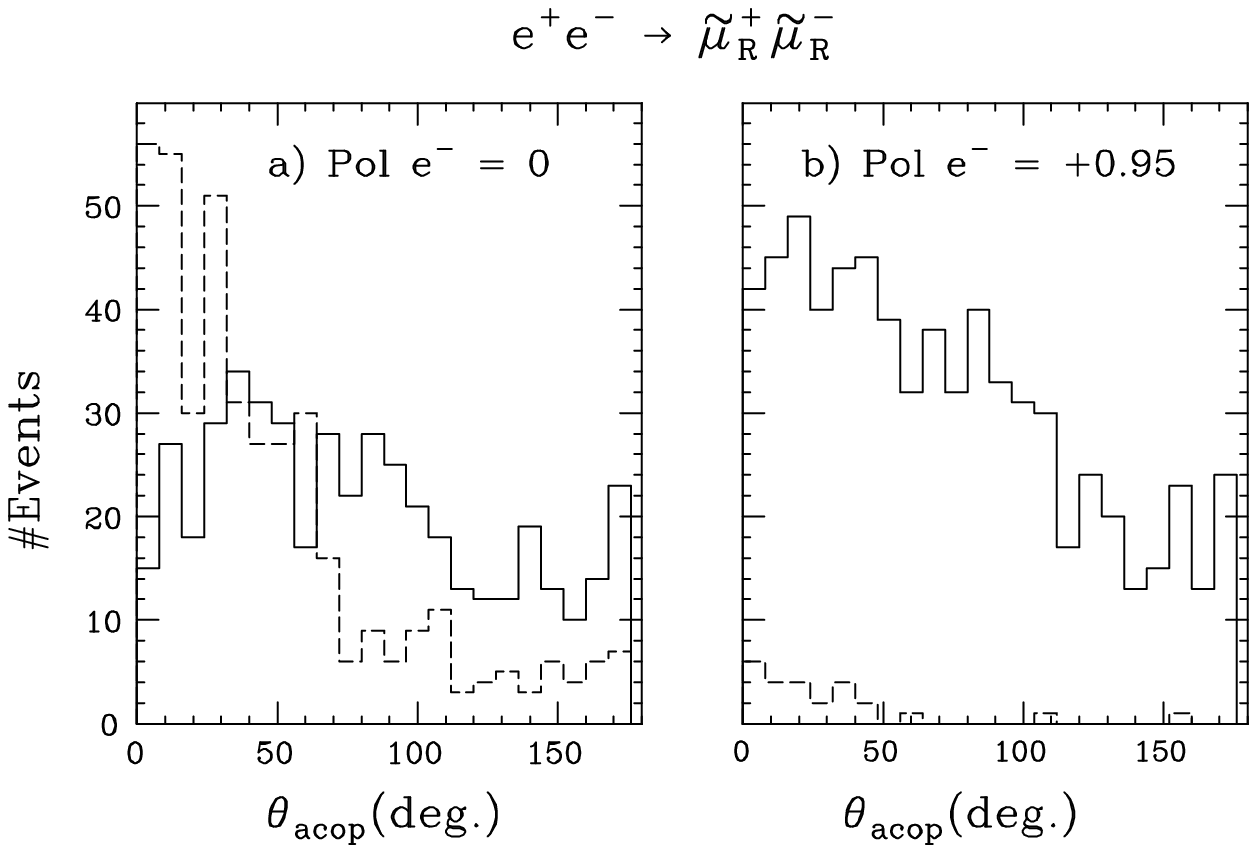,width=8.7cm}}
}\vskip-6pt
\caption{Acoplanarity distribution of muon pair for $E_{CM}=
350~$GeV, $m_{\tilde{l}} = 142$~GeV, $\int {\cal L} dt =
20$~fb$^{-1}$, for the signal (solid) and standard model backgrounds
(dashed).
Fig.~b) shows
the dramatic reduction of the backgrounds using the right-handed
electron beam. Taken from Ref.~\protect\cite{Tsukamoto}.}\label{acop}
\vskip1pc
\centering{%
\centerline{\psfig{file=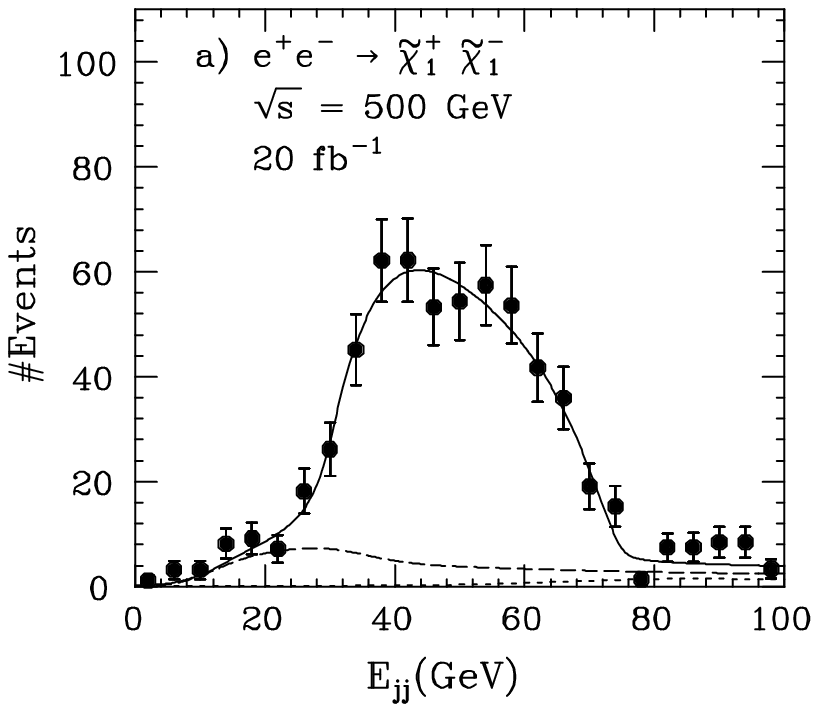,width=6.7cm}}
}\vskip1pc
\centering{%
\centerline{\psfig{file=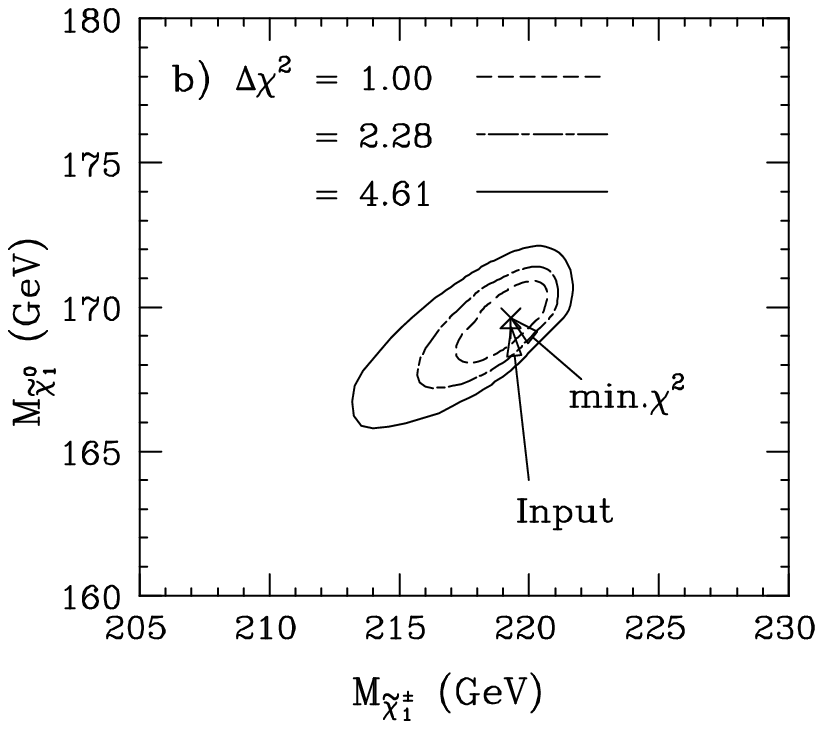,width=6.7cm}}
} \vskip-6pt
\caption{a) Dijet energy distribution from chargino pair
production, for the Monte Carlo generated events (points),
polynomial fit to the Monte Carlo (solid),
contributions from the $W$-pair background (dashed) and
the sum of other standard model backgrounds (dotted).
b) Obtained contour on the masses of chargino and LSP.
Taken from Ref.~\protect\cite{Tsukamoto}.}
\label{chargino}
\end{figure}

We show here the sample results of Monte Carlo analysis with detector
simulation \cite{Tsukamoto} based
on JLC-I detector \cite{JLC-I}.  Fig.~\ref{acop}
shows the power of the polarization to obtain a pure sample ($>$99\%) of
$\tilde{\mu}$ pair signals with high efficiency ($\epsilon > 50$\%). Since the
signature of a slepton pair is an acoplanar lepton pair,
the main background is
$W^+ W^-$, which can be substantially suppressed by employing
right-handed electron beam. 
A kinematic fit to the end points of lepton energy distribution
allows us the determination of both slepton and LSP masses better than
1\%. The typical sensitivity of the chargino and neutralino
mass measurements
are shown in Fig.~\ref{chargino}. Here the mixed hadronic-lepton mode is
used, and the measurement is based on total dijet energy
distribution. Using a technique to combine information from the
tracking detector and
calorimeters, one achieves mass measurements of the chargino and
LSP at the 3\% level.
Similar sensitivities are expected for top and bottom squark signals~%
\cite{Bartl}.

\begin{figure}[htb]
\centering{%
\leavevmode
\centerline{\psfig{file=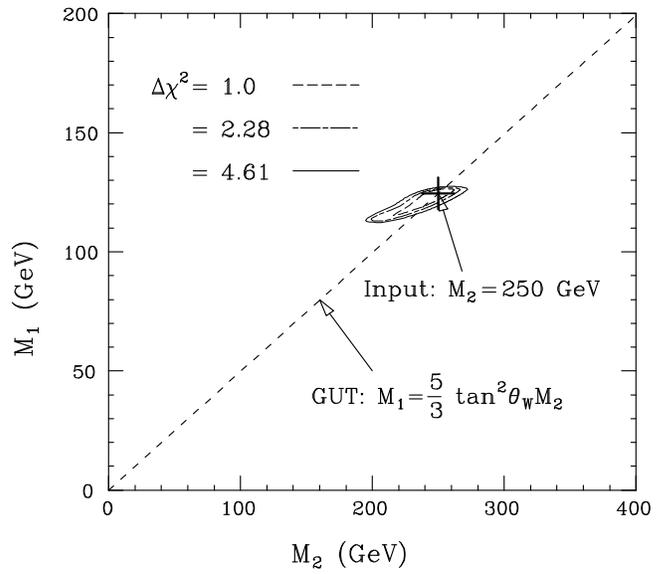,width=8.5cm}
}
\vskip-6pt
\caption{Test of the GUT-relation of the gaugino masses. The
contours are obtained by a four-dimensional fit of $(M_1, M_2, \mu, \tan
\beta)$ plane, using the LSP and chargino masses,
and the production cross sections
for selectrons and charginos with a right-handed electron beam.
Taken from Ref.~\protect\cite{Tsukamoto}.} \label{GUT}
}
\end{figure}

The determination of superparticle masses allows us to test theoretical
assumptions at various levels.
For example, generation independence of slepton masses (universality) can
be tested at the 1\% level. More experimentally challenging is the test
of the GUT-relation among gaugino masses, since the gauginos mix
with higgsinos to
form chargino and neutralino mass eigenstates. However, one can still
test the GUT-relation at a few percent level combining slepton and chargino
signals, based on the measured masses and polarization dependence of the
cross sections (see Fig.~\ref{GUT}).

\subsection{Complementarity}

%
%

Typically, $e^+ e^-$ colliders have lower center-of-mass energies
compared to hadron machines of the same era, but their complementary
character in superparticle searches is easily understood. The
colored superparticles which are the main targets of the hadron
colliders are supposed to be 3--4 times heavier than the colorless ones
under GUT assumptions. For example, the constraints on the
supersymmetric parameter space from experiments at LEP ($\sqrt{s} =
90$~GeV) and the Tevatron ($\sqrt{s} = 1.8$~TeV) are comparable. This
continues to be true for LEP-II and the Tevatron with Main Injector,
especially in the context of models with minimal supergravity
boundary condition constraints \cite{bcmpt,bgkp}. The
discovery reach of an $e^+ e^-$ linear collider with $\sqrt{s} = 1$~TeV
will be again comparable to that at the LHC.
But their complementary character will be more important {\it after}\/
the discoveries of superparticles. Suppose that signals from
squarks and gluinos are seen at the LHC.
Then it would be a formidable task to demonstrate
that they are superparticles due to the simultaneous production of
many new particles and their complicated decay chains.
The $e^+ e^-$ collider might not reach the energies where
squarks could be produced; however, it would study the low-lying
spectrum of
superparticles (the sleptons, neutralinos and charginos) as
described in the previous section.
Analysis of data from the hadron collider with inputs from the
$e^+ e^-$ collider could sort out the complex decay chains,
revealing the mass spectrum and parameters of the heavier
superparticles.  These results
would augment the measurements of the properties of light
superparticles made at the $e^+e^-$ collider,
and would provide additional
tests of GUT and/or supergravity predictions and
test other model assumptions.


\font\tenrm=cmr10
\def\epm{e^\pm}
\def\ep{e^+}
\def\emm{e^-}
\def\wpm{W^\pm}
\def\wmp{W^\mp}
\def\wp{W^+}
\def\wm{W^-}
\def\wl{W_L^{}}
\def\wt{W_T^{}}
\def\pt{p_T^{}}
\def\mww{M^{}_{WW}}
\def\mw2{M_W^2}
\def\mh2{m_H^2}
\def\fbi{{\rm fb}^{-1}}
\def\to{\rightarrow}
\def\gsim{\buildrel {\mbox{$>$}} \over {\raisebox{-0.8ex}{\hspace{-0.05in}
$\sim$}}}
\def\lsim{\buildrel {\mbox{$<$}} \over {\raisebox{-0.8ex}{\hspace{-0.05in}
$\sim$}}}

\section{Strongly Coupled ESB Sector: Model- \hfill\break
Independent Approaches}

If a Higgs boson or techni-rho meson or some other direct indication
of electroweak symmetry breaking is not found at some future accelerator,
one would like to know whether it is still possible to learn something
about  electroweak symmetry breaking by studying $WW$ scattering.  In
this scenario, whatever is responsible for giving the $W$ and $Z$ bosons
their masses occurs at a high energy scale and most probably gives rise to a
strongly interacting sector where perturbation theory is not applicable.
Fortunately, one can still analyze the low energy behavior of the theory
with a generalization of the $\pi\pi$ scattering theorems derived by
Weinberg cast in the form of chiral perturbation theory.

The low energy theorems can easily be generalized from pions to the
longitudinal components of the $W$ and $Z$ gauge bosons.  These scattering
theorems are only valid at low energy, that is $\sqrt{s}<<(M_{\rm Res},
4\pi v)$, where $M_{\rm Res}$ is the mass of the lowest
lying resonance interacting with the longitudinal $W$'s and $Z$'s.
The attraction in this approach
is that it is valid for ${\it any}$ mechanism of electroweak
symmetry breaking.  It is also possible to couple the chiral
Lagrangian describing the low energy interactions of the gauge bosons
with a (color-singlet)  vector meson
(the BESS model-equivalence) or with a scalar particle
(a Higgs scalar type of models).  This description of vector mesons
and scalars coupled to longitudinal gauge bosons relies only on
the symmetry of the theory and so is independent of the details
of specific models.

Lastly, we discuss
a model with a ``hidden'' symmetry breaking sector, one in
which the electroweak symmetry breaking sector
has a large number of particles.
The theory can be strongly interacting and the {\it
total} $W$ and $Z$ cross sections in this model are
large: most of the cross section is for the
production of particles other than the $W$ or $Z$.
In such a model,
discovering the electroweak symmetry breaking sector depends on the
observation the other particles and the ability to associate them with
symmetry breaking.

\subsection{Strongly Coupled ESB Sector at the LHC}

In what follows, we assume the LHC runs at 14 TeV with
a luminosity of 10$^{34}$~cm$^{-2}$~sec$^{-1}$
(an annual integrated luminosity 100 fb$^{-1}$).

\vskip6pt
\leftline{\it Vector Resonance Signal}
\vskip6pt

  A vector meson with a mass  up to
$m_\rho=2.5$~TeV can be found at the LHC through
the  $W^\pm Z$ channel~\cite{bessnew}.
One would need to run the LHC for about 2 years
to collect about 10 leptonic events, after kinematical cuts to effectively
suppress the backgrounds.

\vskip6pt\noindent
{\it Complementarity of  Vector Resonant and Non-resonant
$WW$ Scattering:}
\vskip6pt

Ref.~\cite{chankilg} considered a vector-dominance model
for three sets of mass-width parameters:
($m_\rho, \Gamma_\rho$) = (1.78,0.33), (2.52,0.92), and
(4.0,0.98).
Their criterion for a significant signal ($S$) with a background ($B$) is
$\sigma^{\uparrow}   \equiv  S/\sqrt{B}  \ge  5; \
\sigma^{\downarrow}  \equiv  S/\sqrt{S+B}  \ge  3; \ S \ge B$.
The complementarity here implies that one observes either
a low mass ($I,J$)=(1,1) resonance via  the $W^\pm Z$ channel, or significant
enhancement via the $W^\pm W^\pm$ channel with $I=2$. From Table~\ref{minlum},
one sees that the worst case among those parameter choices
is at $m_\rho=2.52$ TeV, where
an integrated luminosity of  at least 105 fb$^{-1}$---%
slightly more than the LHC annual luminosity---is required for the
$W^\pm W^\pm$ channel to meet their discovery criterion.
ATLAS collaboration \cite{ATLAS}
carried out  a Monte Carlo study for  $W^+ W^+$ channel
including  detector simulations, obtaining
qualitatively similar results.

\begin{table}[htb]
\begin{center}
\caption{\label{minlum}
Minimum luminosity to satisfy observability criterion for
$W^{\pm}Z$ and $W^\pm W^\pm$ channels at the LHC.
Each entry contains ${\cal L}_{MIN}$ in fb$^{-1}$, the number of
signal/background events per 100 fb$^{-1}$.}
\vskip1pc
\setlength{\tabcolsep}{10pt}
\newcommand\STRUT{\rule{0in}{2.5ex}}
\begin{tabular}{|lccc|}
\hline
$m_\rho$ (TeV) \STRUT & 1.78 & 2.52 & 4.0 \\ \hline\hline
    & 44 fb$^{-1}$\STRUT  & 325 fb$^{-1}$ & No \\
$W^\pm Z$  &38/20   & 5.8/3.4 & Signal \\
\hline
   & 142 fb$^{-1}$ \STRUT & 105 fb$^{-1}$ & 77 fb$^{-1}$ \\
$W^\pm W^\pm$  & 12.7/6.0 & 15.9/5.8 & 22.4/8.9 \\
\hline
\end{tabular}
\end{center}
\end{table}
\vskip6pt
\leftline{\it All-Channel Comparison for  $WW$ Scattering}
\vskip6pt

Ref.~\cite{baggeretal} compared all-channel $WW$ scattering at
$\sqrt s=40$ TeV and 16 TeV, and new analysis at 14 TeV is in
progress.  In particular, the signal for a 1 TeV Higgs boson may be
observable through the leptonic channels
$H \rightarrow WW \rightarrow l \nu l\nu$
and  $ZZ \rightarrow l \bar l \nu \bar \nu$,
to add to the clean but low rate $ZZ \rightarrow 4l $ mode.

ATLAS collaboration \cite{ATLAS} also studied a 1 TeV Higgs boson
signal via the decay mode $H \rightarrow WW \rightarrow l \nu jj$
and $ZZ \rightarrow l \bar l jj$.
It was found that after identifying one $W$ in the leptonic mode
and one $W$ in the di-jet mode with $M_{jj} \sim m_W$, with
a tagged forward jet and a veto on all other central jets, the
$WW\rightarrow\ell\nu jj$ channel gives a viable signal with
 statistical significance above 6.8$\sigma$.

\subsection{Strongly Coupled ESB Sector at the NLC}

Our standard NLC has an energy $\sqrt s=$1.5 TeV and an integrated
luminosity of 190~$\fbi$ per year.

\vskip6pt
\leftline{\it Vector Resonance Signal}
\vskip6pt

At the NLC a vector resonance would be probed most effectively
via  $e^+e^- \rightarrow W^+_LW^-_L$. This can be generically
described \cite{peskin}  by an Omn\'es function
with a complex form factor $F_T$.
An infinitely massive vector resonance is equivalent to the case
where the $W^+_LW^-_L$ scattering amplitude is described by the
Low Energy Theorem (LET).

Fig.~\ref{barklowf} shows the results in Ref.~\cite{barklow}
for a technirho-like vector
resonance in the $\ep\emm\to \wp\wm$ process.
In one year at the NLC,
technirhos of arbitrarily large mass
can be distinguished from the weakly-interacting SM.
An integrated luminosity of 225~$\fbi$ (1.2 years at design luminosity)
would yield $7.1\sigma$, $5.3\sigma$, and $5.0\sigma$ signals for
a 4~TeV techni-rho, a 6~TeV techni-rho, and LET, respectively.
As was the case for the LHC of Table~\ref{minlum},
these results assume statistical errors only.
The systematic errors for the
$e^+e^- \rightarrow W^+W^-$ results  will be  small
compared to the statistical error, since the analysis
consists of a background-free maximum
likelihood fit of angular distributions.

%
\begin{figure}[htb]
\centering{%
\vspace{1pc}
\leavevmode
\centerline{\psfig{file=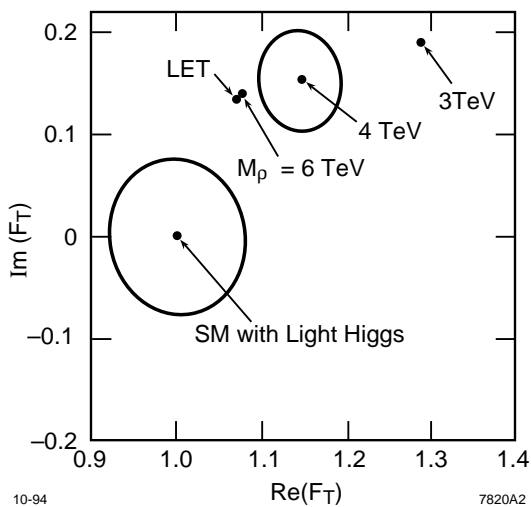,width=7cm}}
}
\vspace{-1pc}
\caption{\label{barklowf}
Confidence level contours for $Re(F_T)$ and $Im(F_T)$ at
c.m. energy 1.5 TeV and ${\cal L}=190~\fbi$.
The contour about the light Higgs value of $F_T=(1,0)$ is 95\%
C.L. and the contour about $m_\rho=4$ TeV point is 68\% C.L..}
\end{figure}

The processes of $\ep\emm \to \wp\wm$ and $\ep\emm \to f \bar f'$ via
the vector resonance have also been studied in the context of the BESS
model and variables such as the
total cross sections, forward-backward and left-right asymmetries
are used to constrain the virtual $V$ effects.
For $\sqrt s \sim m_\rho$, the NLC will be
very sensitive to the BESS model parameters~\cite{dominici}.

\vskip6pt
\leftline{\it $WW$ Fusion Processes}
\vskip6pt

$WW$ fusion processes open more channels to study the strongly coupled ESB
sector
beyond a vector resonance. Ref.~\cite{mine} demonstrates
that the signal for a 1 TeV Higgs boson should be observable at
a 12$\sigma$ level, for $\sqrt s=$1.5 TeV and 200~$\fbi$.
Furthermore, the ratio
$R\equiv \sigma(W^+W^- \rightarrow W^+W^-)/\sigma(W^+W^- \rightarrow ZZ)$
may provide hints that help to distinguish
different strongly coupled ESB sector models, such as a scalar resonance
($R>1$),
a vector resonance ($R \gg1$), or a Low Energy Theorem amplitude ($R < 1$).

Due to the relatively clean environment in $e^+e^-$ collisions, cross
sections for background processes may be measured rather well
for both the normalization as well as the shape of the distribution, so that
statistically significant deviation from the SM background may be easily
identified as new physics signal.  One example is the ``Hidden Sector''
\cite{chivgold}, which may be impossible to observe in hadronic collision
environment but  possibly  feasible in $e^+e^-$ collisions.

\vskip6pt
\leftline{\it Searches at $\gamma\gamma$ Colliders}
\vskip6pt

The heavy Higgs boson and other strong scattering signals in
longitudinal gauge boson channels
can be investigated at the NLC at $\sqrt{s}=1.5$~TeV operating in the
$\gamma\gamma$ collider mode \cite{mberger}.  The processes
considered here involve two gauge boson final states
$\gamma \gamma \to ZZ$,  $\gamma \gamma \to W^+W$, and four gauge
boson final states $\gamma \gamma \to W^+W^-W^+W^-$, $W^+W^-ZZ$
\cite{brodsky}.
The four gauge boson final state processes seem to be more promising in
searching for strongly coupled ESB dynamics as compared with the
two gauge boson final states,  although all the above processes
are plagued by large backgrounds consisting of transversely produced
gauge bosons \cite{mberger}.  This
issue has been addressed recently in Refs.~\cite{jikia,cheung}.
For example, Ref.~\cite{jikia}  finds that
with luminosity of 200 fb$^{-1}$, a 1.5 TeV $e^+e^-$ collider
operating as a $\gamma \gamma $ collider can successfully observe a
heavy Higgs boson with mass up to 700 GeV.
To obtain a reach in the Higgs boson mass
up to 1 TeV, a 2 TeV $e^+e^-$ collider
is required with luminosity of 200 fb$^{-1}$. The results were obtained
taking into account the back-scattered photon luminosity spectrum.
It is also found  that a 2 TeV linear
collider in the $\gamma \gamma $ mode is roughly equivalent to a 1.5
TeV $e^+e^-$ collider in searching for heavy Higgs boson physics
(roughly scales as $\sqrt{s_{\gamma \gamma}}\sim 0.8\sqrt{s_{ee}}$).
The viability of this signal still requires that the decays of the
$W$ and $Z$ bosons be incorporated, and detector simulations be performed
to determine a realistic acceptance.
%

\section{Strongly Coupled ESB Sector: Implications of Models}
%

\subsection{Ingredients of Techni-models}

In theories of dynamical electroweak symmetry breaking (ESB), such as
technicolor \cite{TC}, ESB is due to chiral symmetry breaking
in an asymptotically-free, strongly-interacting, gauge theory with massless
fermions.
%
These theories attempt to understand electroweak symmetry breaking
without introducing any scalar particles such as a Higgs boson.
Instead, the symmetry breaking is modeled after that which
occurs in QCD when the $SU(3)_c$ color gauge group becomes strong.
In QCD, the breaking of the chiral symmetry gives rise to pions.
In the simplest technicolor theory one introduces a left-handed
weak-doublet of ``technifermions'', and the corresponding right-handed
weak-singlets, which transform as
the fundamental [$N$] representation of a strong $SU(N)_{TC}$
technicolor gauge group. The global chiral symmetry of the
technicolor interactions is then $SU(2)_L \times SU(2)_R$. When the
technicolor interactions become strong, the chiral symmetry is broken
to the diagonal subgroup, $SU(2)_{V}$, producing three Nambu-Goldstone
bosons, which  then yield the correct $W$ and $Z$ masses.

However, the symmetry breaking sector must also couple to the ordinary
fermions, allowing them to acquire mass.  In models of a strong ESB
sector there must either be additional flavor-dependent gauge
interactions, the so-called ``extended'' technicolor
(ETC) interactions, or Yukawa couplings to scalars which
communicate the breaking of the chiral symmetry of the technifermions
to the ordinary fermions.  The most popular type of strong ESB model
that attempts to explain the masses of all observed fermions contains
an entire family of technifermions with Standard Model gauge
couplings.

Since these models contain more than one doublet of technifermions, they
have a global symmetry group larger than $SU(2)_L
\otimes SU(2)_R$.  Therefore chiral symmetry breaking
produces additional (pseudo-)Nambu-Goldstone bosons (PNGBs).
Furthermore, the models typically possess a larger
variety of resonances than the one-doublet model.
There will be heavy ($\sim 1$--2~TeV) vector meson resonance
with the quantum numbers of the $\rho,\omega$ etc.
  The phenomenology
of color-neutral  resonances with the quantum
numbers of the $\rho$ is
discussed in Section 5. In the
case of color-neutral  PNGBs the phenomenology is
much like that of the extra scalars in
``two-Higgs'' doublet models. Therefore we will focus
on the colored resonances and PNGBs occuring in most technicolor
models. These
models have a possibly unique signature: resonances associated with
the {\it electroweak} symmetry breaking sector which are {\it strongly
produced}.  Hadron colliders will be especially important for
searching for signatures of such colored particles associated with
ESB.

\begin{table*}[ht]
\begin{center}
\begin{minipage}{13.3cm}
\caption{Discovery reach (in GeV) of different accelerators
for particles associated
with realistic models of a strong ESB sector.}
\label{discovery6}
\vskip6pt 
\setlength{\tabcolsep}{7.6pt}
\renewcommand\arraystretch{1.2}
\begin{tabular}{|lc|c|c|c|c|c|}\hline
&        $(SU(3)_C,$ &          &      &      &        &    \\
Particle& $SU(2)_V)$ &Tevatron  &LHC   &LEP I &LEP II  &NLC-1000\\
\hline
$P^{0 \prime}$&(1,1)&---&110--150$^a$ &8$^{b}$;28$^{b}$ &---$^c$&---$^{c,d}$\\
$P^{0 }$&     (1,3) &---&---          &---           &--- &---$^d$\\
$P^{+} P^{-}$&   (1,3) &---    &400$^e$ &45$^f$ &100$^g$&500$^g$  \\ \hline
$P_8^{0 \prime} (\eta_T)$ &  (8,1) &400--500$^h$ &325$^i$
  &--- &---     &---   \\
$P_8^{0}$d&  (8,3) &10--20$^i$ &325$^{i,j}$   &---     &---   &--- \\
$P^{+}_8 P^{-}_8$&   (8,3)  &10--20$^{i,j}$
&325$^{i,j}$   &45$^f$
         &100$^g$  &500$^g$  \\ \hline
$P^{+}_3 P^{-}_3$&   (3,3)  & ---$^j$ & ---$^j$ &
         &100 GeV$^g$  &500 GeV$^g$  \\ \hline
\end{tabular}
%
\vskip1pc   \footnotesize
$^a$
Decay mode $P^{0 \prime} \rightarrow \gamma \gamma$ ,
similar to a light neutral Higgs \cite{TDR}. \\
$^b$
Decay mode $Z \rightarrow \gamma P^{0 \prime}$,
assuming an one-family model, with $N_{TC} = 7$ and $N_{TC}= 8$ respectively;
no reach for $N_{TC} < 7$; for larger $Z \gamma P^{0 \prime}$ couplings,
the discovery reach extends to $65$ GeV \cite{lep}. \\
$^c$
No reach for one-family model; possibility of reach for the Lane-Ramana~
\cite{lane-ramana}  multiscale model in the processes
$e^+ e^- \rightarrow P \gamma \;\; , \;\; P e^+ e^- $ \cite{lubicz}. \\
$^d$
The analogous discovery reach for a CP-odd Higgs scalar can be greatly
improved using the NLC operating in the $\gamma\gamma$ collider
mode (with polarized photon beams) \cite{GHBBC}. \\
$^e$
Estimated from work on charged Higgs ($g b \rightarrow t H^- \rightarrow
t \bar{t} b$) for $\tan \beta \simeq 1$~\cite{hplus}. \\
$^f$
LEP I kinematical limit.\\
$^g$
Kinematical limits for LEP200 and the NLC-1000. \\
$^h$
Contribution to the $\bar{t} t $ cross section in multiscale models
\cite{EL}. \\
$^i$
QCD pair production of colored PNGBs with decay into $4$ jets
\cite{jet}.\\
$^j$
QCD pair production of colored PNGBs with
decay into $t\bar t$,
$t\bar b$, $\tau^+\tau^-$
or $\tau \nu_\tau$ may allow higher reach in mass.  This has
  yet to be studied.\\
\end{minipage}
\end{center}
\end{table*}

\subsection{Theoretical Challenges for Technicolor Model \hfill\break
Building}

Models incorporating ETC interactions ran into trouble with flavor
changing neutral currents (FCNC's)  early on.  Assuming
technicolor behaved like a scaled-up version of QCD it was found that
it is impossible to generate a large enough mass for the $s$ quark
(much less the $c$ quark) while avoiding FCNC's.  One possible
solution is to make the TC gauge coupling run slower than in QCD
(\ie\ make it `walk').  This makes the
technifermion condensate much larger than scaling from QCD suggests,
so that for a fixed quark or lepton mass, the necessary mass scale for
ETC gauge bosons is increased, suppressing FCNC's.  A second
solution is to build a Glashow-Illiopoulos-Maiani (GIM)
symmetry into the ETC theory (sometimes called
TechniGIM).

A further problem for ETC models is producing a large isospin
splitting for the $t$ and $b$ quark masses without causing large
isospin splitting in the $W$ and $Z$ masses.  Isospin
splitting in the gauge sector, as described by the radiative correction
parameter $T$ is tightly constrained
by experiment.\footnote{The radiative correction parameters,
$S$, $T$ and $U$ are defined in the chapter on
Precision Tests of Electroweak Physics.}

A current challenge for TC models is that models with
QCD-like dynamics tend to produce large positive contributions to the
electroweak radiative correction parameter $S$, while experiments tend
to prefer small $S$. Since the discrepancy grows with the number of
technicolors and the number of technidoublets, one way to evade the
problem is to limit the number of technifermions contributing to ESB.
Alternatively one can invoke mechanisms yielding a negative
contribution to $S$: the technifermions may have exotic electroweak
quantum numbers or the TC dynamics can be sufficiently
unlike QCD so as to invalidate the naive scaling-up of QCD
phenomenology.  Such alternatives usually
also rely on some form of isospin breaking.  In multiscale models
isospin breaking at the lowest ESB scale can produce a negative
contribution to $S$  without infecting the $T$
parameter, which is sensitive to isospin breaking at the highest ESB
scale.

The most recent problem for ETC models arises from the measurement of
the $Z\rightarrow b {\overline b}$ partial width at LEP.
  The experiments find a partial width which is
slightly larger than the Standard Model prediction, while almost all
the ETC models that have been constructed so far produce a correction
which further reduces this width.  The sign of the correction is a
consequence of the almost universal assumption that $SU(2)_L$ commutes
with the ETC gauge group.  The alternative, where $SU(2)_L$ is
embedded in the ETC gauge group, may be interesting, since this
reverses the sign of the correction.

\subsection{Phenomenology of Pseudo-Goldstone Bosons}

Many ETC models are expected to include a spectrum of
pseudo-Nambu-Goldstone bosons ($P$'s) and vector resonances
($\rho_T$'s and $\omega_T$) that may carry color quantum numbers.
These particles may be classified by their $SU(3)_C$ and $SU(2)_{V}$
quantum numbers as indicated (for the $P$'s) in the first column of
Table \ref{discovery6}, using the nomenclature of ref.~\cite{EHLQ}.
Since these particles are strongly produced, they have large cross sections
at the Tevatron and LHC.  The color triplet ($P_8$) and color
octet ($P_3$) pseudo-Nambu-Goldstone bosons will be pair produced from
gluon-gluon initial states in hadron colliders and will give rise
to $4$ jet and $t {\overline t}$ final states, since in many models the
PNGBs couple preferentially to heavy fermions. Since $P_3$ and $P_8$
 couple to the $Z$ boson they can also be pair produced in $e^+e^-$
colliders, which are expected to be sensitive to PNGB masses up
to the kinematic limit.  The LHC will be sensitive
to $P_8$ masses up to $\sim 325$~GeV, while an NLC-1000 would
reach $\sim 500$~GeV.  Since colored PNGBs receive their masses from QCD
interactions and typically have masses in the 100--300~GeV
range, both the LHC and NLC-1000 can probe the interesting region.
The expected discovery potential of various colliders for
the pseudo-Nambu-Goldstone bosons is shown in
Table \ref{discovery6}.

While as yet no completely realistic technicolor model exists, the
models all share the general feature of having a rich spectrum of
scalar and vector particles which can be probed at either the LHC
or a high energy $e^+e^-$ collider.


\def\9{\phantom0}
\def\lsim{\lower3pt\hbox{$\buildrel<\over\sim$}\ }
\def\overtext#1{$\overline{#1}$}
\def\to{\rightarrow}


\section{New Gauge Bosons}


%

\subsection{Introduction}

Extended gauge symmetries and the associated  heavy neutral ($Z'$) and/or
charged ($W'$) gauge  bosons  are a feature of many extensions of the Standard
Model (SM).
We survey and compare
the discovery potential of the experiments that will be performed over the
next decade (Tevatron and LEP-II) and future facilities that are being
planned and considered
for the period beyond.

We present  results for heavy gauge
boson physics for the following representative
models: $Z_\chi$  in $SO_{10}\rightarrow SU_5\times U_{1\chi}$,
$Z_\psi$ in $E_6\rightarrow SO_{10}\times U_{1\psi}$,
$Z_\eta=\sqrt{3/8}Z_\chi-\sqrt{5/8}Z_\psi$  in
superstring inspired models in which $E_6$ breaks directly to a rank 5
group \cite{er5m}, as well as
$Z_{LR}$ and $W'_{LR}$ in the LR symmetric models \cite{mohapatra}.
For simplicity, we assume that the neutrino produced in the
$W'$ decay does not produce visible energy in the detector.

\subsection{Discovery Limits of Extra Gauge Bosons}

Current limits on the mass  of new heavy  gauge  bosons
are relatively weak.
In Table \ref{constraints} the direct bounds on $Z'$ production from
the main production channel $p\bar{p} \rightarrow Z'\rightarrow e^+e^-$
from current Tevatron data
as well as the indirect unconstrained  (no assumption on the Higgs sector)
and constrained (specific assumption  on the Higgs sector)
bounds from a global analysis of electro-weak data
are presented \cite{LL}.  From the non-observation of
$pp  \rightarrow W'\rightarrow e\nu_e$ at the Tevatron,
the CDF Collaboration concludes
that $M_{W'_{LR}}> 652$~GeV \cite{cdfwprime}.  In contrast,
the indirect constrained [unconstrained] Tevatron bounds on $M_{W'_{LR}}$
are in the 1.4 TeV [300 GeV] region.

\begin{table*}[hp]
\begin{center}
\begin{minipage}{10.25cm}
\caption{%
Current constraints on $M_{Z'}$ (in GeV) for typical models
(see ref.~\protect\cite{LL}  
 for details) from direct
production at the Tevatron (${\cal L}_{int}=19.6$ pb$^{-1}$), as well as
indirect limits (95 \% C.L.)
from a global electro-weak analysis.\label{constraints}}
\vskip.5pc
\renewcommand\tabcolsep{12.5pt}
\renewcommand\arraystretch{1.1}
\begin{tabular}{|r|ccc|} \hline
       &      &    Indirect&      Indirect\\[-4pt]
Models &Direct& (unconstrained)& (constrained)\\
\hline
$\chi$ & $425 $ & $330$ & $920$ \\
$\psi$ & 415  & 170 & 170 \\
$\eta$ & $440$ & $210$ & $600$ \\
$LR$ & $445$ & $390$ & $1380$\\ \hline
\end{tabular}
\end{minipage}
\end{center}
\vskip9pt
\begin{center}
\begin{minipage}{14.3cm}
\caption{%
Discovery limits for $M_{Z'}$ (in GeV) for typical models achievable at
proposed hadron and $e^+e^-$ colliders.
At hadron colliders
the discovery limits  for $Z'$  [$W^{\prime +}+W^{\prime -}$] are
for typical models  with  10 events in   $e^+e^-\ +\ \mu^+\mu^-$
[($e^+\nu_e +e^-\bar\nu_e\ $)  +\
($\mu^+\nu_\mu+\mu^-\bar\nu_\mu)$] from Drell-Yan production of the
gauge boson.  For $e^+e^-$ colliders the discovery limits are the
99\% C.L. from a
$\chi^2$ fit of the observables: $\sigma(e^+e^-\to \mu^+\mu^-)$,
$R^{had}=\sigma^{had}/\sigma^\ell$, $A^\ell_{FB}$,
and $A^\ell_{LR}$.\label{bounds}}
\vskip6pt
\renewcommand\tabcolsep{9pt}
\renewcommand\arraystretch{1.1}
\begin{tabular}{|l|cc|rrrrr|}
\hline
 Collider &$\sqrt{s}$ [TeV] &${\cal L}_{int}\  [\hbox {fb}^{-1}]$& $\chi$
& $\psi$ & $\eta$ & $LR$ & $W'_{LR}$ \\
\hline
Tevatron ($p\bar{p})$&      \91.8&\9\91&  775&  775&  795&  825&   920 \\
Tevatron$^*$ ($p\bar{p})$&   2\9 & \910& 1040& 1050& 1070& 1100&  1180 \\
Di-Tevatron ($p\bar{p})$& \94\ \9& \920& 1930& 1940& 1990& 2040& 2225 \\
LHC ($pp)$&               14\ \9 &  100& 4380& 4190& 4290& 4530& 5310 \\
\hline
LEP-II ($e^+e^-$)& \90.2&\9\90.5&  695&  269&  431&  493&  \\
NLC ($e^+e^-$) &   \90.5&\950& 3340&  978& 1990& 2560&  \\
NLC ($e^+e^-$)& \91\ \9 & 200& 6670& 1940& 3980& 5090&  \\
NLC ($e^+e^-$)& \92\ \9 & 200& 9560& 3150& 5830& 7210&  \\
\hline
\end{tabular}
\end{minipage}
\end{center}
\vskip9pt
\begin{center}
\begin{minipage}{10.85cm}
\caption{Values of the ``normalized''  couplings (1)
for the typical models. The statistical error bars indicate
how well the coupling could be
measured at the LHC ($\protect\sqrt s = 14$ TeV,
 ${\cal L}_{int}=100\ \hbox{fb}^{-1}$) for $M_{Z'}=1$ TeV.\label{values}}
\vskip.5pc
\setlength{\tabcolsep}{10.5pt}
\renewcommand\arraystretch{1.1}
\begin{tabular}{|r|cccc|} \hline
&$\chi$&$\psi$&$\eta$&$LR$\\ \hline
$\gamma^\ell_L$&$0.9\pm 0.016$&$0.5\pm 0.02$&$0.2\pm 0.012$&
$0.36\pm 0.007$\\
$\gamma^q_L$&0.1&0.5&0.8&0.04\\
$\tilde{U}$&$1\pm 0.16$&$1\pm 0.14$&$1\pm 0.08$&$37\pm 6.6$\\
$\tilde{D}$ & $9\pm 0.57$ & $1\pm 0.22$ & $0.25\pm 0.16$ & $65\pm 11$ \\
\hline
\end{tabular}
\end{minipage}
\end{center}
\vskip9pt
\begin{center}
\begin{minipage}{15.1cm}
\caption{The value of the couplings  (2) for
typical models and statistical error bars as determined from probes
 at   the NLC ($\protect\sqrt s = 500$ GeV,
 ${\cal L}_{int}=20\ \hbox{fb}^{-1}$).  $M_{Z'} = 1$ TeV.
100\%\  heavy flavor tagging efficiency  and 100\%\ longitudinal
polarization of the
electron beam is assumed for the first set of error bars, while the
error bars in parentheses are for the probes without polarization.
\label{uncertainties}}
\vskip6pt
\renewcommand\arraystretch{1.1}
\begin{tabular}{|r|cccc|} \hline
& $\chi$ & $\psi$ & $\eta$ & $LR$ \\ \hline
$P_V^\ell$ & $2.0\pm0.08\,(0.26)$ & $0.0\pm0.04\,(1.5)$ &
$-3.0\pm 0.5\,(1.1)$ & $-0.15\pm 0.018\,(0.072)$  \\
$P_L^q$ & $-0.5\pm 0.04\,(0.10)$ &  $0.5\pm0.10\,
(0.2)$ &  $2.0\pm0.3\,(1.1)$ & $-0.14\pm
 0.037\,(0.07)$
\\
$P_R^u$ & $-1.0\pm0.15\,(0.19)$ & $-1.0\pm0.11\,
(1.2)$ &  $-1.0\pm0.15\,(0.24)$ & $-6.0\pm1.4\,
(3.3)$\\
$P_R^d$ & $3.0\pm0.24\,(0.51)$ & $-1.0\pm0.21\,(2.8)$
& $0.5\pm0.09\,(0.48)$ & $8.0\pm1.9\,(4.1)$\\
$\epsilon_A$ & $0.071\pm0.005\,(0.018)$
& $0.121\pm0.017\,(0.02)$ &
 $0.012\pm0.003\,(0.009)$ & $0.255\pm0.016\,(0.018)$ \\ \hline
\end{tabular}
\end{minipage}
\end{center}
\end{table*}

Discovery limits for new gauge bosons at future colliders are
shown in Table \ref{bounds} \cite{godfrey95}.
The LHC and a high luminosity 1 TeV
$e^+e^-$ collider have discovery limits for a $Z'$ which are in many
ways comparable.  Both measurements have strengths and weaknesses.
The limits obtained from the LHC are robust, in the sense that they
are obtained from a direct measurement with little background, but
they are, however, dependent on the total width of the $Z'$ which is
sensitive to assumptions on the particle content of the model.
In contrast, limits obtained for the NLC are indirect, based on
statistical deviations from the Standard Model and are therefore
both more model dependent and
dependent on having the systematic errors under control, but do
not depend on the unknown particle content of the model.

If a $Z'$ were discovered, its mass could be measured without any
ambiguity at a hadron collider from the invariant mass of the lepton pairs
which signalled the $Z'$ discovery.  In contrast, at the NLC
it is the ratio $(g/M_Z')^2$ which enters, so that extracting the $Z'$ mass
is very model dependent.  If evidence were found for a $Z'$ at the
NLC, a model
independent determination of $M_{Z'}$ would be imprecise at best, except
for perhaps a $Z'$ not much greater than the collider energy.

\subsection{$Z'$ Diagnostics at   the  LHC and NLC}

An immediate need after a $Z'$  discovery  would be to
determine its origins by measuring  its couplings to quarks and
leptons \cite{ACLI}.
Assuming family universality and
the $Z'$ charge commuting with the $SU(2)_L$
generators, the relevant quantities   to
distinguish between different models are
the fermion charges,
${\hat g}^f_{L2(R2)}$ and the gauge coupling
strength, $g_2$.
  Because
the signs of the charges will be hard to determine at hadron
colliders the following
 four ``normalized'' observables are used:
\begin{eqnarray}
\gamma_L^\ell &=& {{(\hat{g}^\ell_{L2})^2}\over
{{(\hat{g}^\ell_{L2})^2+(\hat{g}^\ell_{R2})^2}}}, \nonumber \\
\gamma_L^q &=& {{(\hat{g}^q_{L2})^2}\over{{(\hat{g}^\ell_{L2})^2
+(\hat{g}^\ell_{R2})^2}}},  \nonumber \\
 \tilde{U}&=& {{(\hat{g}^u_{R2})^2}\over {(\hat{g}^q_{L2})^2}},\nonumber\\
\tilde{D}&=& {{(\hat{g}^d_{R2})^2}\over {(\hat{g}^q_{L2})^2}}\ .
\label{tild}
\end{eqnarray}
The values of these couplings  at the LHC
for the typical models and the corresponding
 statistical uncertainties
are  given in  Table \ref{values} for $M_{Z^{\prime}}$=1 TeV.

For $M_{Z'}\simeq 2$ TeV a reasonable
discrimination between models and determination of
the couplings may still be  possible, primarily from the forward-backward
asymmetry and the rapidity ratio. However, for
$M_{Z'}\simeq 3$  TeV there is little ability
to discriminate  between models.
The LHC can also address $W'$ diagnostics for $M_{W'}\lsim 1$--2
TeV.

If a $Z'$ were produced on shell at the NLC it would be relatively
straightforward to determine its properties.  On the other hand,
if it is far off-shell (a more likely
possibility) its
properties could be deduced through interference effects of the $Z'$
propagator with the $\gamma$ and $Z$ propagators.
 At the NLC  $\sigma ^{\ell},\   R^{had} = \sigma
^{had}/ {\sigma ^{\ell}}$ and $A_{FB}^{\ell}$ will be measured and
assuming  longitudinal polarization of the    electron
beam is available $A_{LR}^{\ell, had},\ \ A_{FB}^{\ell}(pol)$ will
be possible.
Although the above  quantities  can distinguish between different
models  they do not yield  information on all the  $Z'$ couplings.

The following four ``normalized''  charges and  an  overall gauge
coupling strength
divided by the ``reduced'' $Z'$ propagator are  probed at the NLC:
\begin{eqnarray}
P_V^\ell &=& {{\hat g^\ell_{L2} + \hat g^\ell_{R2}}\over
{\hat g^\ell_{L2} - \hat g^\ell_{R2}}}, \nonumber \\
P_L^q &=& {{\hat g^q_{L2}}\over
{\hat g^\ell_{L2} - \hat g^\ell_{R2}}}, \nonumber \\
P_R^{u,d} &=&  {{\hat g^{u,d}_{R2}}\over
{\hat g^q_{L2}}}, \nonumber \\
\epsilon_A &=& (\hat g_{L2}^\ell
- \hat g_{R2}^\ell)^2 {{g_2^2}\over{4\pi \alpha }}{
{s}\over{M^2_{Z'} - s}}.\
\label{eps}
\end{eqnarray}
Recall that  couplings (\ref{tild}) probed by  the LHC, do not determine
 couplings (\ref{eps}) unambiguously.
The couplings with statistical uncertainties are given in
Table \ref{uncertainties}. The $Z^\prime$
charges can be determined to
around 10--20 $\%$.  Relative error bars are about a factor of 2
larger than the corresponding ones at the LHC.

\subsection{Complementarity}

Among existing facilities currently in operation, the Tevatron
offers the highest discovery reach for new gauge bosons
with  masses up to the 700--900 GeV range although
LEP-II can achieve
comparable limits for some models.
In the longer term, hadron colliders, {\it i.e.}, different
upgraded versions of the Tevatron and the LHC, as well as a  high luminosity
$e^+e^-$  collider, {\it i.e.}, the NLC, would significantly improve limits
on the heavy gauge boson masses.
For the typical models such limits are in the 1--2 TeV region for the Tevatron
upgrades, in the 4--5 TeV region for the LHC, and roughly
2--10 times the center-of-mass energy
for the NLC with 50~fb$^{-1}$.

The LHC and  a high
luminosity 1 TeV $e^+e^-$ collider have discovery limits
for $Z'$ which are in many cases  comparable.
However, limits obtained for the LHC and the subsequent measurement
of its mass are robust, in the sense that
they are obtained from a direct measurement with little background.
In contrast, discovery limits and mass measurements at the NLC are
indirect and therefore model specific.

Once discovered,
the LHC and the NLC possess complementary diagnostic power
for the model independent determination of the $Z'$ couplings to quarks and
leptons.
In conjunction, the two machines may allow for
determination of the $M_{Z'}$, an overall $Z'$ gauge coupling strength as well
as a  unique determination of {\it all} the quark and lepton charges with
error bars in the 10--20\% range, provided $M_{Z'}\lsim 1$--2 TeV.




\section{New particles and Interactions}

Many theories beyond the Standard Model (SM) of the electroweak and strong
interactions, such as Grand Unified Theories (GUT) or composite models, require
the existence of new matter particles with the possibility of new interactions
not contained in the SM. These particles can be cast into three categories: new
elementary fermions \cite{er5m,S1,S3,S4,S5}, difermions \cite{er5m,S6,S7} and
excited fermions \cite{S8,S9,S10}.   If any signals are seen, it will be
crucial
to distinguish among the variety of possible new states described in this
section.  In general, total cross-sections, angular distributions and the
polarization of final state particles (resulting from new particle decays) can
be used to discriminate among the different possibilities \cite {S15}.

\subsection{New Elementary Fermions}

The classic examples of new fermions include: sequential fourth generation
fermions (with a right-handed neutrino state to allow for neutrino masses
larger than $m_Z/2$); vector--like
fermions with both left- and right-handed
components in weak isodoublets \cite{er5m}, mirror fermions which have the
opposite chiral properties as the SM fermions \cite{S3} and isosinglet fermions
such  as the SO(10) Majorana neutrino~\cite{S4}.

Fermions that have the usual lepton/baryon quantum numbers but possess
non-canonical SU(2)$_L\times$U(1)$_Y$ quantum numbers are called exotic
fermions.  They occur naturally in GUT models that contain a single
representation into which a complete generation of SM quarks and leptons  can
be
embedded. For instance, in the group E$_6$ each fermion generation lies in the
{\bf 27} representation, which contains 12 new fermions
in addition to the 15 chiral fermions of the SM~\cite{er5m}.

It is conceivable that these new fermions acquire masses not much larger than
the Fermi scale, if these masses are protected by some symmetry.  In fact, this
is necessary if the associated new gauge bosons (which are present in the same
GUT model that contains the new fermions) are relatively light \cite{S11}. In
the case of sequential and mirror fermions (at least in the simplest models
where the ESB pattern is the same as in the SM), unitarity arguments suggest
that their masses should not exceed a few hundred GeV \cite{S12}. These
particles, if they exist, should therefore be accessible at the next generation
of colliders.

Except for singlet neutrinos, the new fermions couple to the photon and/or to
the weak gauge bosons $W/Z$ (and for heavy quarks, to gluons) with full
strength; these  couplings allow for pair production with practically
unambiguous cross sections.
In general, new fermions can mix with SM fermions which have
the same conserved quantum numbers (such as color and electric
charge).
This mixing gives rise to new
currents, which determine the decay  properties of the heavy fermions and allow
for their single production.  However, flavor changing neutral currents
(FCNC's)
will be generated at levels inconsistent with present experiments unless
intergenerational mixing is highly suppressed or absent.  In the latter case,
the mixing pattern simplifies considerably \cite{S14}. The remaining angles are
restricted by LEP and low energy experiment data to be smaller than ${\cal
O}(0.04-0.1)$ \cite{S14}. Note that LEP1 sets bounds of  $\sim m_Z/2$ on the
masses of  these particles \cite{S13} (stronger mass bounds from Tevatron can
be
set for new quarks); masses up to $m_Z$ might be probed at LEP2.

The new fermions decay through mixing into massive gauge bosons plus their
ordinary light partners, $F \to fZ/ f'W$. For masses larger than $m_w(m_Z)$ the
vector bosons will be on--shell. For small mixing angles, $\theta<0.1$, the
decay widths are less than 10 MeV (GeV) for $m_F= 0.1 (1)$ TeV. The charged
current decay mode is always dominant and for $m_F \gg m_Z$, it has a branching
fraction of 2/3.

New fermions can be pair-produced in $e^+e^-$ collisions, $e^+e^- \to
F\bar{F}$,
through $s$--channel gauge boson exchange. The cross sections are of the order
of the point--like  QED cross section and therefore,  are rather large
\cite{S15}. Because of their clear signatures, the detection of these
particles
is straightforward in the clean environment of $e^+e^-$ colliders, and masses
very
close to the kinematical limit can be probed. Charged fermions can also be
pair--produced at $\gamma\gamma$ colliders via $\gamma \gamma \to F\bar{F}$.
The cross-sections are comparable (and may be larger for smaller $F$ masses) to
the corresponding rates in $e^+e^-$ collisions. Heavy quarks can be best
searched for at hadron colliders where the production processes, $gg/ q \bar{q}
\to Q \bar{Q}$, give very large cross sections: at the LHC
with $\sqrt{s}=14$ TeV
and a luminosity of 10 fb$^{-1}$ quark masses up to 1 TeV can be reached
\cite{S16}.

If the mixing angles between the SM and heavy fermions are not too small, then
singly-produced new fermions in association with their light partners may
produce an observable signal. The rate for single production is very model
dependent (in contrast to pair-production rates mentioned above); however, it
has the potential for significantly increasing the new fermion mass reach of a
given collider.  In $e^+e^-$ collisions, single production proceeds only via
$s$--channel $Z$ exchange  in the case of  quarks and second/third generation
leptons, leading to small rates. For the first generation leptons, however, one
has additional $t$--channel exchanges which increase the cross sections by
several orders of magnitude \cite{S15}.  A full simulation of the signals and
the backgrounds has been recently carried out; see ref.~\cite{S5}.

Heavy leptons of the first generation can also be singly produced in $ep$
collisions through $t$--channel exchange. At LEP$\times$LHC with $\sqrt{s}
=1.2$
TeV and $\int {\cal L}=2$ fb$^{-1}$, the $N$ and $E$ mass reach has been
considered in ref.~\cite{S18}.
In Left--Right models, an additional production
mechanism for single $N$ production via $t$-channel $W_R$-exchange must be
included.  A detailed Monte Carlo analysis of this process has been carried out
in ref.~\cite{S19}. In $e^+e^-$ collisions these neutrinos, can be pair
produced
with observable cross-sections if the $W_R$ bosons are not too heavy ($M_{W_R}<
2$ TeV at $\sqrt{s}=0.5$ TeV). One can also search for heavy Majorana
neutrinos,
through their virtual effects in the process
$e^- e^- \to  W_R^-W_R^-$~\cite{S20}.

\subsection{Difermions}

Difermions are new bosons (either spin 0 or spin 1)
that have unusual baryon and/or
lepton quantum numbers. Examples of these particles are leptoquarks (LQ) with
B$=\pm 1/3$ and L$=\pm 1$, diquarks with B$=\pm 2/3$ and L$=0$ and dileptons
with B$=0$ and L$=\pm2$. They occur in some GUT models ({\it e.g.}, in $E_6$,
the color triplet weak isosinglet new particle can be either a LQ or diquark)
and in composite models.

In addition to the usual couplings to gauge bosons, difermions have couplings
to
fermion pairs which determine their decays.  These couplings are {\it a priori}
unknown, although one typically assumes that the couplings between different
generations are negligible in order to avoid generating tree-level FCNC
processes.   As an example, consider the case of LQ's.  A systematic
description
of their quantum numbers and interactions can be made by starting from an
effective lagrangian with general SU(3)$\times$SU(2)$\times$U(1) invariant
couplings and conserved  B and L numbers. This leads to the existence of 5
scalar and 5  vector LQ's with distinct SM transformation
properties \cite{S4}.  In general, present data constrain difermions
to have masses larger than 50--150 GeV.

Leptoquarks can be produced in pairs at $e^+e^-$ colliders through gauge boson
exchange; significant $t$-channel quark exchange can be present in some
channels
if the quark-lepton-LQ couplings are not too small. Depending on  the charge,
the spin and isospin of the LQ, the cross sections can vary  widely \cite{S21}.
LQ's can also be pair produced in $\gamma \gamma$ collisions via $t$-channel
LQ-exchange. Single production of scalar and vector LQs has been also studied
in
the $e^+e^-$, $e\gamma$ and $\gamma \gamma$ modes of the NLC \cite{S22}. The
kinematical  reach is thus extended to $\sim\sqrt{s}$, but the production rates
are suppressed  by the unknown LQ Yukawa coupling to quark--lepton pairs. At $e
\gamma$  colliders, first generation LQ's are observable as long as
the corresponding squared
LQ Yukawa coupling is larger than about $10^{-2}\,e^2$ (where $e^2
\equiv 4\pi\alpha$ is the electromagnetic coupling strength)~\cite{S22}.

Leptoquarks are color-nonsinglets
and hence strongly interacting.  As a result, LQ's
can be produced at hadron colliders with very large rates. At the LHC, pair
production in the $gg/q  \bar{q} \to {\rm LQ}\ \overline{\rm LQ}$
process leads to cross sections
ranging from a few nb for masses ${\cal O}$(0.1 TeV) to a few fb for masses
${\cal O}$(1 TeV). The rate for vector particles (which depends on their
anomalous magnetic  moment $\kappa$) is substantially higher than for scalars.
At the LHC with 100 fb$^{-1}$  the search reach for scalar/vector LQ's is
1.4/2.2(1.8) TeV for $\kappa=1\,(0)$,
if one assumes a branching fraction of unity for the $e^+e^-+$two jet
final state \cite{S23}. Single scalar LQ production, through
$gu\to e^+$LQ and $gd\to \bar \nu$LQ, can also lead to large cross sections for
Yukawa couplings of order unity.    In this case masses up to $\sim$ 1.5 TeV
can
be reached at the LHC \cite{S23}. Vector leptoquarks have larger production
rates and the discovery reach can be extended to $\sim$ 2 TeV. Although LQ's
can
also be singly produced in association with a $W_R$ in alternative Left-Right
models via $gu\to W_R+$LQ, a detailed study of the signal and backgrounds has
shown that the LQ discovery reach in this case is quite limited \cite{S24}. For
first generation leptoquark searches, $ep$ colliders are particularly well
suited.  Such LQ's can be produced as $s$ channel resonances in $eq\to$LQ.  For
example, at LEP$\times$LHC masses up to 1 TeV can be reached, unless
the corresponding LQ Yukawa couplings are very small.

Dilepton production has been considered at the NLC in the three collider modes:
$e^+e^-, e\gamma$ and $\gamma \gamma$. These particles are  accessible up to
masses
close to $\sqrt{s}/2$ in pair production, $e^+e^-/\gamma  \gamma \to
X^{++}X^{--}$:
the rates (especially in $\gamma \gamma$ collisions because of the charge)  are
very large and the signatures (four leptons) are  spectacular. Dileptons can
also be singly produced in the three modes of the collider \cite{S25}. Diquarks
can be pair produced in $e^+e^-$ and $\gamma \gamma$ collisions for masses
smaller
than $\sqrt{s}/2$ with appreciable rates, with a signal consisting of an excess
of 4 jet events. They can be also produced at hadron colliders in pairs (or
singly for the first generation diquark if its coupling to quark pairs is not
too small). However, large QCD backgrounds present a formidable challenge to
diquark searches at the LHC.

\subsection{Excited Fermions}

The existence of excited particles is a characteristic signal of substructure
in
the fermionic sector \cite{S8}: if the known fermions are composite (composed
of
more elementary constituents called {\it preons}), then they should be the
ground state of a rich spectrum of excited states which decay down to the
former
states via a magnetic dipole type de-excitation. Since there is not yet a
satisfactory and predictive dynamical model for composite  fermions, one has to
use phenomenological inputs to study this scenario. For simplicity, one assumes
that the excited fermions have spin and isospin 1/2; their couplings to gauge
bosons are vector--like (form factors and new contact interactions may also be
present).  Furthermore, the coupling which describes the transition between
excited and ordinary fermions is taken to be chiral and inversely proportional
to the compositeness scale $\Lambda$ which is of ${\cal O}$(1 TeV).

Excited fermions can be pair-produced in $f\bar
f$ annihilation via $s$-channel gauge boson exchange ($V=W$, $Z$, $\gamma$ or
gluon), although some suppression due to form factors is
expected \cite{S9}.
The excited fermions decay into gauge bosons and their ordinary partners,
$f^\star \to fV$. The charged leptons have the electromagnetic decay which has
at least a branching ratio of 30\%, and the excited quarks decay most of the
time into quarks and gluons. These two decays constitute a very characteristic
signature and discriminate them from the exotic fermions previously discussed.

In $e^{+}e^{-}$  and $\gamma \gamma$ collisions, the processes and the cross
sections are the same as the ones previously described for
vector--like exotic
fermions (up to possible form-factor suppressions); hence the detection
of $f^*$ masses up to the kinematic limit may be possible.
Excited fermions can be singly produced with
their light partners, but the rates are suppressed
by a factor of $1/\Lambda^2$. At $e^+e^-$ colliders, for quarks and
second/third generation leptons, for which the process is mediated by
$s$--channel boson exchange, the cross sections are very small. But for the
first generation excited fermions, one has substantial  contributions due to
additional $t$--channel diagrams: $W$ exchange for  $\nu_e^*$ and $Z/\gamma$
exchanges for $e^*$, which increase the cross sections by several orders of
magnitude. The excited leptons can be also singly produced in $e\gamma$
collisions ($e^*$ as a resonance and $\nu^*$ in association with a $W$) with
much larger rates. All excited fermions can be singly produced in $\gamma
\gamma$ collisions with appreciable cross sections. For $\Lambda \sim $ few
TeV,
excited quarks and leptons can easily be found in such machines, if
kinematically allowed~\cite{S15}.

Due to the special couplings of the electron to the excited leptons of the
first
generation, single production of $e^*$ through $t$-channel $\gamma$ and $Z$
exchange, and $\nu_e^*$ through $W$  exchange are possible processes in $ep$
collisions. The cross sections are large; requiring a few tens of
events
 to establish a signal, one can probe $\nu^*$ and $e^*$ masses
 up to 800 GeV at LEP$\times$LHC
for $\int {\cal L}=2$ fb$^{-1}$ \cite{S26}. Excited quarks of the
first  generation can also be produced in $ep$ collisions, however background
problems make this possibility less interesting.

Excited quarks can be produced in $pp$ collisions through a variety of
mechanisms \cite{S9}. The dominant production channels are the gluonic
excitation of  quarks $g + q \to q^{*}$ which occurs through the $q^*qg$
magnetic interaction, and the excitation through the preon interactions $qq
\to
qq^{*}$ and $q^{*}q^{*}$.  The latter is also responsible for excited lepton
production, $q\bar{q}  \to ee^{*}$ and $e^{*}e^{*}$. The signatures of excited
quarks are  dijet mass bumps, jet + gauge boson and jet + lepton  pair
combinations. Excited leptons would reveal themselves in leptons + gauge
particles or leptons + quark jets. QCD backgrounds have been studied and have
been shown to be  under control.  At the LHC
with a luminosity of 10 fb$^{-1}$,
a mass range of 5--6 TeV and $\sim 4$~TeV
can be reached for excited quarks and
excited leptons, respectively~\cite{S18}.

\subsection{Synopsis}

The prospects for the discovery of new fermions (of the sequential,
exotic, or excited variety) or di--fermions at future
high--energy colliders can be summarized as follows:
\vspace{1pc}

\noindent {\it $e^+e^-$, $e\gamma$, $\gamma\gamma$, $e^-e^-$
colliders:}

\begin{itemize}

\item[$\bullet$] All new particles considered in this section
(except for weak isosinglet neutrinos)
can be pair produced in $e^+e^-$ collisions
up to the kinematical limit of $\sqrt{s}/2$.  Similarly for new
charged particles in $\gamma\gamma$ collisions.

\item[$\bullet$] Due to their special couplings to the electron, the first
generation of exotic leptons, excited leptons, leptoquarks and dileptons
can be singly produced if the mixing angles, the compositeness scale or the
couplings to the light fermion pairs are not prohibitively small;
one can then reach masses close to the center-of-mass
energy of the collider.  Here, all four potential modes of the
NLC can be useful.

\item[$\bullet$] Due to the clean environment of these machines, one can
perform precision measurements which allow one
to probe the indirect effects
of some of these particles for mass scales much higher than the
center-of-mass energy.

\end{itemize}

\noindent{\it $ep$ colliders:}

\begin{itemize}

\item[$\bullet$] $ep$ colliders are well suited for the production of the
first generation exotic leptons, excited leptons and especially
leptoquarks, if the mixing angles, the compositeness scale or the
couplings to quark--lepton pairs, respectively, are not too small.
At LEP$\times$LHC, masses around 1 TeV can be reached.

\end{itemize}

\noindent {\it $pp$ colliders:}

\begin{itemize}

\item[$\bullet$] Proton colliders are ideal machines for the production of
new color non-singlets. At the LHC, particle masses in the TeV range
can be probed: $\sim 1$ TeV for sequential or
exotic quarks, $\sim$ 1.5---2 TeV for
leptoquarks and 5---6 TeV for excited quarks for reasonable values of the
compositeness scale.

\item[$\bullet$] New color-singlet particles can also be produced, but
with smaller cross sections than for quarks. However the signatures are cleaner
and might compensate for the small rates. For instance, some exotic leptons can
be discovered for masses smaller than $\sim 1/2$ TeV; through contact
interactions, excited electrons can also be produced for masses in the TeV
range.

\end{itemize}

Finally, it should be noted that the effects of new interactions may be
observable even if all the associated new particles beyond the SM are too heavy
to be observed directly.  If the mass scale of the new physics, $\Lambda$, is
assumed to be much larger than $m_Z$, then one can formally integrate out the
heavy particles and remove them from the full theory. The resulting low-energy
effective theory will differ from the SM by new (anomalous) interactions among
SM particles which are suppressed by inverse powers of $\Lambda$. For example,
new physics beyond the SM could generate four-fermion
contact interactions \cite{S27}, which
could be detected as deviations from SM predictions in 2-to-2 scattering
processes involving quarks and/or leptons \cite{S20}.   In this way, one may
have experimental sensitivity to much larger values of $\Lambda$ as compared to
direct searches for new particles. The search for anomalous couplings in the SM
may be particularly fruitful in the precision study of $W$ and $Z$
self-interactions and top-quark interactions.  Since these are the heaviest
particles of the SM, it is plausible that their interactions could be modified
by the dynamics underlying the generation of mass. The phenomenology of
anomalous gauge boson and top-quark couplings is studied in more detail
in the next two chapters.




\section{Anomalous Gauge Boson Couplings}

\subsection{Parametrization of Anomalous Gauge Bosons \hfill\break
Interactions}

One of the most direct consequences of the $SU(2)\times U(1)$ gauge
symmetry of the SM, the non-abelian
self-couplings of the $W$, $Z$, and photon, remains poorly measured to
date. Although information on these couplings can be extracted from
low energy data and high precision measurements at the $Z$ pole, there are
ambiguities and model dependencies in the results. However,
a direct measurement of these vector boson couplings is possible in
present
and future collider experiments, in particular via pair production
processes
like $e^+e^- \to W^+W^-$ and  $p\,p\hskip-7pt\hbox{$^{^{(\!-\!)}}$}
\to W^+W^-,\; W\gamma,\; WZ$. The first and major goal of such experiments
will be the verification of the SM predictions at a quantitative level.

For our discussion of experimental sensitivities we employ a
parameterization of the $WWV$ couplings ($V=\gamma,\, Z$) in terms of
a phenomenological effective Lagrangian~\cite{HPZH}:
\begin{eqnarray}
i{\cal L}_{WWV} &=& g_{WWV} \!
\biggl[ g_1^V \bigl( W_{\mu\nu}^{\dagger} W^{\mu} V^{\nu}
        -W_{\mu}^{\dagger} V_{\nu} W^{\mu\nu} \bigr)  \nonumber \\
&&+ \kappa_V W_{\mu}^{\dagger} W_{\nu} V^{\mu\nu}
+ {\lambda_V \over m_W^2} W_{\lambda \mu}^{\dagger} W^{\mu}_{\nu}
V^{\nu\lambda}
\biggr].
\label{EQ:LAGRANGE}
\end{eqnarray}
Here the overall coupling constants are defined as $g_{WW\gamma}=e$ and
$g_{WWZ}= e \cot\theta_W$. The couplings $g_1^V$, $\kappa_V$, and
$\lambda_V$
need to be determined experimentally.  Within the SM, at tree level, they
are
given by $g_1^Z = g_1^\gamma = \kappa_Z = \kappa_\gamma = 1,\; \lambda_Z =
\lambda_\gamma = 0$. $g_1^\gamma=1$ is fixed by electromagnetic gauge
invariance; $g_1^Z$,
however, may well be different from its SM value 1 and appears at the
same level as $\kappa_\gamma$ or $\kappa_Z$.

The effective Lagrangian of Eq.~(\ref{EQ:LAGRANGE}) parameterizes the
most general
Lorentz invariant and $C$ and $P$ conserving $WWV$ vertex which can be
observed in
processes where the vector bosons couple to effectively massless
fermions. Terms
with higher derivatives are equivalent to a dependence of the couplings on
the vector boson momenta and thus merely lead to a form-factor behavior of
these couplings~\cite{HPZH}. Analogous to the general $WWV$ vertex it is
possible to
parameterize anomalous $Z\gamma V,\, V=\gamma ,Z$ couplings in terms of
two coupling constants, $h_3^V$ and $h_4^V$, if CP invariance is
imposed. All $Z\gamma V$ couplings are $C$ odd; $h^V_3$ corresponds
to a dimension~6, $h^V_4$ to a dimension~8 operator~\cite{HPZH}.

Tree level unitarity uniquely restricts the $WWV$ and $Z\gamma V$
couplings to their SM gauge theory values at asymptotically high
energies. This implies that deviations of $g_1^V$, $\kappa_V$,
$\lambda_V$ and $h^V_i$, $i=3,4$ must be parametrized by form factors
which indicates that the effective Lagrangian describing the
anomalous vector boson self-interactions breaks down at some high
energy scale. SM one-loop effects
lead to corrections of ${\cal O}(\alpha)$ with a form factor scale,
$\Lambda_{FF}$, of order the electroweak symmetry breaking scale.

Models which have been considered so far, such as
technicolor or supersymmetric models, predict rather small
deviations from the SM values, of ${\cal O}(10^{-2})$ or less.
If we regard the Standard Model as an effective field theory, then new
physics can be parameterized in a model independent way by additional
gauge-invariant
operators.  Regardless of whether Higgs physics is strongly or weakly
interacting, there are restrictions on the strength of such
contributions which, however, depend on naturalness assumptions.
Detailed analyses \cite{rujula,HISZ} show that these
restrictions leave a slim chance only that new physics will be
discovered through anomalous couplings at machines which are
less capable than the LHC or NLC.

\subsection{Search for Anomalous Gauge Boson Couplings at Present
and Future Colliders}

The direct measurement of $WWV$ and $Z\gamma V$ couplings in present and
future collider experiments in general makes use of the high energy
behavior of the non-standard contributions to the helicity amplitudes
which grow like a power of  the parton center of mass energy. In
di-boson production, this leads to an excess of
events at large $W$, $Z$ and photon transverse momenta
as well as at large invariant masses of the di-boson system. Furthermore,
the various non-standard couplings contribute to different helicity
amplitudes in the high energy limit. Angular distributions of the $W$ and
$Z$-boson decay products, therefore, are helpful as polarization
analyzers, in particular in $e^+e^-$ collisions. Quantitative bounds on
$\Delta g_1^Z=g_1^Z-1$, $\Delta\kappa_V=\kappa_V-1$ and $\lambda_V$ are
obtained by comparing the shape of the measured and the predicted
distributions.

\begin{table*}[htp]
\begin{center}
\caption{Present and future 95\% CL bounds on $WWV$ and $Z\gamma V$
couplings.}
\label{TABLE1}
\vskip1pc
\renewcommand\arraystretch{1.3}
\begin{tabular}{|lcc|}
\hline
{\bf Experiment} & {\bf Channel} & {\bf Limit} \\
\hline
D\O\ & $p\bar p\rightarrow W^\pm\gamma\rightarrow\ell^\pm\nu\gamma$ &
$-1.6<\Delta\kappa_\gamma<1.8$ \\
 (present limit)
 & $\ell=e,\,\mu$ & $-0.6<\lambda_\gamma<0.6$ \\
\hline
CDF & $p\bar p\rightarrow W^+W^-,\, W^\pm Z\to\ell^\pm\nu jj$  &
$-1.0 <\Delta\kappa_V<1.1$ \\
 (present limit)
 & $\kappa_\gamma=\kappa_Z$, $\lambda_\gamma=\lambda_Z$, $g_1^Z=1$ &
$-0.8<\lambda_V<0.8$ \\
\hline
D\O\ & $p\bar p\rightarrow Z\gamma\rightarrow\ell^+\ell^-\gamma$ &
$-1.9<h^V_{30}<1.8$ \\
 (present limit)
 & $\ell=e,\,\mu$, $\Lambda_{FF}=0.5$~TeV & $-0.5<h^V_{40}<0.5$ \\
\hline
L3 & $e^+e^-\rightarrow\bar\nu\nu\gamma$ & $-0.9<h^Z_{30}<0.9$ \\
 (present limit)
  & & $-2.3<h^Z_{40}<2.3$\\
\hline
Tevatron & $p\bar p\rightarrow W^\pm\gamma\rightarrow e^\pm\nu\gamma$, &
$-0.38<\Delta\kappa_\gamma<0.38$ \\
(CDF sim.) & $\sqrt{s}=2$~TeV, $\int\!{\cal L}dt=1$~fb$^{-1}$ &
$-0.12<\lambda_\gamma <0.12$ \\
\hline
Tevatron & $p\bar p\rightarrow W^+W^-,\, W^\pm Z\rightarrow \ell^\pm\nu
jj,\, \ell^+\ell^-jj$ & $-0.31<\Delta\kappa_\gamma<0.41$ \\
(CDF sim.) & $\int\!{\cal L}dt=1$~fb$^{-1}$, $\Lambda_{FF}=2$~TeV &
$-0.19<\lambda_\gamma <0.19$ \\
\hline
Tevatron & $p\bar p\rightarrow W^\pm Z\rightarrow
\ell_1^\pm\nu_1\ell_2^+\ell_2^-$ & $-0.26<\Delta\kappa_Z<0.70$ \\
(NLO theor.) & $\int\!{\cal
L}dt=1$~fb$^{-1}$, $\Lambda_{FF}=1$~TeV  & $-0.24 <\lambda_Z<0.32$ \\
\hline
Tevatron & $p\bar p\rightarrow Z\gamma\rightarrow e^+e^-\gamma$ &
$-0.105<h^V_{30}<0.105$ \\
(CDF sim.) & $\int\!{\cal L}dt=1$~fb$^{-1}$, $\Lambda_{FF}=1.5$~TeV &
$-0.0064 < h^V_{40} < 0.0064$ \\
\hline
Tevatron & $p\bar p\rightarrow Z\gamma\rightarrow e^+e^-\gamma$ &
$-0.044<h^V_{30}<0.044$ \\
(CDF sim.) & $\int\!{\cal L}dt=10$~fb$^{-1}$, $\Lambda_{FF}=1.5$~TeV &
$-0.0025 < h^V_{40} < 0.0025$ \\
\hline
LEP~II & $e^+e^-\rightarrow W^+W^-\rightarrow\ell^\pm\nu jj$,
 & $-0.13 < \Delta\kappa_\gamma<0.14$ \\
(L3 sim.) & $\sqrt{s}=190$~GeV, $\int\!{\cal L}dt=500$~pb$^{-1}$ &
$-0.13 < \lambda_\gamma < 0.14$ \\
\hline
LEP~II & $e^+e^-\rightarrow Z\gamma\rightarrow \nu\bar\nu\gamma$,
$\sqrt{s}=180$~GeV, & $-0.50< h^Z_{30} < 0.50$ \\
(L3 sim.) & $\int\!{\cal L}dt=500$~pb$^{-1}$, $\Lambda_{FF}=1$~TeV &
$-0.45< h^Z_{40} < 0.45$ \\
\hline
LHC & $pp\rightarrow W^\pm Z\rightarrow\ell_1^\pm\nu_1\ell_2^+\ell_2^-$, &
$-0.006<\Delta\kappa_Z<0.0097$ \\
(NLO theor.) & $\int\!{\cal L}dt=100$~fb$^{-1}$, $\Lambda_{FF}=3$~TeV  &
$-5.3\cdot 10^{-3} <\lambda_Z< 6.7\cdot 10^{-3}$ \\
\hline
LHC & $pp\rightarrow Z\gamma\rightarrow e^+e^-\gamma$ & $-5.1\cdot
10^{-3}<h^V_{30}<5.1\cdot 10^{-3}$ \\
(LO theor.) & $\int\!{\cal L}dt=10$~fb$^{-1}$, $\Lambda_{FF}=1.5$~TeV &
$-9.2\cdot 10^{-5} <h^V_{40}<9.2\cdot 10^{-5}$ \\
\hline
NLC & $e^+e^-\rightarrow W^+W^-\rightarrow\ell^\pm\nu jj$,
 & $-0.0024 <\Delta\kappa_\gamma < 0.0024$ \\
 & $\sqrt{s}=500$~GeV, $\int\!{\cal L}dt=80$~fb$^{-1}$ &
$-0.0018< \lambda_\gamma < 0.0018$ \\
\hline
NLC & $e^+e^-\rightarrow W^+W^-\rightarrow\ell^\pm\nu jj$,
 & $-5.2\cdot 10^{-4} <\Delta\kappa_\gamma < 5.2\cdot 10^{-4}$ \\
 & $\sqrt{s}=1500$~GeV, $\int\!{\cal L}dt=190$~fb$^{-1}$ &
$-3.8\cdot 10^{-4}< \lambda_\gamma < 3.8\cdot 10^{-4}$ \\
\hline
\end{tabular}
\end{center}
\end{table*}

Present and future 95\% confidence level (CL) limits on a variety of $WWV$
and $Z\gamma V$ couplings are listed in Table~\ref{TABLE1}.  In addition,
the limits on the couplings $\Delta\kappa_\gamma$ and $\lambda_\gamma$
are displayed in bargraph form in Fig.~\ref{kaplam}. The present limits
are from di-boson production at the Tevatron~\cite{TEV}
and from single photon production at LEP~\cite{BUS}.
The future limits are estimated for experiments at
the Tevatron, LEP~II, the LHC and a future linear $e^+e^-$ collider (NLC).
In contrast to the bounds on anomalous $WWV$ couplings, the limits on the
$Z\gamma V$ couplings $h^V_{i0}$, $i=3,4$ depend significantly
on the form factor scale $\Lambda_{FF}$.  Here they
are assumed to appear in the form
$h^V_i=h^V_{i0}/(1+\hat s/\Lambda^2_{FF})^{n_i}$, $i=3,4$ with $n_3=3$
and $n_4=4$.

\begin{figure}[htb]
\vspace{1pc}
\leavevmode
\begin{minipage}[b]{7.3cm}
\centering{%
\psfig{file=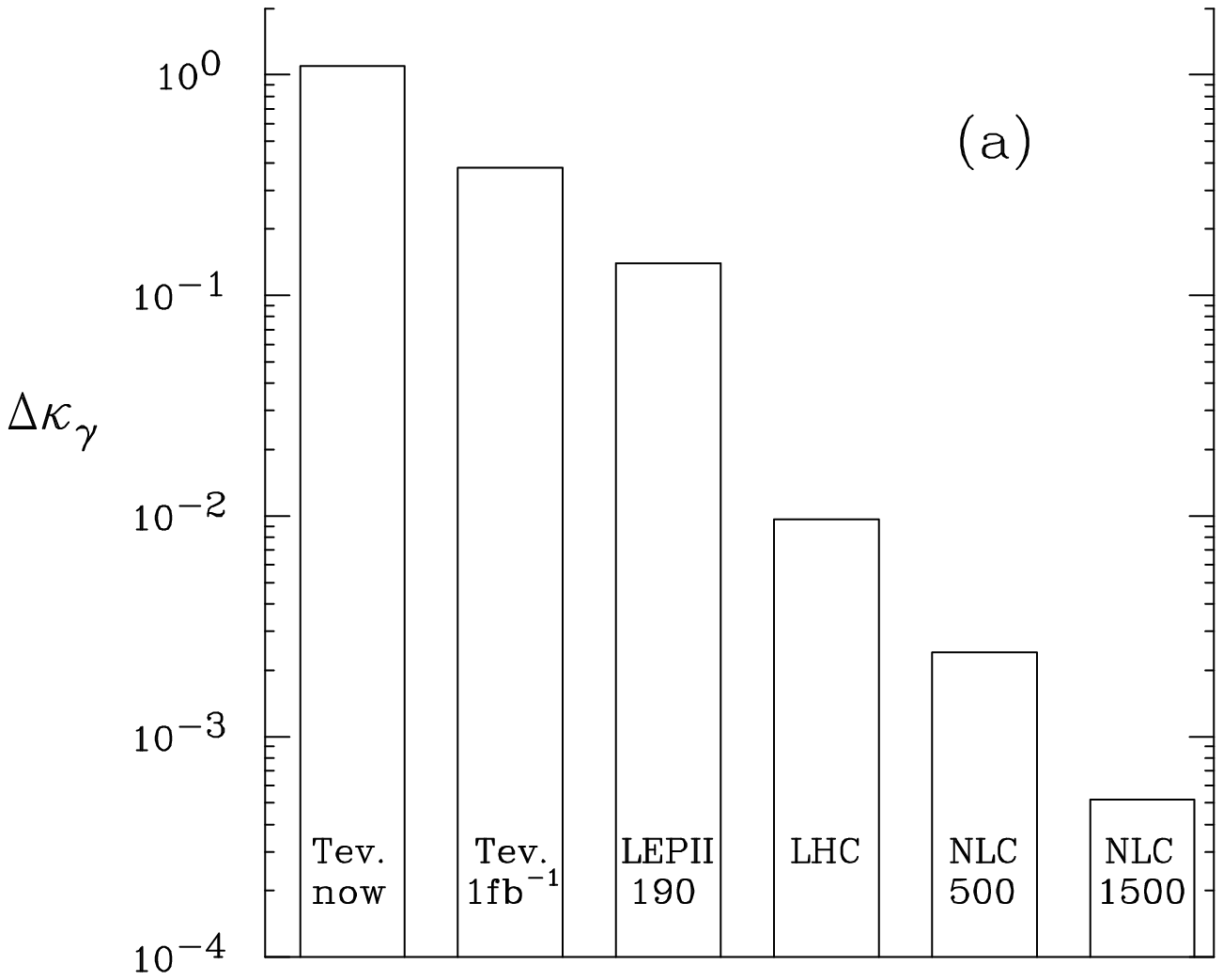,width=\hsize}
}
\end{minipage}
\hfill
\begin{minipage}[b]{7.3cm}
\centering{%
\psfig{file=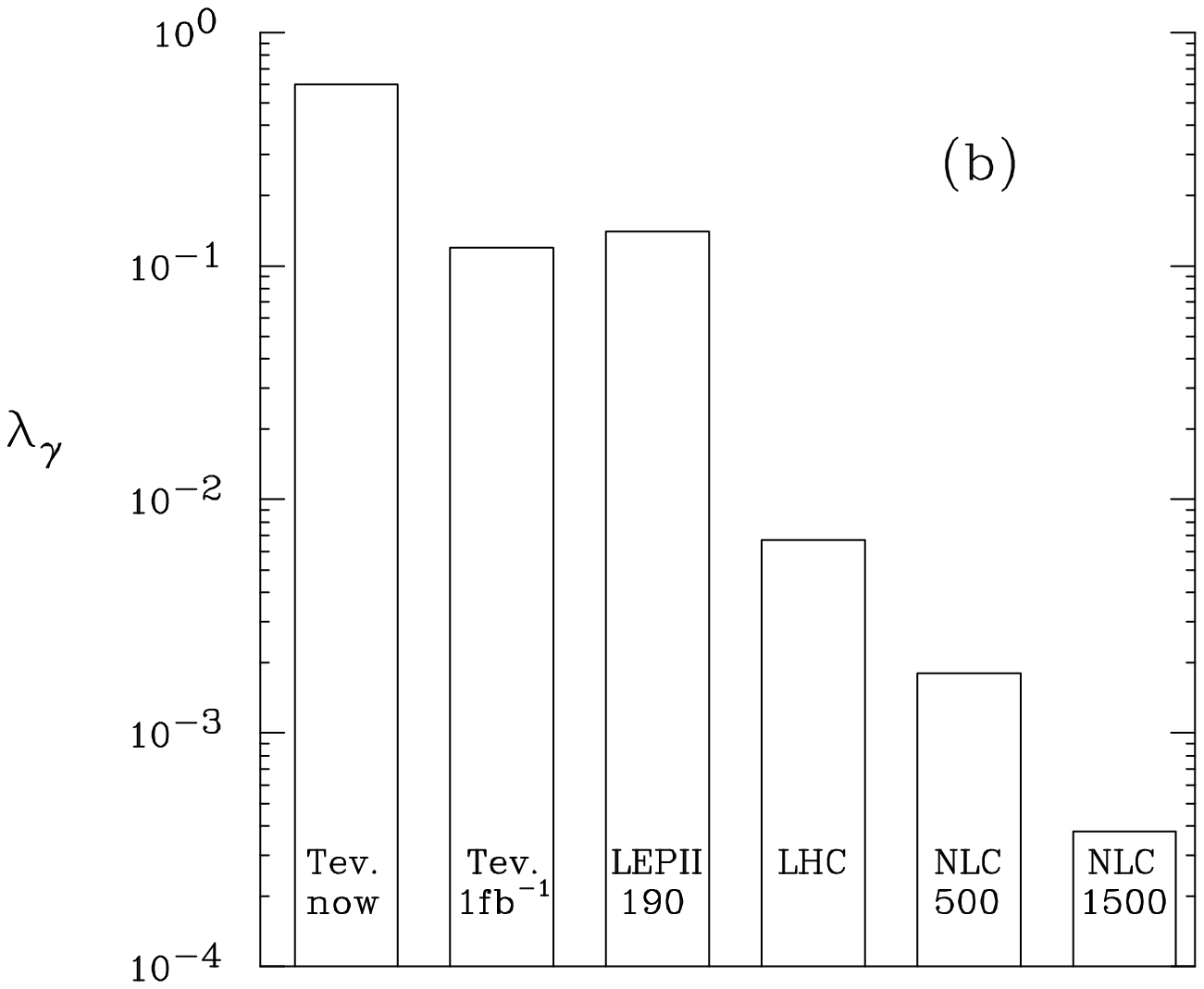,width=\hsize}
}
\end{minipage}
\vspace{-1pc}
\caption{Present and future 95\% CL bounds
on $\Delta\kappa_\gamma$ (a) and $\lambda_\gamma$ (b).}\label{kaplam}
\vspace{1pc}
\end{figure}

A direct measurement of the $WW\gamma$ vertex is achieved via $W\gamma$
production.  The best current bounds on $WWV$ couplings, however, depend
significantly on information obtained from $W^+W^-$ and $WZ$ production at
the Tevatron in the channel $p\bar p\rightarrow W^+W^-,
\, W^\pm Z\rightarrow\ell^\pm\nu jj$. In deriving these bounds,
$\Delta\kappa_\gamma=\Delta\kappa_Z$ and $\lambda_\gamma=\lambda_Z$ is
assumed.

The present limits on anomalous $WWV$ and $Z\gamma V$ couplings can be
improved
by one order of magnitude or more in future Tevatron experiments if an
integrated luminosity of 1~fb$^{-1}$ can be achieved. Raising the
integrated
luminosity to 10~fb$^{-1}$ typically leads to a factor $\sim 2$
improvement over the limits which can be achieved with 1~fb$^{-1}$.
Fitting several different distributions simultaneously might
lead to further improvements. Bounds on
$\Delta g_1^Z$ and $\Delta\kappa_V$ are expected to be comparable to
those which can be obtained at LEP~II. For $\lambda_V$, the limits from
future Tevatron experiments should be better.

The future bounds listed in Table~\ref{TABLE1} for $W^+W^-$ and
$W^\pm Z$ final states were obtained assuming $\lambda_Z
=\lambda_\gamma$, $\Delta g_1^Z=\Delta\kappa_\gamma/2\cos^2\theta_{\rm w}$
and $\Delta g_1^Z+\Delta\kappa_Z=\Delta\kappa_\gamma$. These relations are
motivated by an effective Lagrangian based on a complete set of $SU(2)
\times U(1)$ invariant operators of dimension~6~\cite{HISZ}. If one
assumes $\kappa_\gamma=\kappa_Z$, $\lambda_\gamma=\lambda_Z$ and $\Delta
g_1^Z=0$ instead, the limit on $\Delta\kappa_\gamma$ can be improved by
about 40\%.

At the LHC, with an integrated luminosity of 100~fb$^{-1}$, values of
$\Delta\kappa_Z$, $\Delta g_1^Z$ and $\lambda_Z$ of
${\cal O}(10^{-2})$ can be probed. Similar sensitivities can be
achieved for the $WW\gamma$ couplings. A linear $e^+e^-$ collider with a
center of mass energy of 500~GeV or more in general can probe anomalous
couplings of order $10^{-3}$~\cite{MIYA,barklow}.

The present bounds on $Z\gamma V$ couplings can be improved only by about
a factor~2 at best in $Z\gamma$ production at LEP~II. Future experiments
at the Tevatron will be sensitive to values of $h^V_{30}\sim 0.11$ and
$h^V_{40}\sim 0.006$ (with $\Lambda_{FF}=1.5$~TeV).

Within the next few years experiments at the Tevatron and at LEP are
expected
to confirm the non-abelian self-couplings of the SM at the 10\% level.
The ${\cal O}(\alpha)$ deviations predicted by the SM and most
models of new physics, however,  can only be probed at the LHC or a high
energy linear $e^+e^-$ collider. Of course, if the scale of new physics
is directly accessible at these machines one would rather focus on the
spectrum of new particles and direct measurements of their interactions
with the gauge boson sector.
%



\def\epem{e^+e^-}
\def\ra{\rightarrow}
\section{Top Quarks as a Window to Electroweak \hfill\break
Symmetry Breaking}

Among the various matter fermions, the top quark must
be considered the most mysterious.  It is the heaviest known elementary
particle, with mass 30--40 times that of the next heaviest quark.
It is more strongly coupled to the mechanism of electroweak symmetry
breaking than the weak gauge bosons themselves. It is plausible, in fact,
that the top quark is an essential component of this mechanism.
On a more mundane level, the top quark is the only quark or lepton
whose couplings to the Standard Model are not constrained to percent
accuracy by the LEP experiments.  Thus, the couplings of the top can
easily yield surprises.

  Most of this section is concerned with the search for anomalies in the
couplings of the top quark to the Standard Model gauge bosons.  We will
consider first the possibility of large deviations from the standard
properties of the top quark and then the precision experiments that
could show signs of smaller deviations.  Finally, we will discuss the
possibility that the top quark has decay modes outside the Standard
Model.

\subsection{Anomalous Top Couplings -- s-channel Resonance}

  The original
CDF announcement of evidence for the top quark \cite{CDF,CDFL}
came with an anomaly, the measurement of a production cross section
2--3 times larger than the theoretical prediction.
%
%
The more recent announcements of the observation of the top quark by
CDF \cite{CDFDIS} and D0 \cite{DZDIS} give a production cross section
of 6.5 $\pm1.8$ pb, which is consistent with the Standard Model
unless the top quark mass is above 185~GeV.  Nevertheless,
the possibility of order-1 deviations of the
top quark couplings from the
predictions of the Standard Model is extremely tantalizing and might
well open a new window into physics of flavor.

  At the Tevatron collider, the dominant mechanism of $t\bar t$
production is through $q\bar q$ annihilation; at a top mass of
175 GeV, the $gg$ fusion cross section is a factor 10 smaller.
To dramatically enhance this cross-section, it is natural to consider
models with an
 $s$-channel resonance  in $t\bar t$ production, with a mass of
500--600 GeV, just above the $t\bar t$ threshold.
 Two groups of
authors have proposed explanations of this type based on
pre-existing theoretical models \cite{HP,EL}.

  Higher luminosity running at the Tevatron collider will clarify this
situation.  The clearest prediction of the model is an excess of events
at large $t\bar t$ invarinant mass and high $t$  transverse momentum.
Taking the model  \cite{HP}\ with a 600 GeV resonance for
definiteness, event samples of 0.1 fb$^{-1}$, 3 fb$^{-1}$, and
100 fb$^{-1}$ lead to roughly  10, 300, and 10,000 observed $t\bar t$
events with $m(t\bar t) > 600$ GeV, assuming efficiency (1/10) for
detection, above a background 5 times lower.  These samples correspond
to the goals for the current Tevatron collider run, the new Main
Injector upgrade, and the TeV$^*$ luminosity upgrade.
With the same event
samples, one could put lower
limits on the mass of an $s$-channel
resonance at roughly  700, 1000, 1500 GeV.

 At the LHC, the production cross section for $t\bar t$ is 90\%
$gg$ fusion, but the observed rates are large, of order $10^6$ $t\bar t$
pairs per year.  The cross section measurement at the LHC clearly
distinguishes the $q\bar q$ from the $gg$ model for the $s$-channel
resonance; these models predict cross sections differing by a
factor of 10.  The mass reach of the LHC with 10 fb$^{-1}$ is
about 1000 GeV
 in the $q\bar q$ model and 2000 GeV in the $gg$ model, with stronger
constraints at 100 fb$^{-1}$.

  Since both models involve resonances that do not couple to leptons,
their  direct
effect on $t\bar t$ production is not easily tested at an $e^+e^-$
collider.  However, for both models, an $e^+e^-$ collider with an
energy about 1 TeV could provide important complementary information
on the associated production of the new resonances with $t\bar t$
and on the pair-production of related new particles
with electric charge and exotic color.

\subsection{Anomalous Top Couplings---Effective Lagrangian}

  If the top quark production cross section proves not to be
anomalous at the zeroth order, how well can the top quark
couplings be searched for anomalies at higher order?  An appropriate
framework for discussing this question is an effective Lagrangian
governing the top quark coupling to gauge bosons.  For example,
the coupling of the $W$ boson to $t\bar b$ may be written
\begin{equation}
\Delta {\cal L} =   {g\over \sqrt{2} }F_1^W(q^2) \bar t \gamma^\mu b_L
   W_\mu + {1\over 4m_t} F_2^W(q^2)\bar t\sigma^{\mu\nu} b_L
   F_{\mu\nu}^W .
\end{equation}
Analogous formulae define the $F_1$ and $F_2$  form factors of
$g$, $\gamma$, and $Z$.
Phenomenological couplings of this form have been studied by
many authors \cite{KLYetc}.  Electroweak interactions
typically give contributions to these form factors at the
1\%\ level or below.  However, it is likely that there are larger
contributions from models in which the top quark is strongly coupled
to the symmetry breaking sector.  For example, extended technicolor
models give contributions to the various $F_1$ form factors of $t_L$,
$t_R$, and $b_L$ which are typically of order 5\% \
but whose pattern
depends on the details of the model~\cite{CCST}.

We consider first the measurement of these form factors in
 hadron-hadron collisions.  It is possible that these form factor
corrections are large; in fact, it has been proposed
 \cite{2}  
that a nonzero color anomalous magnetic moment
 $F_2^g \sim 0.25$ could provide an alternative
explanation of the large cross section  reported by CDF.  On the
other hand, if the form factor effect is smaller, one's ability
to measure it quickly becomes limited by the 15\%\ QCD uncertainty
in the $t\bar t$ production cross section.
In a one-parameter fit based on the total cross section,
the Tevatron could bound the $F_2$ of the $ttg$ coupling
to $\pm 0.15$ (95\% conf.) with 0.1 fb$^{-1}$ and reach the
systematics limit of $\pm 0.10$ with 0.5 fb$^{-1}$ \cite{AKR}.
The same systematics limit appears at the LHC.
Yuan has suggested that the $F_1$
coefficients of the $Wt\bar b$ couplings  can be constrained by
measuring the cross section for single top production: $g W\ra
t\bar b$; this cross section, however, has a
theoretical uncertainty of order 30\% \cite{Ytb}.
 One parameter of $t\bar t$ production at hadron colliders
 which seems very promising for a
precision  constraint is the measurement of the
longitudinal/transverse polarization of $W$ bosons from top
decay; a 3 fb$^{-1}$ event sample at the Tevatron should
bring the statistical uncertainty below 4\% \cite{Amidei}.

   The environment offered by $\epem$ colliders offers many
advantages for a precision study of the top couplings.  In addition
to low backgrounds and the ability to reconstruct $t\bar t$ in the
6-jet mode, top production is characterized by order-1
forward-backward and polarization asymmetries which reflect
sizeable interference of the $\gamma$ and $Z$ exchange diagrams.
Top quarks produced in the forward direction carry the electron
polarization and thus provide a polarized sample for studies of the
decay form factors.

For comparison with the estimate given above, the total
cross section measurement at an $\epem$ collider with 20 fb$^{-1}$
should bound a
common $F_2$ for the $tt\gamma$ and $t\bar t Z$ couplings, in a
one-parameter fit, at $\pm 0.03$.
However, much more detailed studies are possible, at least with
large event samples.  For example, consider a variation in the
$Zt\bar t$ coupling leaving the $t\bar t \gamma$ coupling fixed.
The left- and right-handed form factors of $Zt\bar t$ coupling
can be constrained
in a two-parameter fit for an event sample of
100 fb$^{-1}$ at 500 GeV, to deviations of  $\pm 0.08$ (95\%\
conf.).   (The systematic contribution to this
error should be small, and will be analyzed carefully for the
final report.)
 For comparison with the parameter of top decays discussed above,
  Fujii
has found that the fraction of longitudinal $W$ bosons in $t$
decay can be determine with a statistical
 precision of $2\%$ with 10 fb$^{-1}$ of data at 350 GeV.
By studying top pair production with an extra jet, one can
obtain \cite{4}
a direct limit on the color anomalous magnetic moment $F_2^g$,
 to be compared
with that from hadron colliders; with 200 fb$^{-1}$, the constraint
is $\pm 0.16$ (95\%~conf.).

  The $F_1$ associated with the $Wt\bar b$ vertex can be obtained
from a measurement of the total width of the top quark, assuming
that no exotic decay modes are seen.  This quantity can be measured
at the threshold for $t\bar t$ production by three techniques:
the width of the 1S resonance or shoulder, the momentum
distribution of reconstructed top quarks decaying to $Wb$, and
the forward-backward asymmetry of top production due to the
interference of overlapping $S$ and $P$ wave states.  Fujii,
Matsui, and Sumino \cite{FMS} have found that, by combining
these techniques, the width of the $t$ can be determined
with 100 fb$^{-1}$ of data
to
an accuracy of 4\%, plus an error from the  uncertainty in
$\alpha_s$ that boost this number to about 10\%. We should
also note that,
in models with a
relatively light Higgs boson, of mass about 100 GeV,
the height
of the 1S enhancement at the $t\bar t$ threshold measures the
$t \bar t$ Higgs coupling constant,  to about 25\%\ accuracy.

 In addition to  magnetic dipole  ($F_2$) couplings, the top quark
may also have electric dipole couplings which violate CP.  Direct
measurements of the form factors give limits on the CP-violating
couplings similar to  those cited above.  This is probably not
adequate for those models of CP violation involving
CP-violating Higgs bosons coupling strongly to the top quark.
However, Schmidt and Peskin have suggested observing the
energy asymmetry in $e^+$ vs. $e^-$ decay
products of $t\bar t$.  Combined with the high statistics
available at the LHC, this technique could reasonable reach
the CP asymmetries of $10^{-3} - 10^{-4}$   predicted by the
models \cite{SP}.

\subsection{Exotic Decays of the Top Quark}

The top quark may also connect to exotic states by its decays.
Many exotic decay modes of top have been considered in the
literature.  Here are the most important ones:

$t \ra H^+ b$:  This mode is interesting because it is the
best  way to find $H^+$ at hadron colliders.  In addition, the
ratio of $t$-quark branching ratios to $H^+b$ and $W^+b$
depends on $\tan\beta$, providing one of the best ways
to measure this parameter of an extended Higgs  sector.
The decay mode should be found at LHC with $10^{33}$ luminosity,
or with a  500 GeV $\epem$ collider with 1 fb$^{-1}$ of data.

$t \ra \widetilde\chi^0 \widetilde t$:  This is a supersymmetric
decay of top which occurs in a reasonable volume of parameter
space, with a branching ratio of roughly 5\% if $\widetilde t$
is much lighter than $t$ \cite{EE}.

$t\ra c h^0$, $t \ra c Z^0$, $t \ra c g$, {\it etc.}:
  These flavor changing neutral current
modes are extremely rare in the Standard Model, with branching
ratios $< 10^{-10}$. If these modes are dramatically
 enhanced by new physics,
they might be observed at the LHC.

$t\ra s W^+$.  This mode is expected to be at the 0.1\%\ level
in the Standard Model.  If this mode is enhanced to the few-percent
level, it should be observable at an $\epem$ collider \cite{Hild}.



\def\to{\rightarrow}
\def\Slash{\hskip -.6em/}
\def\Zbb{Zb\bar{b}}
\def\glb{g_L^b}
\def\grb{g_R^b}
\def\dglb{\delta g_L^b}
\def\dgrb{\delta g_R^b}
\def\ds{\delta s^2}
\def\da{\delta\alpha_s}
\def\drho{\delta\rho}
\def\seff{\sin^2\theta_{\rm eff}}
\def\sig0{\sigma_{\rm had}^0}
\def\Gamhad{\Gamma_{\rm had}}
\def\Gamb{\Gamma_{b\bar{b}}}
\def\Gamc{\Gamma_{c\bar{c}}}
\def\Game{\Gamma_{e^+e^-}}
\def\delb{\delta_b}
\def\xib{\xi_b}
\def\zetab{\zeta_b}
\def\beq{\begin{equation}}
\def\eeq{\end{equation}}
\def\beqa{\begin{eqnarray}}
\def\eeqa{\end{eqnarray}}
\def\ie{{\it i.e.}}
\def\eg{{\it e.g.}}
\def\etc{{\it etc}}
\def\etal{{\it et al.}}
\def\ibid{{\it ibid}.}
\def\tev{\,{\rm TeV}}
\def\gev{\,{\rm GeV}}
\def\mev{\,{\rm MeV}}

\section{Virtual Effects of New Physics}

\subsection{Overview}

The study of virtual effects can open an important
window on electroweak symmetry breaking and physics
beyond the Standard Model (SM). The
examination of indirect effects of new physics in higher order processes offers
a complementary approach to the search
for direct production of new particles at high
energy accelerators.  In fact, tests
of loop induced couplings can provide a means of probing the detailed
structure of the SM at the level of radiative corrections where
Glashow-Iliopoulos-Maiani (GIM) cancellations are important.
In some cases the constraints on new degrees of
freedom via indirect effects surpass those obtainable from collider searches.
In other cases, entire classes of models are found to be incompatible.
Given the large amount of high luminosity data which will become available
during the next decade, this approach to searching for physics
beyond the SM will become a valuable tool.

\subsection{Precision Electroweak Measurements}

Virtual effects from new physics are
easily detected in precision electroweak measurements.
These include the measurements of the $Z$ line--shape and
asymmetries at SLD/LEP, deep inelastic muon--neutrino scattering,
and atomic parity violation.
Because of the accurate theoretical predictions in
the electroweak sector of the SM,
especially in purely leptonic processes that are free of
QCD corrections, any observed
deviations must be explained as
due to new physics not encompassed within
the SM.

Radiative corrections from new physics are often classified
into two groups; the `oblique' and the `direct'
corrections.
`Oblique' refers to
the vacuum polarization
corrections to the gauge boson propagators. They are independent
of the external particles and thus are universal to all electroweak
processes.
`Direct' refers to
the vertex and box corrections which are specific to
each process.
Because of the universality of oblique corrections, it is possible
to express them in a model independent fashion with a few
phenomenological parameters.  Under the two assumptions that the
electroweak gauge group is the standard $SU(2)_L \times U(1)_Y$,
and that the scale of new physics is large compared to the electroweak
scale, the virtual effects of new physics can be expressed
in terms of 3 parameters, often called $S$, $T$, and $U$ \cite{PT}.
Using a set of measurements to which the direct corrections
from new physics
are expected to be small, one can place tight
constraints on the values of $S$ and $T$.
These include the purely leptonic processes at SLD/LEP,
and atomic parity violation measurements.
Currently, the strongest constraint on $S$ and $T$ comes from
the SLD/LEP measurement of $\sin^2\theta_w^{\rm eff}$ and
$\Gamma(Z\rightarrow \ell^+\ell^-)$.

If one relaxes the second assumption that the scale of
new physics is large compared to the electroweak scale,
then it becomes necessary to introduce three additional
parameters called $V$, $W$, and $X$ \cite{MBL}.
If one relaxes the first assumption and extends the
electroweak gauge group to $SU(2)_L \times U(1)_Y \times U(1)_{Y'}$,
$SU(2)_L \times SU(2)_R \times U(1)_{B-L}$,  $SU(3) \times U(1)$, etc.
then it becomes necessary to take into account the
exchange of the additional gauge bosons as well as their
virtual loop effects so these models cannot be treated within
the $S$--$T$ formalism and
each case must be analyzed separately.

A process where the direct corrections from new physics is
expected to be sizable is $Z\rightarrow b\bar{b}$.
This is due to the fact that the $b$--quark is the isospin
partner of the $t$--quark so the same mechanism which generates
the large $t$--quark mass will also generate a large vertex
correction to this process.  This is true in the SM
which has a vertex correction proportional to $m_t^2$, and also in
Technicolor models and SUSY models.
If such new vertex corrections exist, then one
should be able to see
a deviation from the SM prediction
after the universal oblique correction effects have been folded out.
Such a deviation at the $2\sigma$-level is presently observed
at LEP in the ratio
$R_b = \Gamma_{b\bar{b}}/\Gamma_{\rm had}$.

\subsection{Rare Processes in the Quark Sector}

Rare $K$ processes have played a strong and historical role in constraining
new interactions.  The box and penguin diagrams which typically
mediate these processes receive
contributions from the SM, as well as
potentially sizeable contributions from new physics.
For example, the strongest limit (albeit assumption dependent)
on the mass of a right-handed $W$-boson is derived from its
contributions to $K^0$--$\bar K^0$ mixing, and the requirement of
near degeneracy of squark masses results from a super-GIM mechanism imposed
in the $K$ sector.
Rare $K$ decay experiments at Brookhaven and
KTEV at the Tevatron will provide further
probes in the near future.  The decay
$K^+\to\pi^+\nu\bar\nu$ is an interesting
mode as it is theoretically clean and experiment will reach the SM level
during the next decade.
$K^0_L\to\mu^\pm e^\mp$ is forbidden in the SM and
provides a probe of
lepton number violation, leptoquark exchange, and family symmetries.

Short distance SM contributions to rare charm processes are small
due to the effectiveness of the GIM mechanism (as there are no large quark
mass splittings in the loops).  Long distance effects usually dominate and
are difficult to reliably calculate.
Observation of $D\to\rho\gamma$
would provide an excellent test of the  techniques used in
these calculations, and
could be used to reliably evaluate the related long distance contributions
to $B\to K^*\gamma, \rho\gamma$.
However, there is a window for the clean observation of possible new physics
contributions in some processes, such as $D^0$--$\bar D^0$ mixing,
$D\to X_u\ell^+\ell^-$, and $D\to\mu\mu$.
The types of non-standard scenarios which give the largest
contributions to these reactions are (i) two-Higgs-doublet models in
the large $\tan\beta$ region, (ii) multi-Higgs doublet models with
flavor changing couplings (iii) new $Q=-1/3$ quarks, (iv) supersymmetric models
with non-degenerate squarks, and (v) variants of left-right symmetric models.
Several high volume charm experiments are planned for the future,
with $10^{7-8}$ charm mesons expected to be reconstructed.
$D^0$--$\bar D^0$ mixing can then be probed
another 1--2 orders of magnitude below present sensitivities.

A large amount of data in the $B$-meson system will be acquired during the
next decade at SLC/LEP, CESR, the Tevatron, and the SLAC and
KEK B-factories.  Here, one-loop processes occur with sizeable
rates in the SM, due to large top-quark contributions.
Many classes of new models can give significant
and testable contributions to rare $B$ processes.
The inclusive process $B\to X_s\gamma$ and the related exclusive decay
$B\to K^*\gamma$ has been observed by CLEO and
is the first direct observation of a penguin mediated process.  It has
provided strong restrictions on parameters in several theories beyond the
SM \cite{jlh}, with the bounds being competitive with
those from direct collider searches.  For example, it constrains the mass of
a charged Higgs boson to be $>260\gev$ for $m_t=175$ GeV, and provides
complementary bounds to those obtained at colliders on an anomalous $WW\gamma$
vertex.
The inclusive processes $b\to s\ell\ell, s\nu\bar\nu$ are also excellent
probes of new physics.
They are sensitive to the form of
possible new interactions since they allow measurements of various
kinematic distributions as well as the total rate.
Techni-GIM models predict $B(b\to s\mu^+\mu^-)\sim {\cal O}(10^{-4})$ and
appear to already be in conflict with the present experimental limits.

CP violation in the $B$ system will be examined
at the SLAC and KEK B-Factories.
Signals for new sources of CP violation include, (i) non-closure of the
3 generation unitarity triangle, (ii) non-vanishing CP asymmetries for the
channels $B^0_d\to \phi\pi^0, K_S^0K_S^0$, (iii) inconsistency of separate
measurements of the angles of the unitarity triangle,
and (iv) a deviation of CP rates from SM predictions.
The large data sample and asymmetric configuration of the
B-factories will provide a series of unique consistency tests of the quark
sector and will challenge the SM in a new and quantitatively precise manner.

Loop-induced top-quark decays are small in the
SM due to the effectiveness of the GIM mechanism.   Branching fractions for
$t\to c\gamma, cZ, cg, ch$ lie in the range $10^{-12}$--$10^{-8}$.
Long distance
effects are expected to be negligible.  Contributions from new physics can
enhance these rates by 3-4 orders of magnitude.
The tree-level decay $t\to ch$, if kinematically accessible,  is possible
in multi-higgs models without natural flavor conservation.
$t\to \tilde t\tilde\chi^0$ can occur in SUSY if the stop-squark is light.
These would provide
clean signals for new physics and may be observable at a high luminosity
Tevatron.

\subsection{Electric Dipole Moments}

Electric dipole moments (EDMs) of atoms, molecules, and the neutron
are the most sensitive test of low
energy flavor conserving CP violation.
Such CP violation from the CKM angle
is highly suppressed (by at least 3 loops), while
other sources of CP violation are generally not so suppressed.
Atoms with an unpaired electron (such as $^{133}$Cs)
are generally most sensitive
to the electron EDM, while
the neutron and atoms with paired electrons (such as $^{199}$Hg)
are sensitive
to CP violation in the strongly interacting sector.
Although no EDM has yet been observed, the current
bounds already put stringent constraints on CP-violating
extentions of the SM.
For example in supersymmetric theories the bound
of $1.3 \times 10^{-27}$ e cm for the EDM of
$^{199}$Hg  places a limit on new CP-violating phases
that can arise in gaugino
mass terms and in off-diagonal squark masses and squark interactions.
These new phases must be less than a few tenths of a percent
for $M_{\rm SUSY}\simeq 100$~GeV.
This bound also limits the CP-violating
phases in the Higgs potential in multi-Higgs theories
to be less than roughly $10^{-1}$ for
$m_{\rm Higgs} \simeq 100$ GeV.
The bound from the neutron gives a slightly less stringent
limit.
Improvement in most experiments is anticipated over the
next few years.  New techniques, such as atomic traps,
may allow improvement of up to a few orders of magnitude.
If positive measurements are eventually made the
QCD vacuum angle contribution can
be distinguished
from new physics by comparing atoms or molecules
with paired and unpaired electrons.

\subsection{Double Beta Decay}

Neutrinoless double beta decay requires the violation of
lepton number and is thus a clear signal for
physics beyond the SM.
Such a decay could arise from a nonzero Majorana mass
for the electron neutrino.
In theories with spontaneously broken lepton number,
the decay could also be accompanied by the
emission of a Majoron.
The current half life limit for $^{76}$Ge
gives a bound on the effective Majorana neutrino mass
of $m_{\nu} < 1$--2 eV.
The neutrino-Majoron coupling is also constrained to
be $g_{\nu,\chi} < 10^{-4}$.
The current generation of enriched Ge experiments could
reach the 0.3 eV level in $m_\nu$.
A large experiment, NEMO III, using $^{100}$Mo,
is being designed to reach the 0.1 eV level.




\setcounter{section}{11}
\section{Experimental Issues at Hadron Colliders}

The frontiers of particle physics have been significantly
pushed forward by discoveries made at colliders with the highest
available center-of-mass energies.
Since the early 1980's, hadron colliders have become the tool of choice for
extending our knowledge of
the high energy frontier.
The ability of collider detectors to cover a broad range of new phenomena over
a wide range of mass scale allows many of the most important measurements
to be made and limits to be set in an economical way.
A mix of both ``discovery'' and precision measurements have been successfully
carried out at such  machines over the last ten years.
Progress has depended upon a mix of both theoretical and experimental
developments.

During the last ten years, the theoretical community has been able to refine
our knowledge of hadron-hadron interactions to enable calculations with
increasing precision.
The soundness of the theoretical framework puts the hadron-hadron
collider on a firm footing for reliable estimation of event rates and
backgrounds for a variety of new phenomena.

The experimental community has likewise made important advances during this
period.
Operating experience at hadron colliders
with precision tracking detectors has been gained.
These devices open up the possibility of tagging jets containing $b$-quarks,
a capability that
considerably broadens the physics reach of collider detectors as the recent CDF
top quark evidence has shown.
In addition, the broad-based program of detector R\&D for SSC and LHC detectors
has deepened our understanding of how to handle the high occupancies and
radiation fluxes at high luminosity hadron colliders.
The knowledge gained from the detector R\&D and design studies
gives us confidence that the theoretically predicted physics capabilities of
the hadron collider machines can be exploited in detectors that are practical
to build and operate.

\vskip1pc
\subsection{Machine Environment}

The LHC program is designed to study physics beyond the
Standard Model over a wide mass range for all types of hypothetical particles.
To achieve a substantial reach for the production and observation of new
particles, LHC plans to run at a luminosity of $10^{34}$cm$^{-2}$s$^{-1}$
with a 25 ns bunch spacing.
There will be on average
25 interactions per crossing contributing considerable complexity to the
detection of interesting phenomena.
Future luminosity upgrades at Fermilab may also generate conditions
comparable to these.
The challenge met by the SSC and LHC R\&D programs was to create detector
designs that could survive in these harsh conditions of high occupancy and
high radiation, while providing adequate resolution and robust triggering
capability.

The general results of the R\&D program showed that existing detector
technologies could be successfully employed to meet the SSC/LHC requirements.
The higher
\par\eject
\noindent
occupancies expected are accommodated in first order by more
finely segmenting the detector.
The main R\&D challenges then lay in the area of electronics and DAQ design.
The widespread use of custom integrated circuits in the detector front-end
electronics allows practical designs to handle the high rates and fine
segmentation at finite cost.
Modest extension of existing multi-level
trigger designs are able to handle the expected event rates.

The high radiation levels in high-luminosity hadron machines have been a
cause for concern for both the SSC and LHC experimental groups.
Again, the R\&D program has shown that radiation resistant versions of
the most sensitive detector elements can be fabricated with only modest
extension of existing capability.
In particular, many front end electronic circuits that would be
exposed to the harsh radiation conditions have been fabricated in
radiation-hard versions and shown to work.
Also, the level of thermal neutron flux inside the inner tracking cavity and
within the detector hall is brought within acceptable limits for detector
operation by appropriate shielding design, especially in the region around
the machine final focusing quads.
Thus, while the problem of radiation resistance remains an ongoing area of
further development for the experimenters, solutions to all the most critical
problems appear to be in hand.
For more specific details
on proposed solutions, see the SSC and LHC detector design reports.

\vskip2pc
\subsection{Detectors}

Two large general purpose detectors are being designed for the LHC, namely
ATLAS and CMS.
Both experiments have been designed to
detect all hypothesized particles accessible to the LHC.
Many studies have been carried out by these collaborations, and
are detailed in the ATLAS and CMS letters of intent.
Generally, to fully exploit the potential physics reach of future hadron
colliders, detectors must feature good charged particle tracking, precision
electromagnetic calorimetry,
full angular coverage for the hadronic calorimetry,
and a robust muon detection system.

How physics processes set the requirements  on the detector design has been
fairly thoroughly explored by the
SSC and LHC experimental groups.
As examples, detection of the signal $H\rightarrow\gamma\gamma$ for
low mass Higgs requires shower counter resolution in the range
$\Delta E/E=10\%/\sqrt E\oplus.7\%$, measurement of photon position with
accuracy in the range $\sim 5$ mm/$\sqrt E$, and timing resolution
adequate for tagging from which bunch the photon originated.
Measurement of high mass like-sign $W$ production requires measurement of the
sign of the charge of the $W$ decay lepton out to $p_\perp\sim500$ GeV/$c$.
Angular
coverage for both electrons and muons must extend to \unskip\break
\noindent
rapidities of
$\sim 2.5$ for good angular acceptance for the $H\rightarrow4\ell$ signals.
Many other physics processes have been studied to set detailed design
parameters for the detectors.

SSC and LHC R\&D programs were required to show that detectors with the
requisite capabilities are practical to construct.
The SSC and LHC proposals have shown that detectors with resolutions
comparable to existing detectors are adequate for the task.
In large measure, the higher event rates expected at future machines are
handled by finer segmentation of the detector elements,
as mentioned earlier.

The existence of multiple interactions per crossing provides a further
complication of this problem, mainly by degrading the pattern
recognition capabilities of the detector.
Each physics process must be studied with a detailed detector simulation
to quantify the capability of  the design to detect the signal of interest.
Such studies have been carried out in detail, with full event
reconstruction, for many of the important Standard Model Higgs detection modes.
Numerous other physics processes have also been studied at this level of
detail.
However,
the full extent of the capabilities of the designs as they currently stand
has not been explored.
In several areas, preliminary parton-level estimations  of detector
capability will be supplemented by more detailed simulations later.
Areas of active investigation by the LHC collaborations include b-jet tagging
capability at high luminosity, forward jet tagging capability, and alternate
modes for MSSM Higgs detection.

\subsection{Outlook}

During the last five years, important progress has been made in understanding
the experimental challenges of hadron colliders.
This progress includes improved theoretical understanding of the basic
Standard Model physics processes which occur at such machines, improved
experimental technique through the operating experience gained mainly at
Fermilab, and improved instrumentation mainly through the program of SSC
and LHC R\&D.
Over the next five years, we expect to maintain a program of continual upgrade
at FNAL, and continue preparations for LHC.
New limits will be set at FNAL during this period on physics beyond the
Standard Model.
In addition, measurement of the top and $W$ masses with improved precision
will provide indirect information on the electroweak symmetry breaking sector.
Within the next ten years, we hope to see the turn-on of the LHC with a first
real chance for direct production of more massive Higgs particles, and
considerable risk of discovery for physics beyond the Standard Model.




\def\mum{\mu{\rm m}}

\section{Experimental Issues at \protect\boldmath$e^+e^-$ Linear Colliders}

\subsection{The Physics Environment}

Experimental studies at $e^+e^-$ colliders over the past two decades
have provided key observations and insights to the
nature of the fundamental particles and interactions of the Standard
Model.
Electron-positron collisions
yield events with simple and transparent structures.
Annihilation events carry the full energy of the beam, and
produce final states with few partons.
Searches for new phenomena give complete and unambiguous results, and
precision studies of strong and electroweak interactions are made
with a minimum of bias and background.
The physics environment at TeV-scale colliders will continue to be as
ideally suited to the exploration of particle physics \cite{exp}.

The Standard Model final states that will predominate at a
$\sqrt{s}=500$ GeV
$e^+e^-$ collider are $udscb$ pair production (9 units of $R$),
$W^+W^-$ (20 units of $R$), $ZZ$ (1.2 units of $R$),
and $t\bar t$ (1 unit of $R$), where $R=87$ fb/s if
s is in TeV$^2$.   These processes
correspond to relatively simple final states, and,
with a typical large solid-angle detector,
event structures can be sorted to reduce
the 50 or so pions, kaons, and photons
to a few 4-vectors that accurately describe the underlying partons.
With the partons reconstructed with high precision,
topological and kinematic constraints can be utilized
to take complete advantage of the simplicity of the
physics environment at $e^+e^-$ colliders.

Polarization of the electron beam in a linear collider is a unique
tool that provides new and important views of particle physics.
It can be utilized to isolate longitudinal W boson states,
probe the helicity structure of the interactions of the top
quark, reveal the underlying symmetry of SUSY, and determine the
gauge structure of new interactions that might be found in high
energy collisions - to name just several targets of opportunity.
Sources of intense bunches of polarized electrons are presently
used at the SLC where beam polarizations of 80\% are routinely
delivered to the collision point.
Development of sources is continuing and it is reasonable
to expect polarizations of 90-95\% in the future.
Instrumentation has been developed to measure beam polarization
with accuracies of a per cent or better.

\subsection{Detectors}

The analysis of events at future $e^+e^-$ colliders
requires good tracking of high-momentum
leptons and good reconstruction of quark jets.
The isolation of heavy quarks with precision vertex detectors
will continue to play an important role in the analysis of events.
Particles that go unseen in the detector,
such as neutrinos and beamstrahlung photons, can be fully
reconstructed in many instances by applying energy-momentum and mass
constraints.

Most studies of physics at future electron-positron colliders
have used Monte Carlo simulations of detectors with capabilities
that have already been achieved at LEP and SLC.
The parameters of a ``Standard Detector''
are listed in Table~\ref{TABEE1}.
We see that the
charged-particle tracking and electromagnetic and hadronic
calorimetery does not extend any
technology beyond that now in existence.

\begin{table}[htb]
\vskip-1pc
\begin{center}
\caption{``Standard Detector'' Parameters.
All momenta and energies are in GeV.  The symbol $\oplus$ means
quadratic sum.}
\label{TABEE1}
\setlength{\tabcolsep}{12.5pt}
\renewcommand\arraystretch{1.2}
\begin{tabular}{|ll|}
\hline
Calorimetry  & \\
\quad     $\sigma_E/E$ (electromagnetic)     & $8\% /     \sqrt{E}
                                            \oplus 2\%$    \\
\quad     $\sigma_E/E$ (hadronic)            & $60\% /    \sqrt{E}
                                            \oplus 2\%$    \\
\quad     Cell size  (electromagnetic)     & $2^\circ$    \\
\quad     Cell size  (hadronic)            & $4^\circ$    \\
\hline
Tracking   & \\
\quad      $\sigma_{P_t}/{P_t}$           & $10^{-3} \cdot P_t$  \\
\quad      Vertex resolution (impact)
 & $ 20\mum    \oplus \bigl({{100 \mum}\over{P}}\bigr) $
 \\ [3pt]
\hline
Hermiticity   &   \\
\quad      Calorimetery and tracking
                          & $\theta_{e^\pm} > 10^\circ$ \\
\hline
\end{tabular}
\vskip1pc
\caption{``Improved Detector'' Parameters.
All momenta and energies are in GeV.  The symbol $\oplus$ means
quadratic sum.}
\label{TABEE2}
\vskip6pt
\begin{tabular}{|ll|}
\hline
Calorimetry   & \\
\quad      $\sigma_E/E$ (electromagnetic)     & $8\% /     \sqrt{E}
                                            \oplus 1\%$    \\
\quad      $\sigma_E/E$ (hadronic)            & $35\% /     \sqrt{E}
                                            \oplus 2\%$    \\
\quad      Cell size  (electromagnetic)     & $1^\circ$    \\
\quad      Cell size  (hadronic)            & $2^\circ$    \\
\hline
Tracking   & \\
\quad      $\sigma_{P_t}/{P_t}$
    & $2\times 10^{-4} \cdot P_t$  \\
\quad      Vertex resolution (impact)
    & $  5\mum    \oplus \bigl({{50 \mum}\over{P}}\bigr) $
     \\ [3pt]
\hline
Hermiticity   &   \\
\quad      Calorimetery and tracking
                          & $\theta_{e^\pm} > 10^\circ$ \\
\hline
\end{tabular}
\end{center}
\end{table}

We can anticipate that
improvements in the techniques and technologies used in the design
and construction of particle detectors will occur.
At center of mass energies of several hundred GeV, most applications
of electromagnetic and hadronic calorimetery do not suffer from lack
of shower statistics, but it is increasingly important to minimize
systematic errors in the reconstruction of the
energies and positions of neutral hadrons and photons.
The charged particles of greatest interest are similarly energetic,
so that the scattering of tracks in the material of the detector becomes
less problematic.
The more aggressive set of goals shown in Table~\ref{TABEE2}
might be realizable and certainly
would enhance the physics output of the experiment.
More systematic and quantitative study is needed to arrive
at a complete set of design parameters, but it is clear that existing
detectors are not far from ideal.

\subsection {Event Rates and Time Structures}

The total $e^+e^-$ annihilation cross section at high energies is
approximately 30 units of $R$,
or $10^{-35}$cm$^2$ at 500 GeV.    Colliders at these energies are
designed to generate
luminosities typically $5 \times 10^{33}$cm$^{-2}$s$^{-1}$,
with an interesting event produced every 10 seconds or so.
Small-angle Bhabha scattering of electrons and positrons into the
region covered by detector elements (above 10 degrees polar angle)
occurs with a cross section 5-10 times the annihilation rate. This
serves as a good monitor of the luminosity and energy spectra of
collected data samples.

Cross sections for
backgrounds produced by the coherent beam-beam interaction and
low-energy peripheral two-photon reactions are also
larger than those of annihilation events.
For example, numerically the total cross section for the
reaction $ee \rightarrow eeee$ is $\approx 10^{-26}$cm$^{-2}$
(at beam energy of 500 GeV).
Most of this cross section
corresponds to production of $e^+e^-$ pairs near threshold, but the
total rate can be sufficiently large that care must be taken in the
design of the accelerator and detector to minimize the probablility
that soft particles from these interactions will overlap with
the desired annihilation events.
Accelerator and detector designs have evolved that account for these
processes.

The luminosity per bunch crossing and the time interval between
crossings varies considerably from one machine design to
another.
Those based on superconducting linacs yield the lowest luminosity
per crossing and spacing between bunches of up to a microsecond.
The duty cycle and time structure of these machines are similar
to those found at LEP for example, and place similar demands on
detector data acquisition systems.
Colliders that utilize linacs with
high frequency rf (at X-Band for example),
create brighter collisions in short trains of typically 100 bunches
pulsed at repetition rates of 100--200 Hz.
Bunches within a train will be spaced by a few nanoseconds.
Detector trigger and data acquisition systems will have long periods
to analyze and format information from each repetition cycle, but
it will be advantageous for detectors to be able to assign observed
charged and neutral particles to their proper bunch-bunch interaction.
Detectors have been built
with this capability, and these issues have been considered
in the evaluation of physics analyses at colliders with
center of mass energies as high as 1.5 TeV.

\subsection {Beamstrahlung}

Another property of high-energy linear colliders
is that of ``beamstrahlung''
--- the radiation emitted by beam particles as they pass through
the electromagnetic field of the opposing bunch.
Corrections for the effects of initial-state radiation in
$e^+e^-$ collisions has long been a well-understood process, and
similar techniques will be needed at future colliders to
account for the smearing of annihilation
center of mass energies by beamstrahlung.
All proposed $e^+e^-$ machines
are designed so that the smearing of
the center of mass energy by beamstahlung is never greater than,
and often less than, the smearing by initial state bremsstrahlung.

The consequences of these effects on the analysis of data have
been studied by many groups, and techniques to
correct data, and even individual events, for radiation of
energy by the incident particles prior to collision have been developed.
It is extremely rare that more than one hard photon is emitted in any
event, and those that are radiated travel along the incident beam
direction,  so it is often possible to simply include
the radiation as an unknown parameter in a likelihood fit to the
event topology.
This works well for analyses of exclusive final states such as
$W^+W^-$, $Z^0H^0$, and $t \bar t$ pairs.

\subsection {Machine-Induced Backgrounds}

While the physics environment created by electron-positron annihilation
is remarkably neat and orderly, the single-pass nature of a
linear collider poses a very special challenge to the experimenter
and machine designer.
Unlike a storage ring in which the circulating beam is quickly
reduced to only those particles that are well-contained within
the phase space of the physical and dynamic apertures of the
machine, a pulsed collider will transmit particles that are very
near to its aperture limit.
Such particles will create backgrounds in experimental detectors
either through the emission of synchrotron radiation as they
pass through magnetic fields near the detector, or by the
creation of secondary debris when they strike physical elements
along the beamline.
The detector must be properly masked, and the beam must be
properly collimated to remove beam extremeties,
or ``tails'', well upstream of the interaction region.
It is important that the design of the collider include accomodation
for special optical sections to allow for complete collimation of the
phase space of the beam.

The experience of the MARK II
and SLD detectors \cite{bobj} at the SLC
has been an invaluable guide to the problems that will be encountered
by experiments at future linear colliders.
Successful operation of these experiments and the lessons that have
been learned from the SLC provide a good basis for design and
operation of experiments at the NLC.


\section{Conclusions}

Exploration of the TeV-scale requires a new generation of colliders
and detectors beyond those currently in operation, and
is essential for the long-term vitality of particle physics.
In this report, we describe an in-depth study of the phenomenology of
electroweak symmetry breaking and quantify the
``physics reach'' of present and future colliders.  Our investigations
focus on the Standard Model (with one Higgs doublet) and beyond:
models of low-energy supersymmetry, dynamical electroweak
symmetry breaking, and other approaches
containing new particles and interactions.
Implications for future experiments at hadron and $e^+e^-$ colliders
are considered.  On the basis of this work,
we present the following conclusions and recommendations.

The participation of the United
States in the LHC program should be vigorously supported.
An extensive study of the electroweak symmetry breaking sector
and the elucidation of physics beyond the Standard Model will
require a robust experimental program at the LHC.
The LHC
possesses an
impressive discovery potential for Higgs bosons, low-energy
supersymmetry, and a large
variety of beyond the Standard Model phenomena, with a mass reach
in many cases five to ten times the corresponding mass reach at
existing facilities.
Of course, with
the LHC approximately ten years from its initial run, one should
not overlook the possibility that initial evidence for
physics associated with electroweak symmetry breaking and/or physics
beyond the Standard Model could be uncovered at upgrades of existing
facilities.

Detailed simulations of high energy particle physics processes
at hadron colliders
are often required to obtain an accurate assessment  of the new physics
discovery capabilities of future experiments.  These include
realistic treatments of hadronic jets (beyond parton-level Monte
Carlo analyses) and full detector simulations.
An active collaboration between experimentalists and theorists is
crucial to the further development of strategies to extract evidence
for electroweak symmetry breaking
and beyond the Standard Model physics at future colliders.

There are many instances of the complementarity of
future hadron colliders and future $e^+e^-$ colliders.
A high energy $e^+e^-$ collider will provide a unique low-background
environment for the study of the production and decay of heavy gauge
bosons and top quarks and of new physics beyond the Standard Model.
A Next Linear Collider (NLC) at
0.5~TeV is an ideal laboratory for the study of intermediate-mass
Higgs bosons, while at 1.5~TeV, it can begin to probe the physics of
a strongly-coupled electroweak symmetry breaking sector.
Experimental studies at the NLC would provide outstanding opportunities
for discovery, while significantly
enhancing the ability of the LHC to interpret evidence for new
physics.

Other possible future collider
scenarios are beginning to be examined.  Physics studies at a
number of hypothetical Tevatron upgrades beyond the Main Injector
have been initiated.  A full study of the scientific potential of such
machines is necessary to provide input into any future cost/benefit
evaluation that might be undertaken in the future.

In plotting out the course for future colliders, the US must work
closely with its international partners.  By avoiding the
duplication of resources and taking into account the complementarity
of various approaches, it should be possible to propose a
set of future facilities that can fully explore the TeV
energy scale and elucidate the physics of electroweak symmetry breaking.
In doing so, we can push the frontiers of particle physics forward
and keep our field intellectually vital and exciting.

\def\MPL #1 #2 #3 {Mod.~Phys.~Lett.~{\bf#1},\  #2 (#3)}
\def\NPB #1 #2 #3 {Nucl.~Phys.~{\bf#1},\  #2 (#3)}
\def\PLB #1 #2 #3 {Phys.~Lett.~{\bf#1},\  #2 (#3)}
\def\PR #1 #2 #3 {Phys.~Rep.~{\bf#1},\ #2 (#3)}
\def\PRD #1 #2 #3 {Phys.~Rev.~{\bf#1},\  #2 (#3)}
\def\PRL #1 #2 #3 {Phys.~Rev.~Lett.~{\bf#1},\  #2 (#3)}
\def\RMP #1 #2 #3 {Rev.~Mod.~Phys.~{\bf#1},\  #2 (#3)}
\def\ZP #1 #2 #3 {Z.~Phys.~{\bf#1},\  #2 (#3)}
\def\IJMP #1 #2 #3 {Int.~J.~Mod.~Phys.~{\bf#1},\  #2 (#3)}
\def\waikoloa{%
{\it Proceedings of the 2nd International Workshop on
``Physics and Experiments with Linear $\epem$ Colliders''},
Waikoloa, HI, 1993,
eds. F. Harris, S. Olsen, S. Pakvasa and X. Tata, Waikoloa,
(World Scientific, Singapore, 1994)}

\end{document}